\documentclass[fleqn,usenatbib]{mnras}

\usepackage{newtxtext,newtxmath}

\usepackage[T1]{fontenc}

\DeclareRobustCommand{\VAN}[3]{#2}
\let\VANthebibliography\thebibliography
\def\thebibliography{\DeclareRobustCommand{\VAN}[3]{##3}\VANthebibliography}

\usepackage{graphicx}	
\usepackage{amsmath}	

\title[Multi-wavelength POL-2/HAWC+ polarimetry of N2071]{The JCMT BISTRO Survey: Multi-wavelength polarimetry of bright regions in NGC 2071 in the far-infrared/submillimetre range, with POL-2 and HAWC+}

\author[L. Fanciullo et al.]{\parbox{\textwidth}{
Lapo Fanciullo$^{1}$\thanks{E-mail: lfanciullo.astro@gmail.com},
Francisca Kemper$^{2,1}$,
Kate Pattle$^{3}$,
Patrick M. Koch$^{1}$,
Sarah Sadavoy$^{4}$,
Simon Coud\'{e}$^{5}$,
Archana Soam$^{32,5}$,
Thiem Hoang$^{6}$,
Takashi Onaka$^{7,8}$,
Valentin J.~M.~Le~Gouellec$^{5,9}$,
Doris Arzoumanian$^{10,27}$,
David Berry$^{11}$,
Chakali Eswaraiah$^{12,33}$,
Eun Jung Chung$^{13}$,
Ray Furuya$^{14}$,
Charles L.~H.~Hull$^{15,16,17}$,
Jihye Hwang$^{6,18}$,
Douglas Johnstone$^{19,20}$,
Ji-hyun Kang$^{6}$,
Kyoung Hee Kim$^{6}$,
Florian Kirchschlager$^{21}$,
Vera K\"{o}nyves$^{22}$,
Jungmi Kwon$^{8}$,
Woojin Kwon$^{23,24}$,
Shih-Ping Lai$^{25}$,
Chang Won Lee$^{6,18}$,
Tie Liu$^{26}$,
A-Ran Lyo$^{6}$, 
Ian Stephens$^{34,35}$,
Motohide Tamura$^{8,27,30}$,
Xindi Tang$^{28}$,
Derek Ward-Thompson$^{22}$,
Anthony Whitworth$^{29}$,
Hiroko Shinnaga$^{31}$
}
\\
\\
\textit{The authors' affiliations are shown in Appendix~\ref{affiliations}.}
}

\date{Accepted XXX. Received YYY; in original form ZZZ}

\pubyear{2021}

\begin{document}
\label{firstpage}
\pagerange{\pageref{firstpage}--\pageref{lastpage}}
\maketitle

\begin{abstract}
Polarized dust emission is a key tracer in the study of interstellar medium and of star formation. The observed polarization, however, is a product of magnetic field structure, dust grain properties and grain alignment efficiency, as well as their variations in the line of sight, making it difficult to interpret polarization unambiguously. The comparison of polarimetry at multiple wavelengths is a possible way of mitigating this problem. We use data from HAWC+/SOFIA and from SCUBA-2/POL-2 (from the BISTRO survey) to analyse the NGC 2071 molecular cloud at 154, 214 and 850~$\mu$m. The polarization angle changes significantly with wavelength over part of NGC 2071, suggesting a change in magnetic field morphology on the line of sight as each wavelength best traces different dust populations. Other possible explanations are the existence of more than one polarization mechanism in the cloud or scattering from very large grains. The observed change of polarization fraction with wavelength, and the 214-to-154 $\mu$m polarization ratio in particular, are difficult to reproduce with current dust models under the assumption of uniform alignment efficiency. We also show that the standard procedure of using monochromatic intensity as a proxy for column density may produce spurious results at HAWC+ wavelengths. Using both long-wavelength (POL-2, 850~$\mu$m) and short-wavelength (HAWC+, $\lesssim 200\, \mu$m) polarimetry is key in obtaining these results. This study clearly shows the importance of multi-wavelength polarimetry at submillimeter bands to understand the dust properties of molecular clouds and the relationship between magnetic field and star formation.
\end{abstract}

\begin{keywords}
ISM: individual objects (NGC 2071) -- polarization -- ISM: magnetic fields -- ISM: clouds -- submillimetre: ISM
\end{keywords}



\section{Introduction}
\label{sec:intro}


The thermal emission of interstellar dust has an essential role in modern astrophysics, as a tracer of the interstellar medium (ISM) and its physical conditions. Dust polarization, in particular, is essential for the study of interstellar magnetic fields and their role in star formation. Non-spherical dust grains tend to be aligned with their axes of maximum inertia (i.e. their shorter axes) parallel to the local magnetic field lines \citep{Davis+Greenstein51}. This follows from two phenomena: an ``internal alignment'' which aligns a grain's axis of maximum inertia with its rotation axis \citep[][]{Purcell79}; and an ``external alignment'' between the rotation axis and the magnetic field lines \citep[e.g.][]{Davis+Greenstein51, Lazarian+Hoang07}. Dust thermal emission is preferentially polarized along the long axes of the grains, and therefore can be used to trace magnetic field orientation \citep[e.g.][]{Hildebrand88, PlanckIR_XXXV}. Spatial variations in the dust polarization angle can also be used to estimate the magnetic field strength via the Davis-Chandrasekhar-Fermi method \citep{Davis51, Chandra+Fermi53}. 

In addition to the polarization angle, an important observable is the polarization fraction $P$. The value of $P$, however, is determined by several factors which are degenerate and hard to disentangle: the orientation and degree of disorder in the magnetic field, the grain alignment efficiency, and the optical properties of dust itself. The polarization fraction for a uniform environment, i.e. one with a uniform dust population and with a constant intensity of the interstellar radiation field, can be expressed as \citep[e.g.][]{Lee+Draine85}:
\begin{equation}
    P = P_{\rm intr} \, R \, F_{\rm dis} \, \cos^2{\gamma},
	\label{eq:P_decomposition}
\end{equation}
where $P_{\rm intr}$ is the ``intrinsic'' polarization determined by the dust properties, $R~=~\frac{3}{2} \langle \cos^2{\beta} \rangle - \frac{1}{2}$ is the Rayleigh reduction factor that accounts for imperfect alignment \citep{Greenberg68}, $\beta$ is the angle difference between magnetic field lines and the symmetry axis of the grains, $F_{\rm dis}~=~\frac{3}{2} \langle \cos^2{\phi} \rangle - \frac{1}{2}$ is another reduction factor that accounts for magnetic field disorder (assuming that the field can be decomposed into the sum of a uniform and a random component, $\phi$ being the variable angle between the two), and $\gamma$ is the angle between the magnetic field lines and the plane of the sky. 
This degeneracy between dust properties, alignment, and magnetic field structure is the reason why the measurement of $P$ on its own provides limited information. However, $F_{\rm dis}$ and $\gamma$ from equation~\ref{eq:P_decomposition} are independent of wavelength to first order, and  they do not influence the wavelength dependence of $P$. Therefore, the shape of the polarization spectrum is mainly determined by the optical properties and alignment of dust grains as long as that the condition of validity for equation~\ref{eq:P_decomposition} (uniform environment) is met.

The interpretation $P$ is important for instance in the estimation of dust alignment efficiency. The polarization fraction decreases with column density in molecular clouds \citep[e.g.][]{whittet+08, Alves+14, Jones+15}. This has been interpreted as supporting the Radiative Torques or RATs model of grain alignment \citep{Lazarian+Hoang07, Hoang+Lazarian16}, according to which grain alignment is driven by an asymmetric radiation field and is less efficient at the center of dark clouds. However, the decrease of $P$ with column density is partly caused by the magnetic field changing orientation on unresolved scales in dense regions. In the terms of equation~\ref{eq:P_decomposition}, dense regions have a low value of $F_{\rm dis}$ in addition to a low value of $R$. Being able to distinguish the effects of $F_{\rm dis}$ from those of $R$ would increase the usefulness of $P$ as a test of grain alignment models. 

A further complication is that equation~\ref{eq:P_decomposition} is only valid in uniform environments. This is generally not a good approximation for very dense environments such as dark cores or clouds with embedded sources. We should therefore expect that at least in the densest parts of molecular clouds the polarization spectrum is no longer independent of the magnetic field geometry. Even in this scenario, however, multi-wavelength polarimetry has the potential to break degeneracies in the interpretation of data. For instance, a change of polarization angle with wavelength would imply the existence of at least two dust populations with different temperatures and different orientations (due to a change in either the magnetic  field morphology or in the alignment mechanism along the line of sight). This information could help quantify the relative abundances of these dust populations with more precision than unpolarized dust emission.

This article combines data from the HAWC+ instrument \citep{hawc+ref} on board the Stratospheric Observatory for Infrared Astronomy \citep[SOFIA,][]{sofia_ref} and from the B-fields in Star-Forming Region Observations survey \citep[BISTRO,][]{W-T+2017} at the James Clerk Maxwell Telescope (JCMT). Our purpose is to test the potential of spatially resolved, multi-wavelength, multi-instrument polarimetry to explore dust properties, grain alignment and dust temperature effects in molecular clouds. The use of multi-wavelength polarized emission to explore dust and cloud properties has been pioneered by \citet{Hildebrand+99}, who observed the far-infrared (FIR) polarized spectra (at 60, 100 and 350~$\mu$m) in the envelopes of the M17 and Orion clouds. Their results show a decrease of polarization fraction with wavelength. Since the simplest dust models were expected to produce flat spectra, they proposed that this wavelength dependence may be caused by the presence on the line of sight of two different dust populations -- one warmer and strongly aligned, the other colder and poorly aligned. The support for this ``heterogeneous cloud'' model in subsequent studies has been mixed: polarization spectra observed from the ground \citep[e.g.][]{Vail_02, Vail+08, Vail+12} tend to show a polarization minimum at $\sim 350\, \mu$m that is consistent with the existence of at least two dust population of different temperature. On the other hand, no such minimum was found in the balloon-based observations of the Vela cloud  \citet{Gandilo+16}. It has been observed that these discrepant observations were taken at different resolution and cover different column density regimes (ground-based observations being limited to high column density). 
Another difficulty is that these observations are averaged over whole clouds, thus losing spatial resolution.\footnote{While the cited articles did use and analyse data in the form of maps, their cloud-averaged results -- e.g. the existence of a polarization minimum at $\sim 350\, \mu$m -- are usually the most cited.}

More recently, stratospheric observations from SOFIA have provided multi-wavelength, high-resolution polarimetry of molecular clouds and provided a more detailed picture of polarized dust emission. \citet{Santos+19} studied HAWC+ polarimetry of $\rho$ Oph at 89 and 154~$\mu$m (bands C and D), focusing on the ratio of the polarization fraction in the two bands: ${\mathcal R }_{DC} = P_D/P_C$. This ratio decreases with increasing column density, which is consistent with dust in dense regions being less well aligned. \citet{Michail+21} have performed a similar study in OMC-1, using HAWC+ polarimetry in all four available bands \citep[53, 89, 154 and 214~$\mu$m; observed by][]{Chuss+19}. They find that the polarization spectrum is flat at high temperature and has a negative slope at low temperature, while showing no clear trends with column density.  Further, they suggest that this is consistent with a heterogeneous cloud model where cold dust is poorly aligned, as is predicted by the RATs alignment model \citep[e.g.][]{Andersson+15}. Note that in the region studied by \citet{Santos+19} temperature and column density could not be disentangled. Covering a different wavelength range, \citet{Lyo+21} compared NGC 2071 polarimetry at 450 and 850~$\mu$m from BISTRO. They find that the ratio of polarization fractions $P_{850}/P_{450}$ is generally smaller than 1 in the cloud core, consistent with RATs alignment. However, this ratio increases slightly at the highest column density, which may indicate a change in grain properties.

While they provide a spatially-resolved view of the polarization spectrum, these recent studies are limited to a single instrument. Consequently, they explore a relatively narrow wavelength range, around the wavelength peak of dust emission (HAWC+, $\sim50 - 200\, \mu$m) or in the Rayleigh-Jeans regime (BISTRO, 450 and 850~$\mu$m). Combining BISTRO and HAWC+ data significantly increases the available wavelength range, and with it our ability to extract information from the data. For instance, experimental measurements show that the wavelength dependence of opacity for dust analogues changes markedly around 300-500 $\mu$m depending on their chemical composition \citep[e.g.][]{Demyk+17}. To check whether this has an effect on polarization we need to observe at both shorter and longer wavelengths. In cases where a strong temperature gradient is present, short and long wavelengths provide a complementary picture, since they are more sensitive to warm and cold dust respectively.

The present work focuses on NGC 2071 (hereafter N2071), a low-mass star-forming region in Orion B. We chose this region because it is one of the brightest observed in the BISTRO survey (after OMC-1 and $\rho$ Oph), which makes HAWC+ follow-up observations easier, while remaining for the most part optically thin, which simplifies the data interpretation. Furthermore, it is relatively nearby \citep[$417\pm5$ pc][]{Kounkel+18}, allowing to obtain a good spatial resolution.
The region contains at least eight embedded infrared sources identified as IRS 1 to 8 \citep{Walther+93}. It is also characterized by several outflows originating from the IR sources, the largest of which is aligned along the NE-SW direction and associated with IRS 3 \citep{Bally+82, Eisloeffel00}. The present paper uses BISTRO observations of N2071 reduced by \citet{Lyo+21} and combines them with HAWC+ observations of the same region commissioned by our team, which have a similar resolution ($\sim14\arcsec - 18\arcsec$). 

The paper is organized as follows: Section~\ref{sec:data} describes the the data and its observation, reduction and selection. Section~\ref{sec:res} examines how both the polarization angle $\theta$ and the polarization fraction $P$ change across bands in the observational data. Section~\ref{sec:discussion} discusses the results and presents a preliminary comparison with dust models. Finally, Section~\ref{sec:conclusions} summarizes our conclusions.

\section{Observations, data reduction and selection}
\label{sec:data}

\subsection{JCMT data}
\label{sec:jcmt}

The observation and reduction of BISTRO data for N2071, taken with the POL-2/SCUBA-2 system on JCMT \citep[][]{Dempsey+13, Holland+13}, has been described in detail in \citet{Lyo+21}, and so we will give only a brief overview here. The region was observed twenty times between 8 September 2016 and 11 November 2017 for a total of 13.3 hours of integration time. The observations took place simultaneously at 450 and 850~$\mu$m. We elected to use only 850~$\mu$m polarimetry in the present paper due to the lower signal-to-noise ratio (S/N) and the uncertainties on the instrumental polarization of the 450 $\mu$m data. The area covered by the POL-2 observational field is shown in Figure~\ref{fig:map_all}.

The 850~$\mu$m data consist of maps of the Stokes parameters $I$, $Q$ and $U$ maps for the region centred on R.A. = 05$^h$47$^m$05.040$^s$, Dec. = 00${^\circ}$21'51\hbox{$\,.\!\!^{\prime\prime}$}7 (J2000), with a spatial resolution of 14\hbox{$\,.\!\!^{\prime\prime}$}1 (corresponding to $\sim$6000 AU or 0.03 pc at 417 pc). The maps were corrected for $^{12}$CO (3--2) contamination and instrumental polarization using \textit{Starlink}, as described in \citet{Lyo+21} and \citet{Parsons+18}. The original 4$\arcsec$ pixel maps used in \cite{Lyo+21} were binned to 8$\arcsec$ pixels (Table~\ref{tab:band_properties}). At this pixel size, the typical uncertainties at 850~$\mu$m are $\sim$0.006 mJy arcsec$^{-2}$ for Stokes $I$ and $\sim$0.004 mJy arcsec$^{-2}$ for Stokes $Q$ and $U$ (Table~\ref{tab:band_noise}), estimated from the median uncertainties in the central 6\hbox{$^{\prime}$} of the map. Outside of the central 6\hbox{$^{\prime}$} the data quality degrades rapidly, so we excluded that region from our analysis. The map of $I$ and polarization vectors at 850~$\mu$m is shown in the bottom row of Figure~\ref{fig:map_all}.

Since the notation for polarized quantities is not always consistent across the literature, we provide the following definitions of the symbols we use:
\begin{itemize}
    \item $I$ is the total intensity.
    \item $I_{\rm p}$ is the polarized intensity, equal to $\sqrt{Q^2 + U^2}$ before debiasing (see Section~\ref{sec:debias}).
    \item $P = I_{\rm p} / I$ is the polarization fraction, which we will often express as a percentage. When referring to $P$ for a specific band, we will add the band wavelength as a subscript; e.g., $P_{850}$.
    \item $\theta = 0.5 \arctan(U, Q)$ is the polarization angle as measured from north to East (i.e. the angle is positive moving counter-clockwise from North).
\end{itemize}

\subsection{HAWC+ observations}
\label{sec:sofia}

\begin{figure*}
\includegraphics[width=.99\columnwidth]{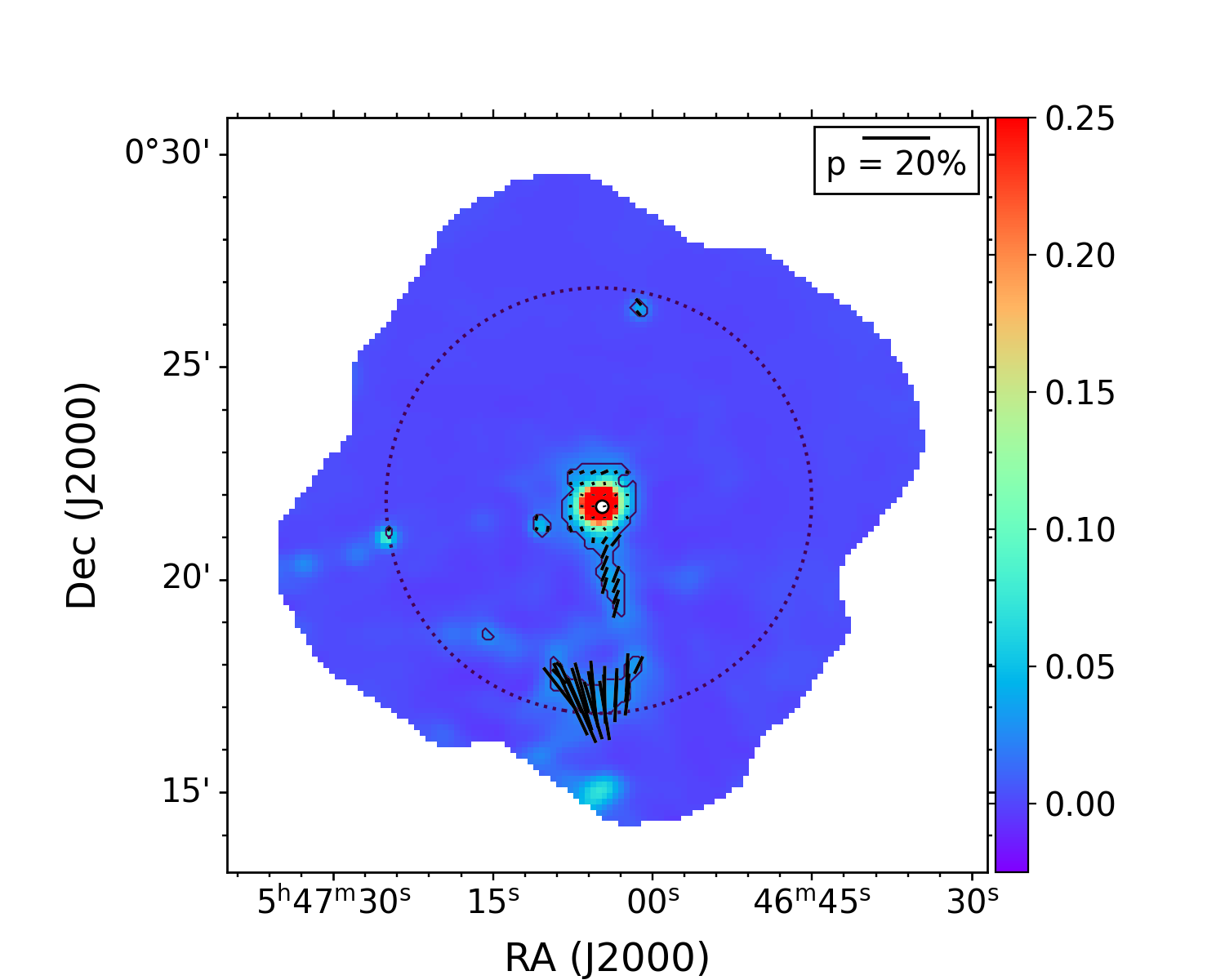}
\includegraphics[width=.99\columnwidth]{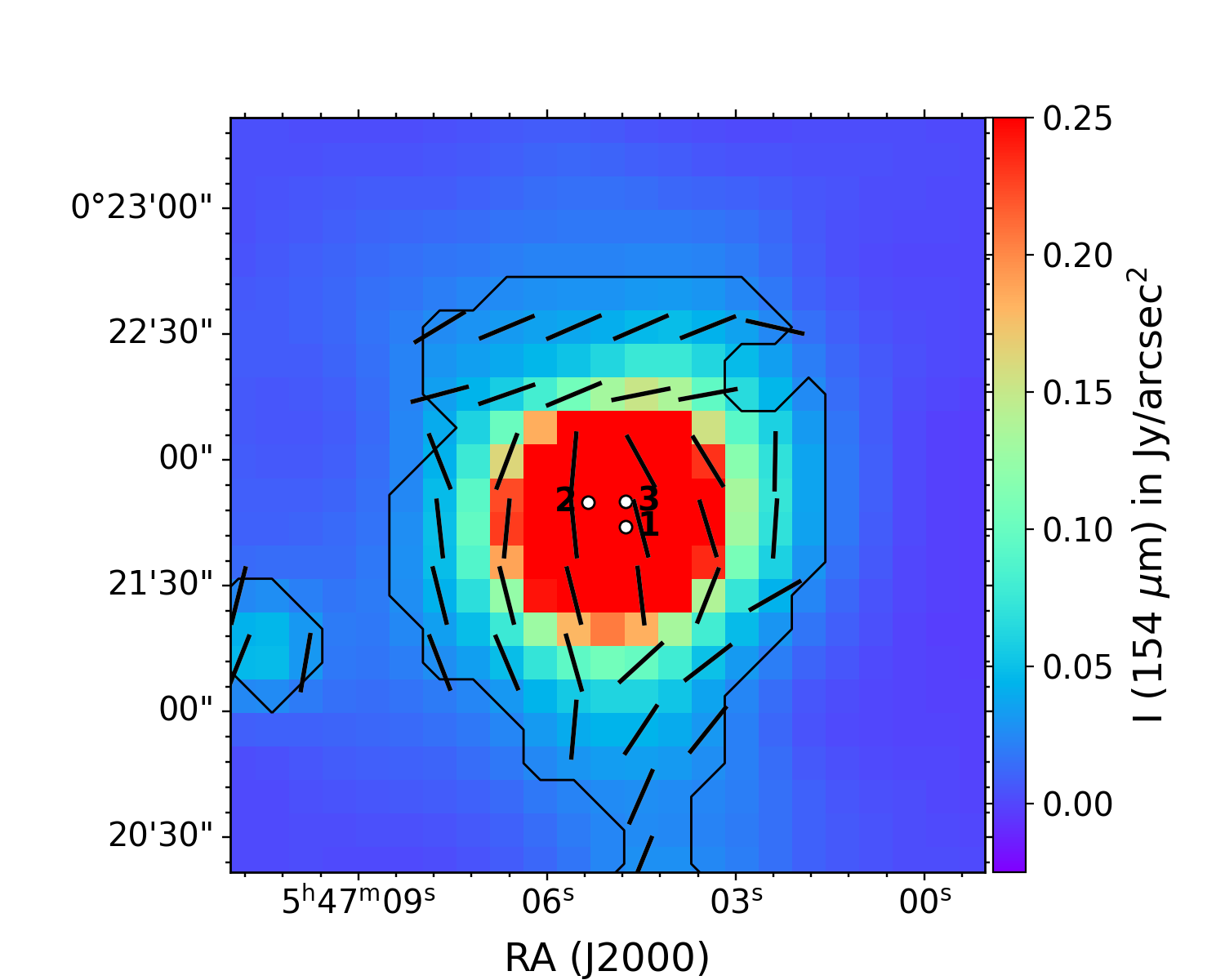}
\includegraphics[width=.99\columnwidth]{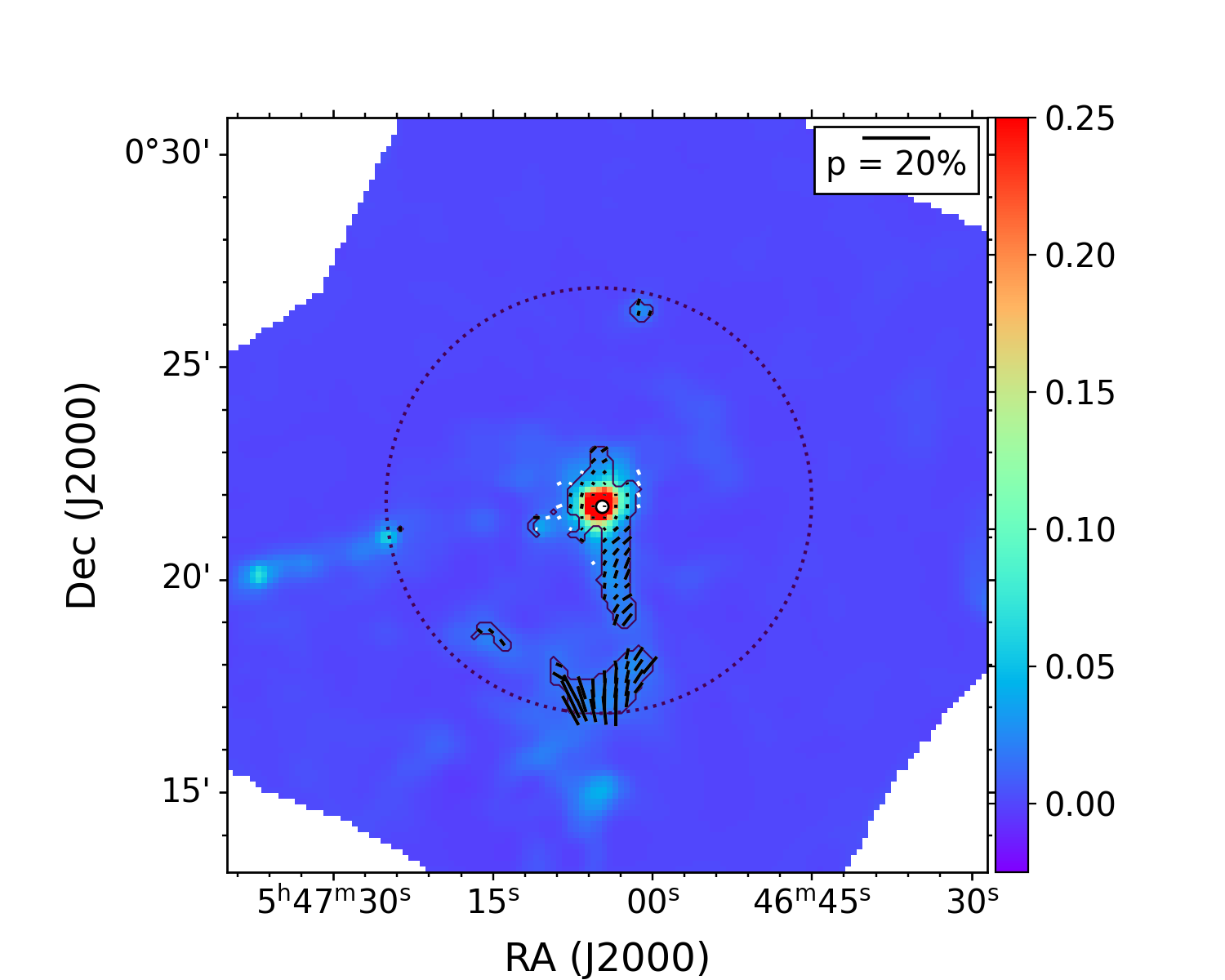}
\includegraphics[width=.99\columnwidth]{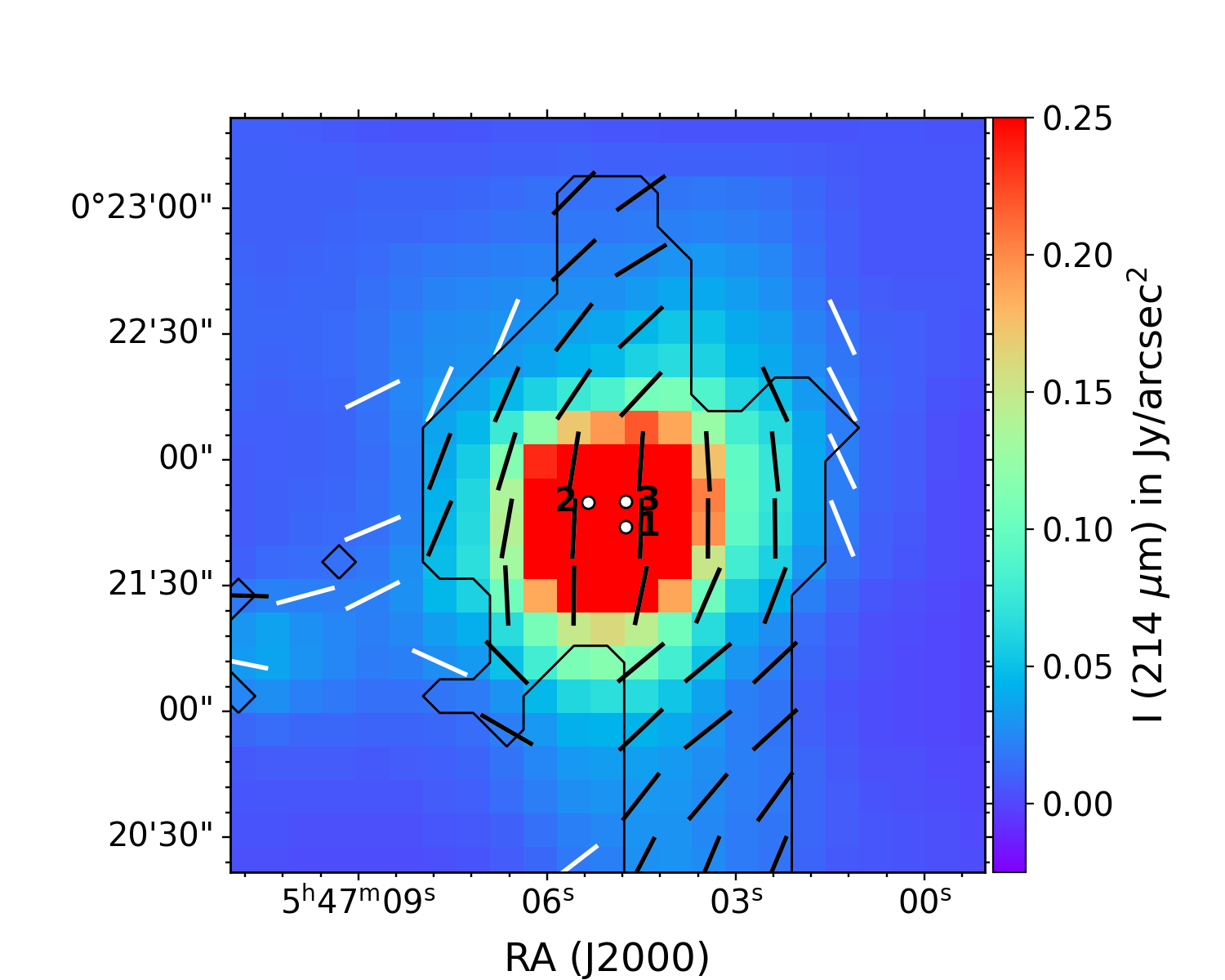}
\includegraphics[width=.99\columnwidth]{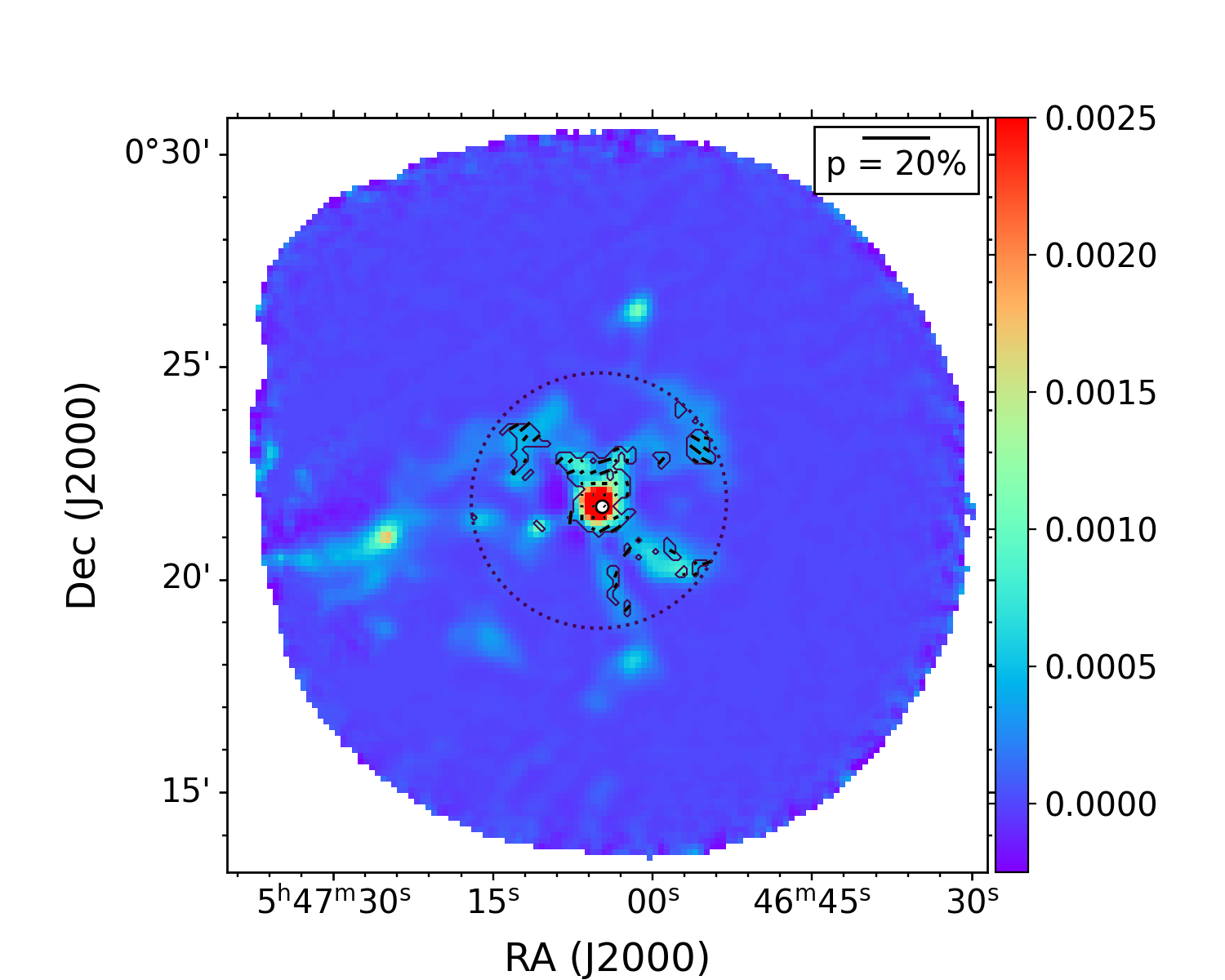}
\includegraphics[width=.99\columnwidth]{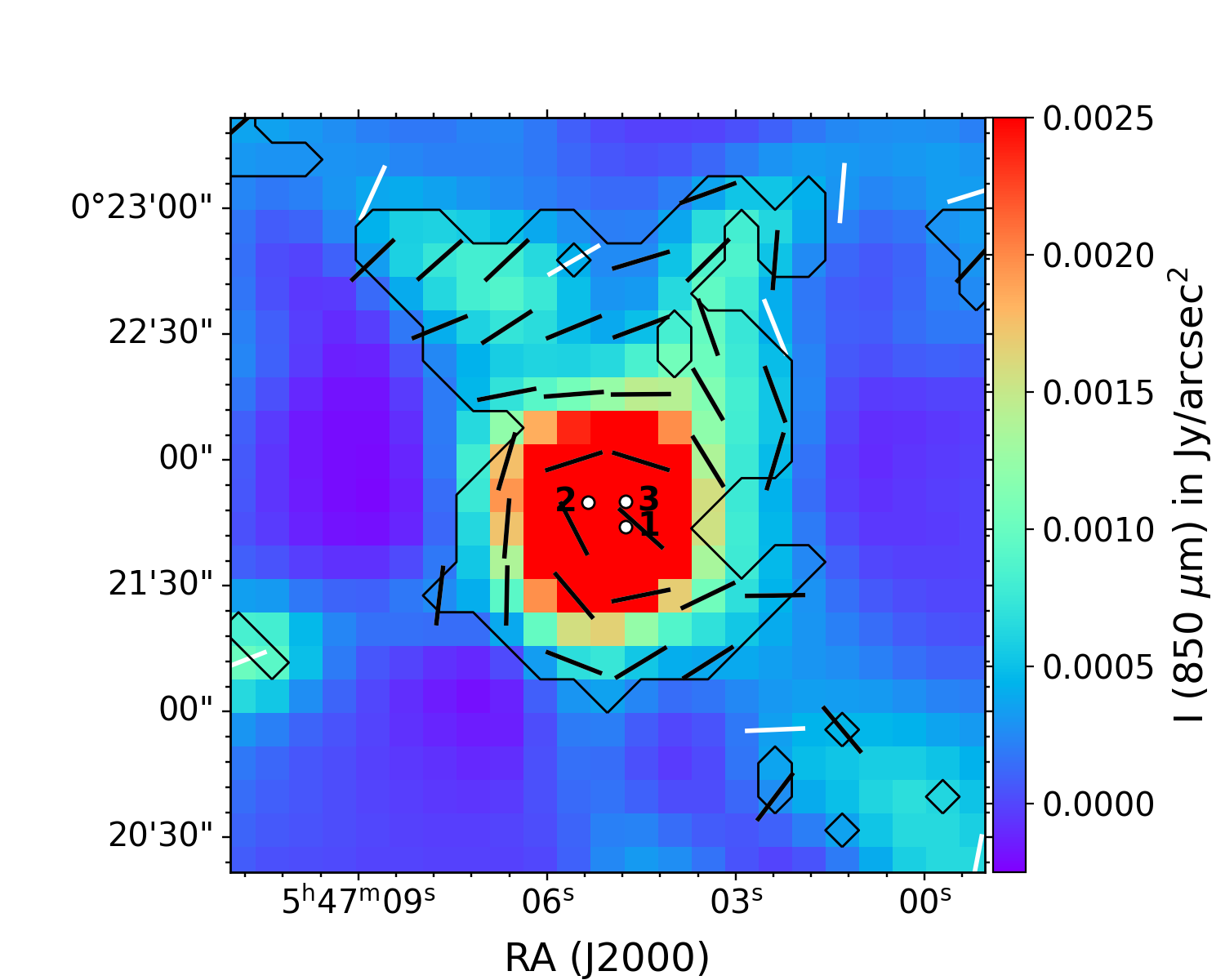}
\caption{
Maps of the N2071 Stokes $I$ and polarization vectors in the three available bands: HAWC+ D (top, 154 $\mu$m), HAWC+ E (middle, 214 $\mu$m) and POL-2 (bottom, 850~$\mu$m). The left column shows the full observational fields, with polarization vector length proportional to $P$. The right column shows a zoom on the central 3$\arcmin$ region, with polarization vectors of constant length. Black vectors have $P/\delta P > 3$, white vectors have $2 < P/\delta P < 3$. The absence of white vectors in band D is due to the fact that $P/\delta P > 3$ everywhere in the chosen area. The black contours show the area selected for analysis (see Section~\ref{sec:selection}). Dotted circles show the central 10$\arcmin$ (HAWC+ bands) and central 6$\arcmin$ (POL-2 band). All maps are centred on R.A.~=~05$^h$47$^m$05.040$^s$, Dec.~=~00${^\circ}$21'51\hbox{$\,.\!\!^{\prime\prime}$}7 (J2000). The side of the maps measures 15$\arcmin$ in the left column and 3$\arcmin$ in the right column. 
}
\label{fig:map_all}
\end{figure*}

Polarimetry of the area centred on the coordinates R.A. = 05$^h$47$^m$05.040$^s$, Dec. = 00${^\circ}$21'51\hbox{$\,.\!\!^{\prime\prime}$}7 (J2000) was observed on 2 October 2019 using HAWC+/SOFIA during flight HA\_615, as part of proposal 07\_130. The resulting data are published for the first time in the present article. We observed the region in bands D (central wavelength 154 $\mu$m) and E (214 $\mu$m), which have angular resolutions of, respectively, 14\hbox{$\,.\!\!^{\prime\prime}$}2 ($\sim$6000 AU) and 18\hbox{$\,.\!\!^{\prime\prime}$}9 ($\sim$8000 AU) (see Table~\ref{tab:band_properties}). Note that these beam sizes are slightly larger than the nominal values for bands D and E (13\hbox{$\,.\!\!^{\prime\prime}$}8 and 18\hbox{$\,.\!\!^{\prime\prime}$}2, respectively) because of additional polarimetry data smoothing in the data reduction process. The observations were performed using the on-the-fly-mapping configuration, which was offered for the first time as shared-risk observations in Cycle 7 for HAWC+ polarimetry. The instrument's pixel size on the sky is 6\hbox{$\,.\!\!^{\prime\prime}$}90 in band D and 9\hbox{$\,.\!\!^{\prime\prime}$}37 in band E, although the level 4 (science-ready) maps are regridded to pixel sizes of 3\hbox{$\,.\!\!^{\prime\prime}$}40 and 4\hbox{$\,.\!\!^{\prime\prime}$}55, respectively. At this pixel size, typical uncertainties on the Stokes parameters are 0.084 mJy arcsec$^{-2}$ (0.148 mJy arcsec$^{-2}$) for $I$ ($Q$ and $U$) at 154 $\mu$m and 0.063 (0.085) mJy arcsec$^{-2}$ for $I$ ($Q$ and $U$) at 214 $\mu$m (Table~\ref{tab:band_noise}). These values have been estimated from the median of the central 10$\arcmin$ of the uncertainty map to exclude low-quality data at the map edges. The maps of $I$ and polarization vectors at for the two bands are shown in the two top rows of Figure~\ref{fig:map_all}.

The level 4 data for the HAWC+ observations also include the polarization intensity $I_{\rm p}$, polarization angle $\theta$ and polarization fraction $P$ (including a debiased version, see Section~\ref{sec:debias}) and the respective uncertainties \citep{Gordon+18}. However, to keep the smoothing and regridding procedure as close as possible to that for POL-2 data (see Section~\ref{sec:smoothing}), in the present paper we use the Stokes vectors to recalculate $I_{\rm p}$, $\theta$ and $P$ on our own.

\subsection{Data treatment and selection}
\label{sec:data_final_treatment}

\subsubsection{Smoothing and resampling}
\label{sec:smoothing}

The POL-2 and HAWC+ data is in the form of $I$, $Q$ and $U$ maps in three different bands, centred at wavelengths of 154~$\mu$m, 214~$\mu$m and 850~$\mu$m. Each map has a different beam and pixel size (see Table~\ref{tab:band_properties}). Therefore, to make a pixel-to-pixel comparison the maps need to be smoothed to a common resolution and regridded to a common pixel size. This was done using the \textit{Starlink} software \citep{starlink_ref}. The maps were resampled to the largest pixel size (that of POL-2, 8$\arcsec$) using the \textit{wcsalign} routine. This also regridded the HAWC+ bands to the same World Coordinate System as the POL-2 observations. The resampling used the \textit{Nearest} method, which gives the most reliable values for the resampled variance. The parameter \textit{conserve} was set to \textit{no} to preserve the units of Jy arcsec$^{-2}$. After resampling, we converted all bands to the largest beam size (that of HAWC+ E at 214 $\mu$m, with an 18\hbox{$\,.\!\!^{\prime\prime}$}9 FWHM) using the \textit{Gausmooth} routine. The routine convolved the maps using a Gaussian kernel with FWHM~=~$\sqrt{beam_{\rm final}^2 - beam_{\rm init}^2}$, with beams expressed in pixels. All procedures were performed on the $I$, $Q$, $U$ maps and those of their covariances. The \textit{Gausmooth} routines treat noise as uncorrelated between pixels. This is a reasonable first-order approximation when the noise is a product of receiver electronics. Furthermore, we only applied the smoothed after increasing the pixel size to 8$\arcsec$, i.e. approximately Nyquist-sampled, to avoid oversmoothing.

\begin{table}
\centering
\caption{Properties of the three bands used in this study.}
\label{tab:band_properties}
\begin{tabular}{lccc} 
\hline
Instrument (Band) & Central $\lambda$ & Beam FWHM & Pixel size\\
 & ($\mu$m) & (arcsec) & (arcsec)\\
\hline
HAWC+ (D) & 154 & 14.16 & 3.40\\
HAWC+ (E) & 214 & 18.89 & 4.55\\
POL-2 (850) & 850 & 14.1 & 8.00\\
\hline
\end{tabular}
\end{table}

\begin{table}
\centering
\caption{
Median uncertainties for the three bands, before and after smoothing to a common beam of 18\hbox{$\,.\!\!^{\prime\prime}$}9 and resampling to 8$\arcsec$. Estimated from the median uncertainty on each band. Units are mJy arcsec$^{-2}$.
}
\label{tab:band_noise}
\begin{tabular}{lccc} 
\hline
Instrument (Band) & Stokes  & Original & Smoothed \\
 & parameter &  & and resampled\\
\hline
HAWC+ (D, 154~$\mu$m)  & $I$    & 0.084  & 0.035\\
           & $Q, U$ & 0.148  & 0.052\\
HAWC+ (E, 214~$\mu$m)  & $I$    & 0.063  & 0.063\\
           & $Q, U$ & 0.085  & 0.085\\
POL-2 (850~$\mu$m) & $I$    & 0.0062 & 0.0027\\
           & $Q, U$ & 0.0042 & 0.0018\\
\hline
\end{tabular}
\end{table}

One effect of the smoothing and regridding was to significantly decrease the noise on the maps. The new typical uncertainties in $I$ ($Q$ and $U$) maps are $\sim$0.035 ($\sim$0.052) mJy arcsec$^{-2}$ for HAWC+ at 154~$\mu$m and $\sim$0.0027 ($\sim$0.0018) mJy arcsec$^{-2}$ for POL-2 at 850~$\mu$m, showing a noise reduction of a factor $\sim$2-3. The uncertainties on the 214 $\mu$m map are virtually unchanged, since it is unsmoothed (Table~\ref{tab:band_noise}). 

In the rest of this paper, when we mention the data, we will be referring to the smoothed and resampled version unless otherwise specified.

\subsubsection{Debiasing of polarized quantities}
\label{sec:debias}

The main quantities of interest in the study of polarization are the polarization fraction $P$ and the polarization angle $\theta$. One particularity of $P$ is that, being the result of a sum in quadrature, it is positively biased, i.e. the measured value is larger than the real one.\footnote{This is most evident for the case $P = 0$. Assuming measurement errors $\delta Q$ and $\delta U$ on Q and U respectively, the measured value of $P$ will be the non-zero quantity $\sqrt{(\delta Q)^2 + (\delta U)^2}$.} It is common to correct, or ``debias'', the polarization fraction to obtain a non-biased estimator of polarization. One of the simplest estimators is the one proposed by \citet{Wardle+Kronberg74}: $P_{\rm db} = \sqrt{P^2 - \delta_P^2}$. This is the default debiasing method for BISTRO data and the one used in the present paper. All mentions of polarization will refer to the debiased quantity unless otherwise indicated. 

\subsubsection{Selection criteria}
\label{sec:selection}

We apply a quality selection to the data by adopting only the polarization vectors with $I/\delta I \geq 20$, $P/\delta P \geq 3$ and $\delta P \leq 5\%$ on each band. Additionally, only the central 6$\arcmin$ of the POL-2 map are used, since the data quality significantly decreases outside of it. The quality of the HAWC+ maps is more uniform, but it still drops near the map edges, so we limit our analysis of the HAWC+ data to the central 10$\arcmin$. The final selection in all three bands is shown in Figure~\ref{fig:map_all}. 

An important issue when analysing multi-instrument data is the difference in zero-points, which severely complicates the comparison in low-intensity regions. Polarimetric observations by POL-2 are characterized by the loss of large-scale structures, mainly affecting regions of low-intensity, extended emission. While HAWC+ does not suffer from the same losses, its observations are  background-subtracted, which also affect extended emission in a different way. An additional complication with low-intensity regions is that a significant fraction of the 154~$\mu$m map has unusually high values of $P$ ($\gtrsim$30\%), which may be due to having $I \sim 0$ at the denominator of $P = I_{\rm p}/I$. We have therefore decided to leave the analysis of extended emission for a follow-up and limit the present paper to bright compact regions. We do this by applying an intensity threshold to the data and, in the case of 850~$\mu$m, an additional spatial selection. We chose a threshold of $25$ mJy arcsec$^{-2}$ in the HAWC+ band D at 154 $\mu$m, which is several times higher than both the median (4.0 mJy arcsec$^{-2}$) and the mean intensity (6.1 mJy arcsec$^{-2}$) of the map. Compact emission at 154~$\mu$m is about twice as bright than at 214~$\mu$m and two orders of magnitude brighter than at 850~$\mu$m, as measured at the intensity peak. We therefore applied thresholds of 15 mJy arcsec$^{-2}$ to HAWC+ band E at 214 $\mu$m and 0.25 mJy arcsec$^{-2}$ to POL-2 at 850~$\mu$m. These thresholds cuts out most of the extended emission while retaining some filaments in addition to the central hub in each band. In the rest of this paper, if not otherwise specified, we will be referring to the data selected as described above.

Since the thresholds cuts on $I$ still leave long filaments in the 850~$\mu$m map (see Figure~\ref{fig:map_all}), we decided to perform an additional spatial selection when comparing HAWC+ and POL-2 data. For an estimate of the scales where POL-2 flux loss becomes important, we refer to \citet{Kirk+18}, who processed synthetic sources using SCUBA-2 data reduction to estimate losses as a function of source intensity and size. For sources that peak at 20 times the value the local rms noise, losses become important above sizes of $75 - 100\arcsec$. Our 850~$\mu$m $P$ map has a maximum S/N of $\sim26$ at its peak. We therefore limit comparisons between HAWC+ and POL-2 to within 45$\arcsec$ from the location of the 850~$\mu$m emission peak, R.A. = 05$^h$47$^m$05.040$^s$, Dec. = 00${^\circ}$21'43\hbox{$\,.\!\!^{\prime\prime}$}7 (J2000),\footnote{The R.A. of the peak and that of the observational field center coincide on our map because they are less than one pixel distant.}
i.e. to a 90$\arcsec$ wide disc. Our choice of radius is based on the assumption that the results of \citet{Kirk+18} for unpolarized observations remain valid for polarized structures. Preliminary attempts by the authors to apply similar processing to polarized data, as described in Appendix~\ref{sec:filtered}, seem to support this assumption.

\subsection{Ancillary data}
\label{sec:ancillary}

To explore the correlation between observed polarization and local ISM conditions we acquired maps of the molecular hydrogen column density $N_{\rm H_{2}}$ and the line of sight averaged dust temperature $T_{\rm d}$ for Orion B, as obtained from the \textit{Herschel} Gould Belt Survey archive \citep[HGBS;][]{Andre+10}. 
Maps of $N_{\rm H_{2}}$ and $T_{\rm d}$ in N2071 were produced by \citet{Konyves+20} from a modified blackbody fit to HGBS data, assuming a mean molecular weight of 2.8 and a dust opacity per unit mass (dust + gas) of 0.1 cm$^{-2}$ g$^{-1}$ at 300 $\mu$m, with a $\lambda^{-2}$ dependence on wavelength \citep[see appendix A of][]{Palmeirim+13}. 
We used these values to convert $N_{\rm H_{2}}$ back into dust optical depth ($\tau$) at 154, 214 and 850 $\mu$m. The value of $\tau$ at each wavelength is therefore a re-scaling of the molecular hydrogen column density, with $N_{\rm H_{2}} = 10^{22}\, {\rm cm}^{-2}$ corresponding to $\tau = 1.8 \times 10^{-2}$ at 154 $\mu$m, $\tau = 9.2 \times 10^{-3}$ at 214 $\mu$m and $\tau = 5.8 \times 10^{-4}$ at 850~$\mu$m.
The maps have a resolution of 36\hbox{$\,.\!\!^{\prime\prime}$}3 for $T_{\rm d}$ and 18\hbox{$\,.\!\!^{\prime\prime}$}2 for $N_{\rm H_{2}}$ (equivalent to the HAWC+ band E resolution), so smoothing was not necessary. However, both maps have been converted to 8$\arcsec$ pixels. 

\begin{figure}
\includegraphics[width=\columnwidth]{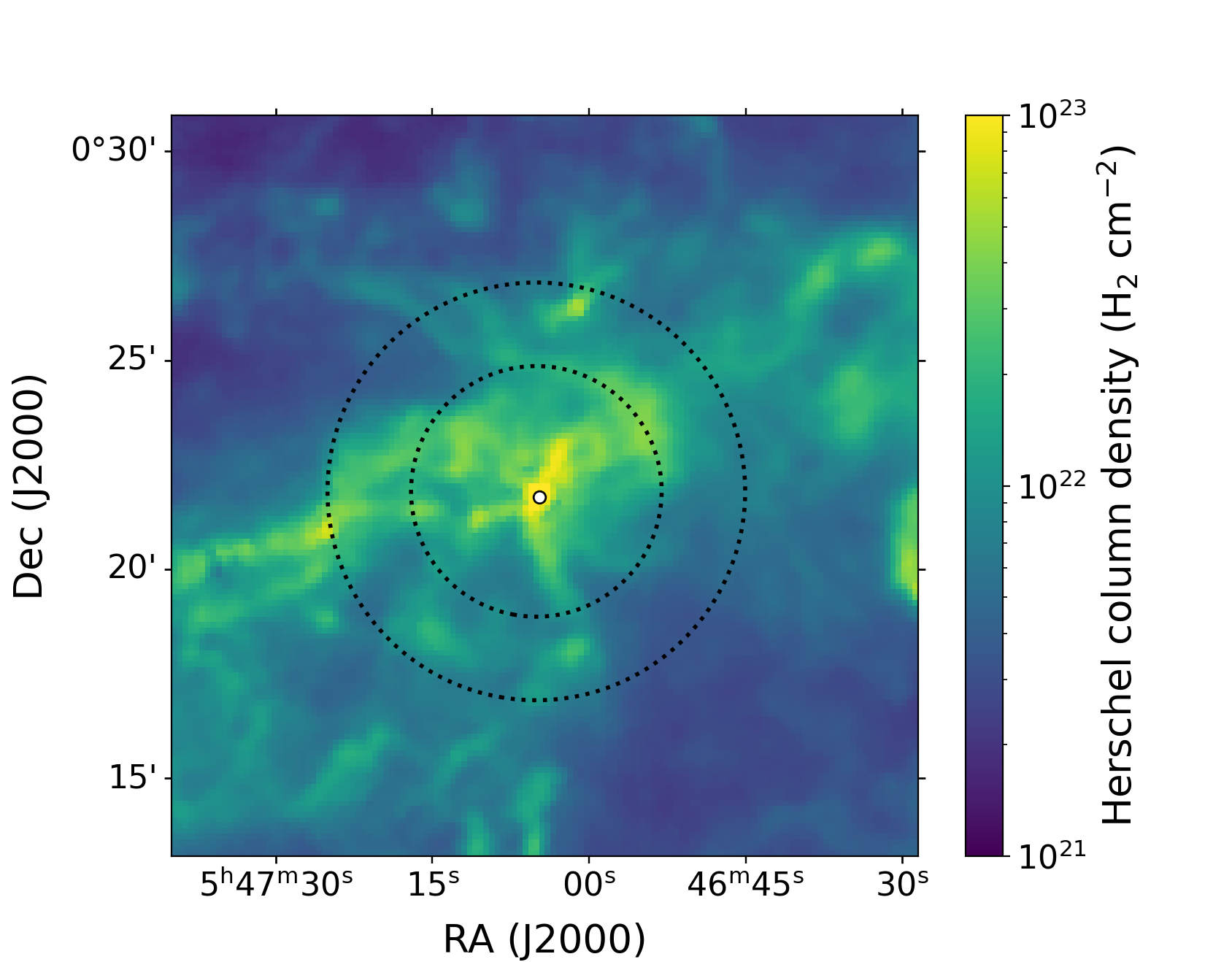}
\includegraphics[width=\columnwidth]{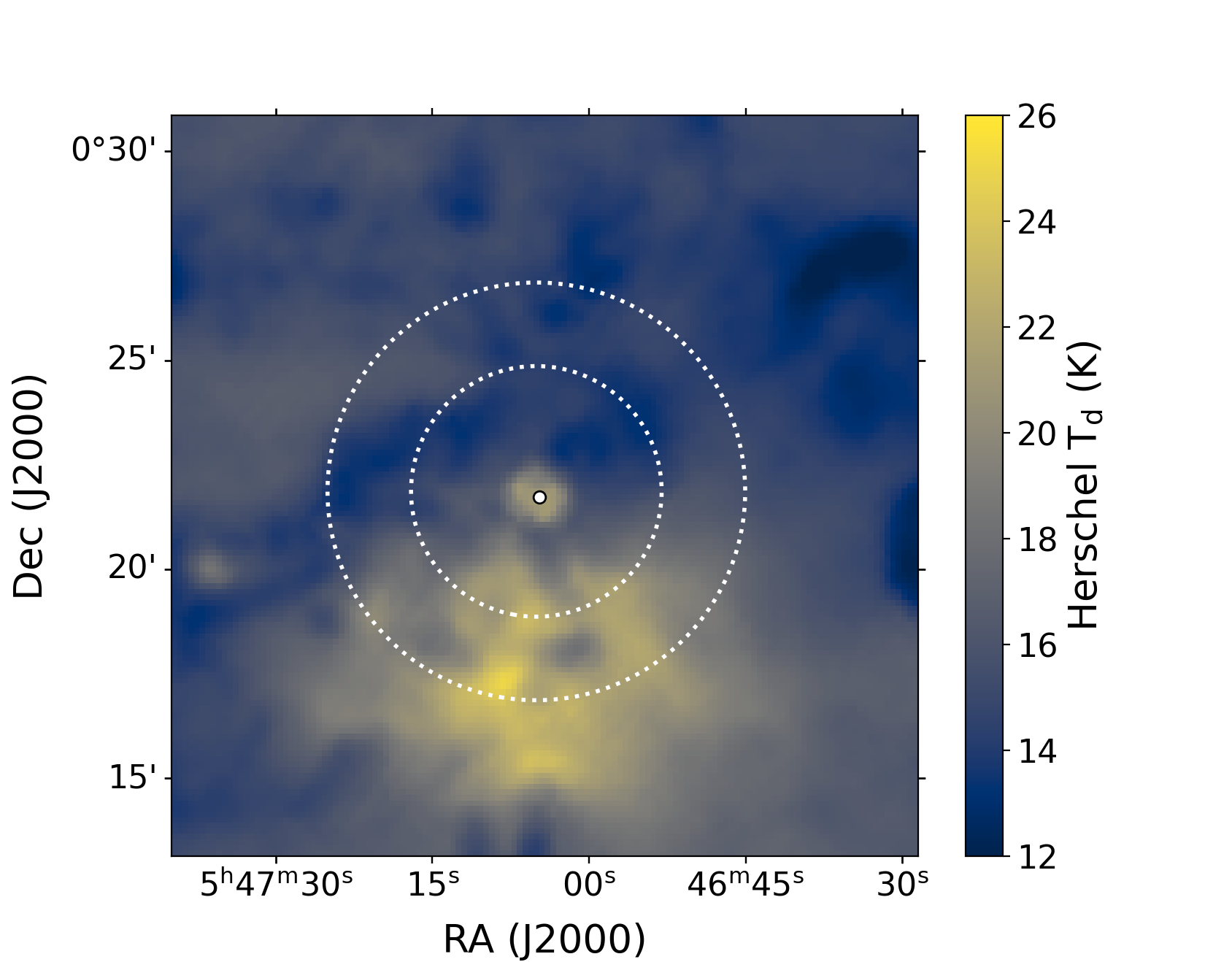}
\caption{
Maps of N2071 column density (top) and temperature (bottom) from HGBS. Dotted circles show the central 10$\arcmin$ and 6$\arcmin$ regions, to provide a comparison with Figure~\ref{fig:map_all}. The white dot shows the position of the IR source IRS 1 (IRS 2 and 3 would overlap with the symbol on this scale). 
}
\label{fig:selected_areas}
\end{figure}

While $N_{\rm H_{2}}$ and $T_{\rm d}$ are mostly anticorrelated in this region, the central column density peak appears to be warmer than the surrounding environment, as can be intuitively expected from the presence of embedded infrared sources (see Figure~\ref{fig:selected_areas}). Note, however, that a single-temperature, constant-power-law fit is unlikely to be an accurate representation of dust near the center of the cloud. Therefore, the values of $T_{\rm d}$ and $N_{\rm H_{2}}$ should be taken as a first approximation.

\section{Results}
\label{sec:res}

\subsection{Maps}
\label{sec:maps}

The final N2071 maps are shown in Figure~\ref{fig:map_all}. All maps have been smoothed to an 18\hbox{$\,.\!\!^{\prime\prime}$}9 FWHM beam and resampled to 8$\arcsec$ pixels. The HAWC+ band D and E maps are plotted with the same scale in intensity, capped to 250 mJy arcsec$^{-2}$. This is lower than the peak emission, since otherwise structure would be too faint to be visible in the figure. In the case of the POL-2 map, because of its lower intensity, we used a cap of 2.5 mJy arcsec$^{-2}$. Black polarization vectors are shown only where the data satisfies all the selection conditions presented in Section~\ref{sec:selection}. White vectors show data where the condition on polarization is $2 < P/\delta P < 3$ rather than $P/\delta P > 3$. The vectors show the polarization angle $\theta$. While it is common procedure to rotate $\theta$ by 90$^\circ$ in the plot to show the inferred direction of the magnetic field, we decided to show our polarization vectors unrotated. This is because $\theta$ is not identical in all bands (see Section~\ref{sec:polangle}), and it is not a priori evident which band's $\theta$ would be most representative of the magnetic field orientation. Since the map pixel size is smaller than the beam, we only show every other polarization vector to avoid oversampling and giving a false impression of smoothness in the observed field. The distance between polarization vectors in Figure~\ref{fig:map_all} is therefore 16$\arcsec$, close to the FWHM of the beams. See also Appendix~\ref{sec:3bdpol} for a map comparing polarization angles in all three bands. 

All maps show a strong central emission peak in $I$, but the structure surrounding it changes significantly between the HAWC+ and POL-2 bands. At 154 and 214~$\mu$m the main polarized structure visible outside of the central peak is a southward-pointing filament. At 850~$\mu$m, several more filaments are visible around the peak, which is revealed as a hub. This may be a consequence of dust in the centre of N2071 being warmed by the IR sources, which would enhance the central emission, especially at short wavelengths. Therefore, the polarized structures visible in all three bands can be classified as a ``central hub'' and a ``southward-pointing filament''.

\subsection{Change of the polarization angle $\theta$ with wavelength}
\label{sec:polangle}

In this section we conduct a comparison of the polarization angle across all three bands.
For each line of sight, if we are observing the same dust at all wavelengths and the grains are in the Rayleigh regime (small grain,  which is what we would expect of sub-$\mu$m or $\mu$m grains at submillimeter wavelengths), the polarization angles should be the same in all bands. 
If the polarisation angles differ significantly between bands, then this result would indicate that the dust grains are much larger and not in the Rayleigh regime, or that we are not tracing the same dust populations in the three wavelengths. Different dust populations could be aligned with differently-oriented magnetic fields, or be aligned with something other than the magnetic field \citep[as is the case for k-RATs alignment:][see also Section~\ref{sec:discussion_polangle}]{Lazarian+Hoang07, Tazaki+17}. Anisotropic radiation could also cause polarized thermal emission independently of the magnetic field \citep[PTEAR,][Section~\ref{sec:discussion_polangle}]{Onaka+95, Onaka+00}. It is possible that along some lines of sight, the three bands will have similar polarisation angles by coincidence even though they are tracing different dust populations.  Nevertheless, we feel that this possibility is unlikely given the complex magnetic field structure of molecular clouds.

\begin{figure}
    \centering
    \includegraphics[width=\columnwidth]{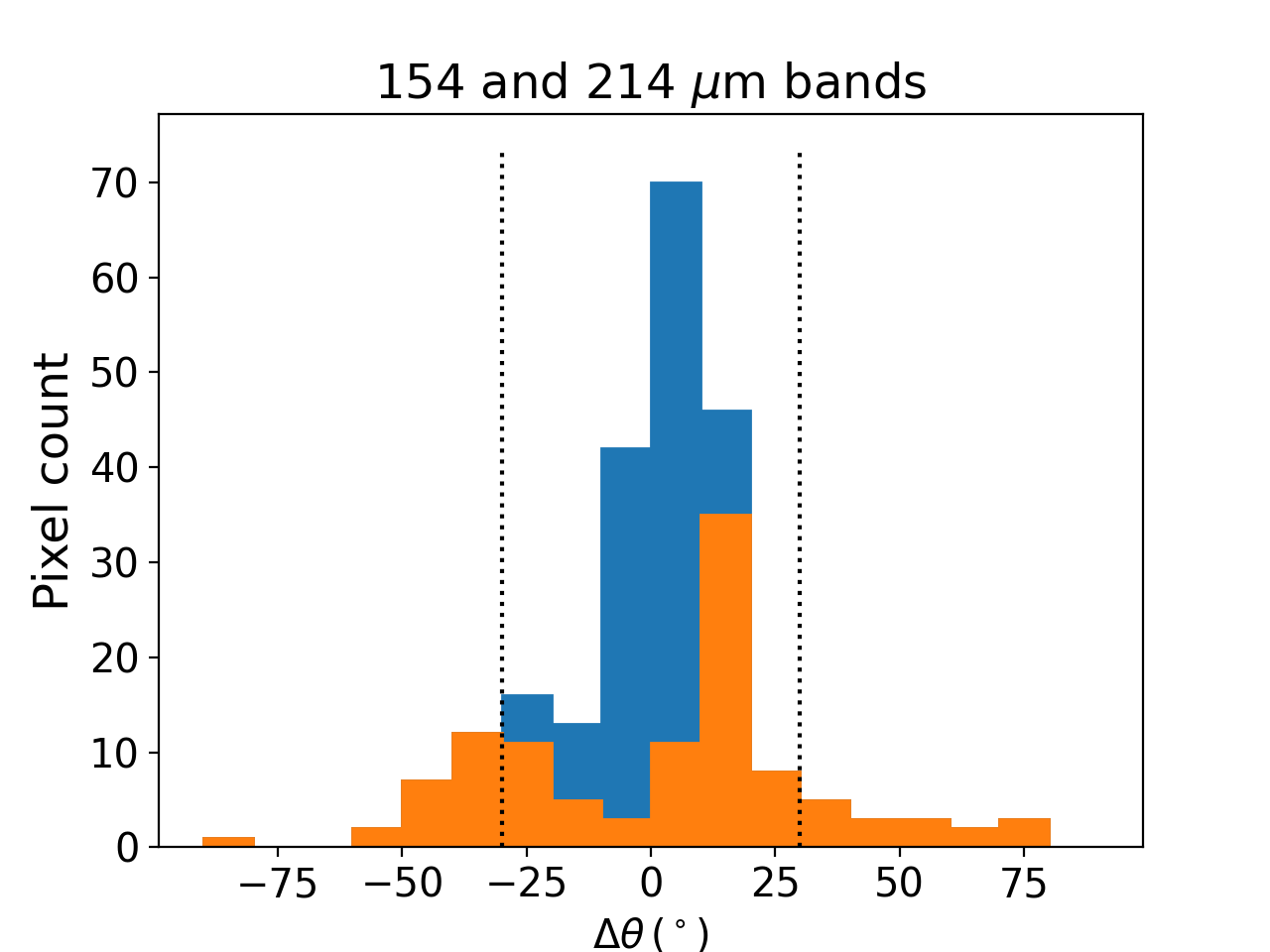}
    \includegraphics[width=\columnwidth]{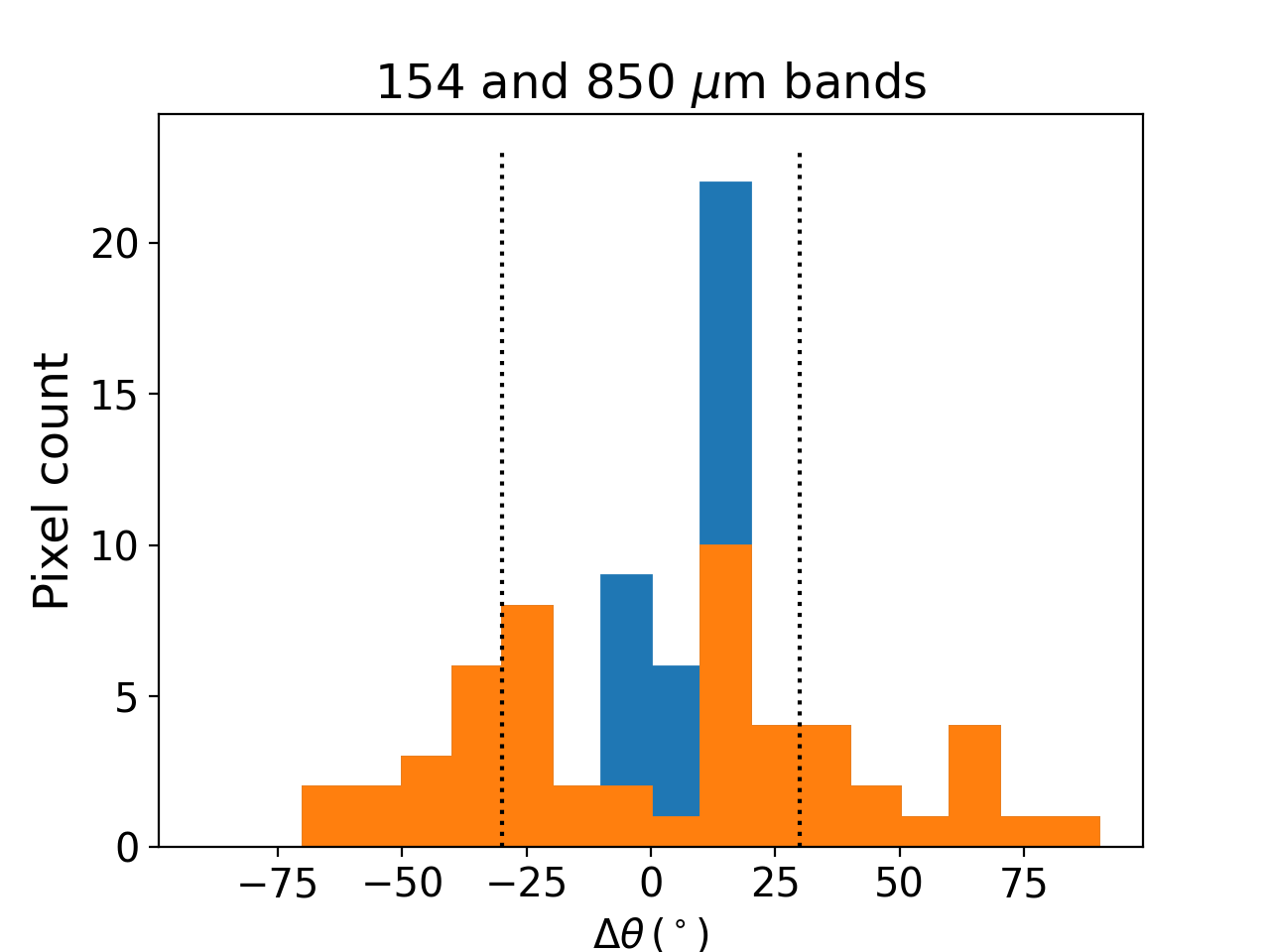}
    \includegraphics[width=\columnwidth]{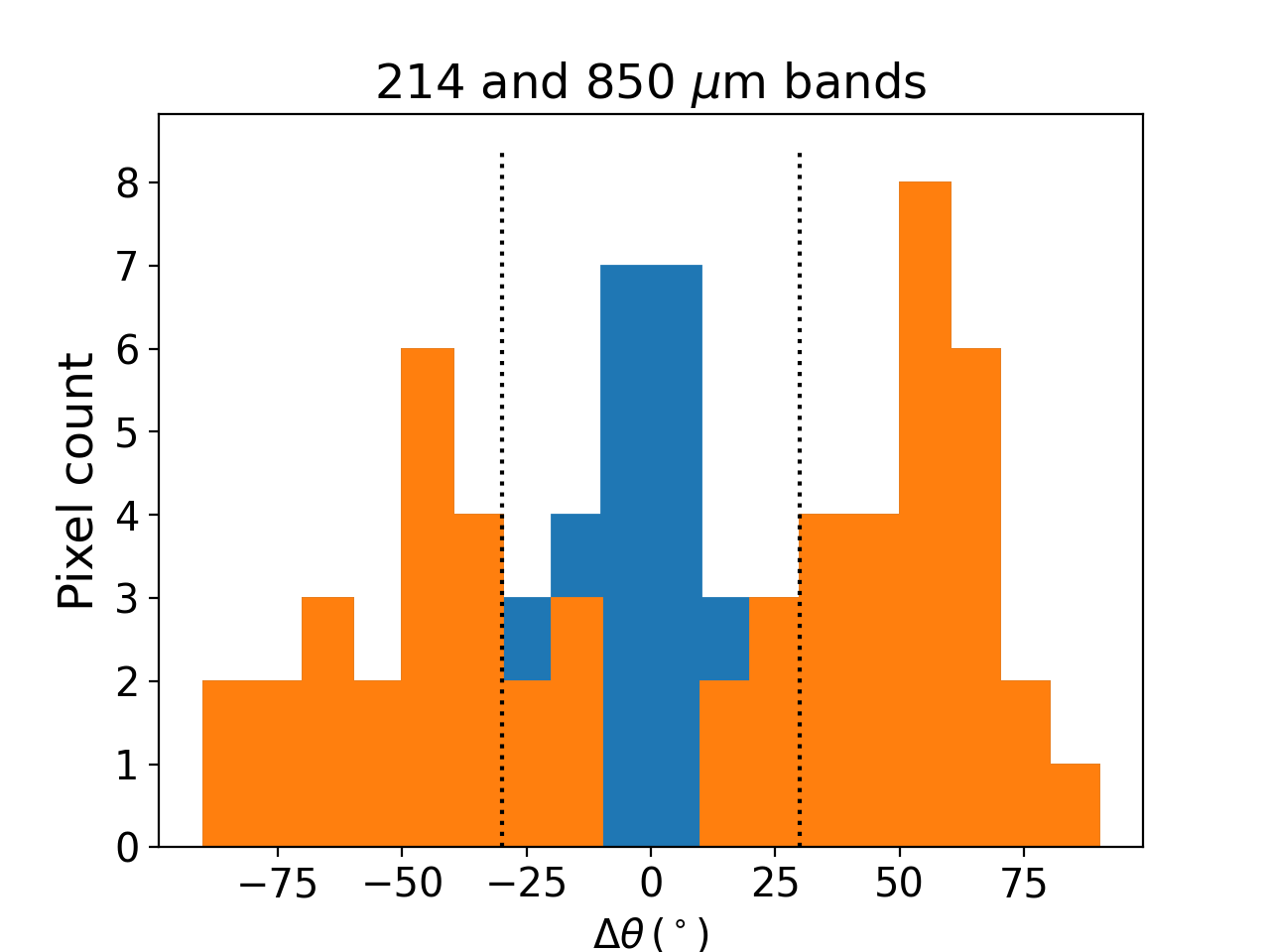}
    \caption{
    Histogram of the difference in polarization angles between 154 and 214 $\mu$m (top),  between 154 and 850~$\mu$m (middle) and between 214 and 850~$\mu$m (bottom). Blue bars show data where the angles agree within $3\sigma$ ($\Delta \theta / \delta \Delta \theta < 3$), orange bars show data where angles differ by $3\sigma$ or more. The two bottom histograms only include data from the 90$\arcsec$ region centered on the emission peak. Vertical dotted lines enclose the $\pm 30^\circ$ region. 
    }
    \label{fig:theta_compare}
\end{figure}

\begin{figure}
    \centering
    \includegraphics[width=\columnwidth]{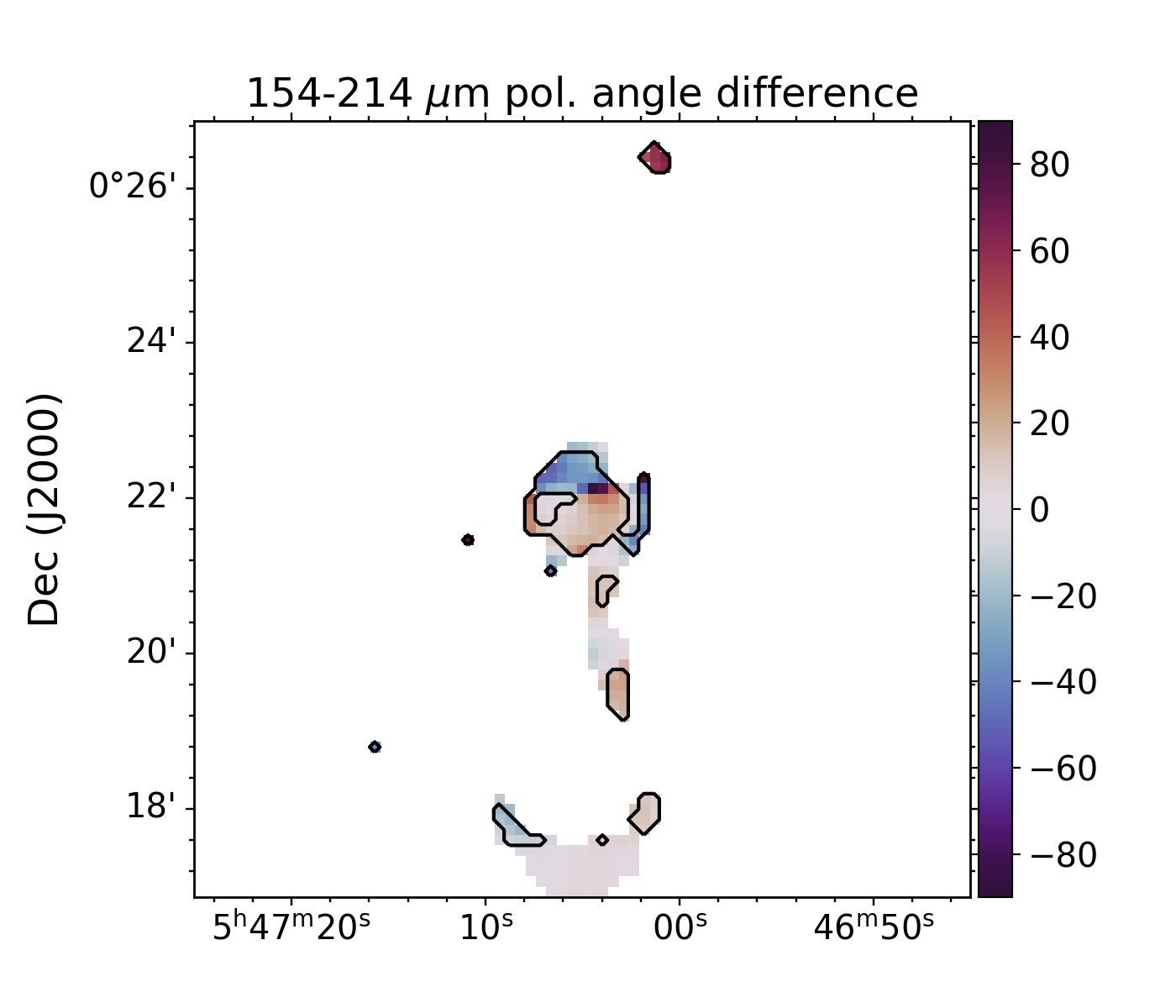}
    \includegraphics[width=\columnwidth]{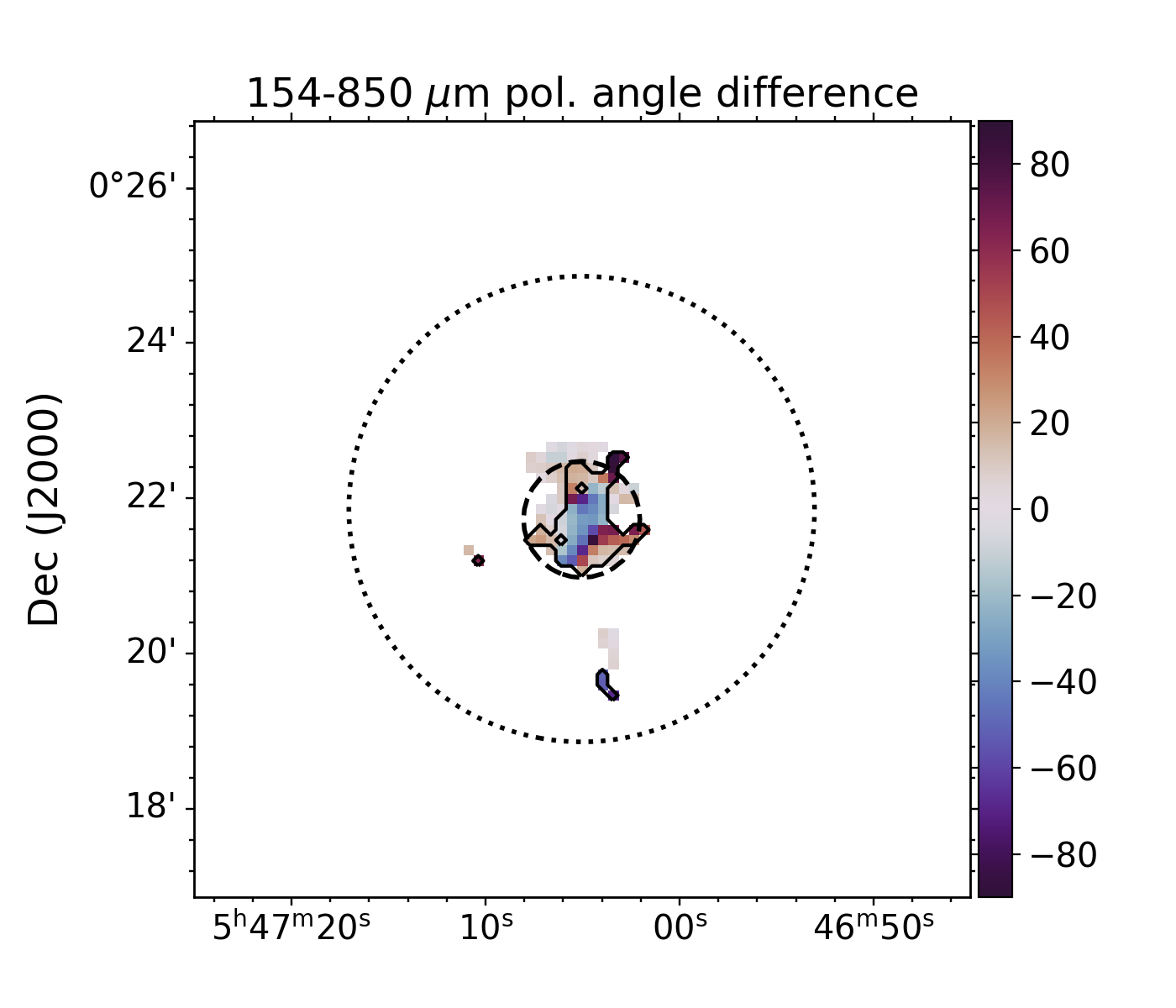}
    \includegraphics[width=\columnwidth]{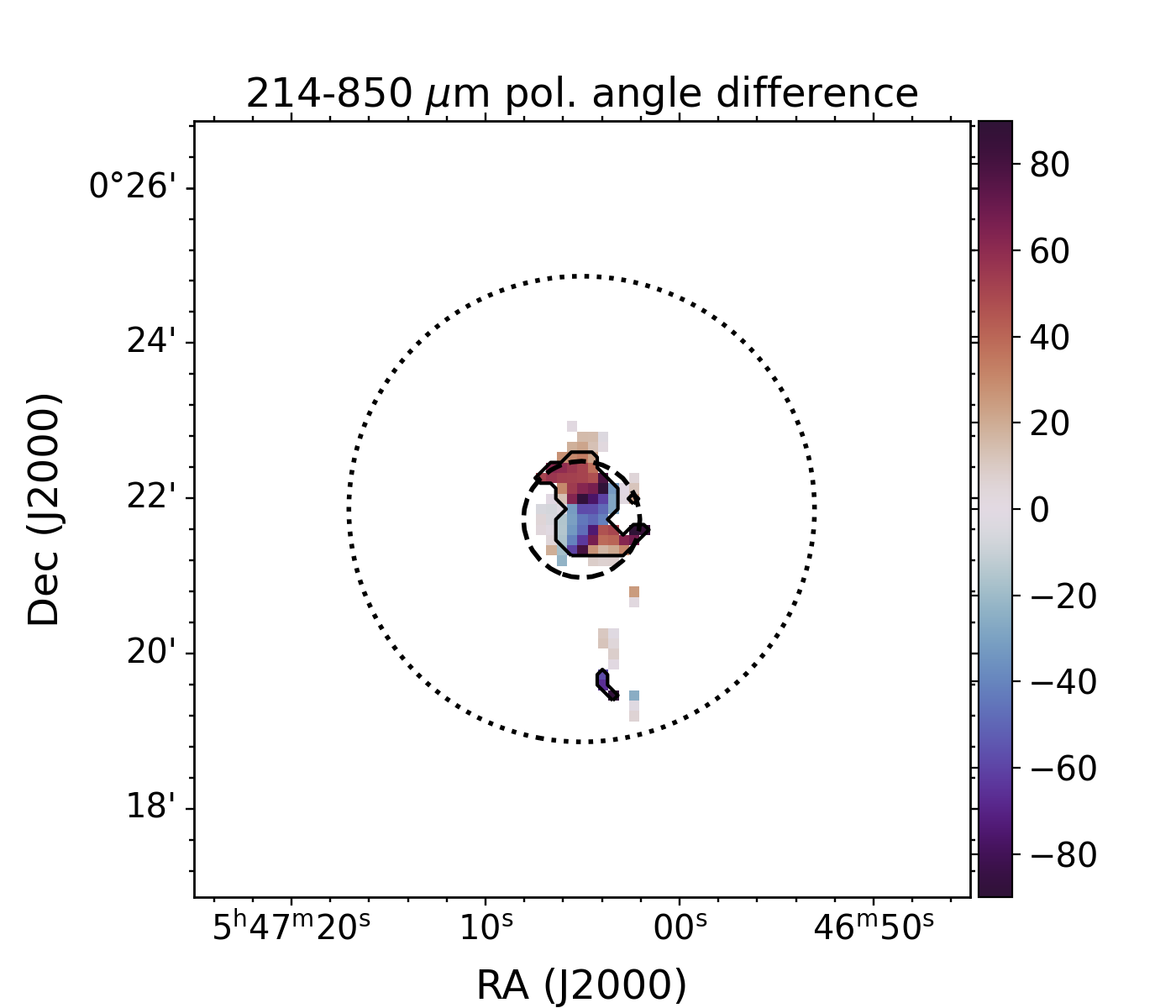}
    \caption{
    Spatial distribution of $\Delta \theta_{154-214}$ (top), $\Delta \theta_{154-850}$ (middle), and $\Delta \theta_{214-850}$ (bottom). The color shows the value of $\Delta \theta$ in the ($-90^\circ, \, 90^\circ$) range, while the solid contours show the regions where $\Delta \theta$ is significant (i.e. $>3\sigma$). The dotted circles on the two bottom maps show the central 6$\arcmin$ area; the smaller dashed circle shows a 90$\arcsec$ diameter centered on the emission peak.
    }
    \label{fig:theta_diff_locations}
\end{figure}

The polarization angle difference between bands, $\Delta \theta$, is portrayed in Figure~\ref{fig:theta_compare}, which shows the difference between 154 and 214 $\mu$m (top), that between 154 and 850~$\mu$m (middle) and that between 214 and 850~$\mu$m (bottom). The data used {for the 154-to-214 $\mu$m comparison} are from inside the selection contours in Figure~\ref{fig:map_all}, with each pixel corresponding to a data point. For the 154-to-850 and 214-to-850 comparison, we further restricted the data to 90$\arcsec$ from the 850~$\mu$m emission peak (see Section~\ref{sec:selection}). The angle differences are in the ($-90^\circ, \, 90^\circ$) range. The sign provides information on the direction of the difference, i.e. which of the two angles is larger. Data visualisation using a ($0^\circ, \, 90^\circ$) range for $\Delta \theta$ can be found in the supplementary online material. The uncertainty on the angle difference $\Delta \theta_{xy}$ between two bands $x$ and $y$ can be expressed as the sum in quadrature of the uncertainty on each angle: $\delta \Delta \theta_{xy} = \sqrt{(\delta \theta_{x}^2 + \delta \theta_{y}^2})$. In the following, we will refer to the angle differences as $\Delta \theta_{154-214}$ when measured between the 154 and 214 $\mu$m bands, $\Delta \theta_{154-850}$ when measured between 154 and 850~$\mu$m, $\Delta \theta_{214-850}$ when measured between 214 and 850~$\mu$m, and simply $\Delta \theta$ when not referring to specific bands. 

As can be seen from Figure~\ref{fig:theta_compare}, the distribution of $\Delta \theta_{154-214}$ peaks at small angles ($0 - 10^\circ$), meaning that most lines of sight have similar polarization angles in the two HAWC+ bands. However, the distribution has wide tails and a significant fraction of pixels shows angle differences larger than $3\sigma$ (i.e., $\Delta \theta_{154-214} / \delta \Delta \theta_{154-214} > 3$). Therefore, the polarization angles in HAWC+ bands D and E disagree on a significant fraction of the maps. 
The distribution of $\Delta \theta_{154-850}$ shows relatively larger tails as well as a larger fraction of pixels with significant ($\Delta \theta_{154-850} > 3 \sigma$) angle differences, even though it still shows a peak at small angles ($0 - 20^\circ$). Therefore, the polarization angles are not as well correlated between 154 and 850~$\mu$m than between 154 and 214~$\mu$m, at least for the 90$\arcsec$ of the map centered around the emission peak. Finally, the distribution for $\Delta \theta_{214-850}$ is wider and has no evident primary peak. This polarization angles at 214 and 850 $\mu$m seem mostly uncorrelated in the selected region.

The spatial distribution of $\Delta \theta$ and that of the areas with $\Delta \theta \geq 3 \sigma$ are shown in Figure~\ref{fig:theta_diff_locations}. The color scale indicates $\Delta \theta$, with reds being positive values and blue being negative values. The color becomes darker as $\Delta \theta$ gets farther from $0^\circ$. Solid contours delineate the area where $\Delta \theta > 3 \sigma$. Most of the pixels with a significant $\Delta \theta$ are concentrated in what we dubbed the central hub region. We note that the central region is likely to have a complex structure influenced by the outflow and embedded infrared sources. In the southern filament, instead, $\Delta \theta_{154-214}$ is mostly consistent with $0^\circ$.
In the case of $\Delta \theta_{154-850}$ and $\Delta \theta_{214-850}$ we cannot make a comparison in the southern filament, which lies outside of the 90$\arcsec$ region centered on the emission peak. However, within these 90$\arcsec$, most lines of sight do show $\Delta \theta > 3 \sigma$. In regions with significant $\Delta \theta$ either the two bands are not observing the same dust along the line of sight, or dust grains are not in the Rayleigh regime, or both. 

\begin{figure}
\includegraphics[width=\columnwidth]{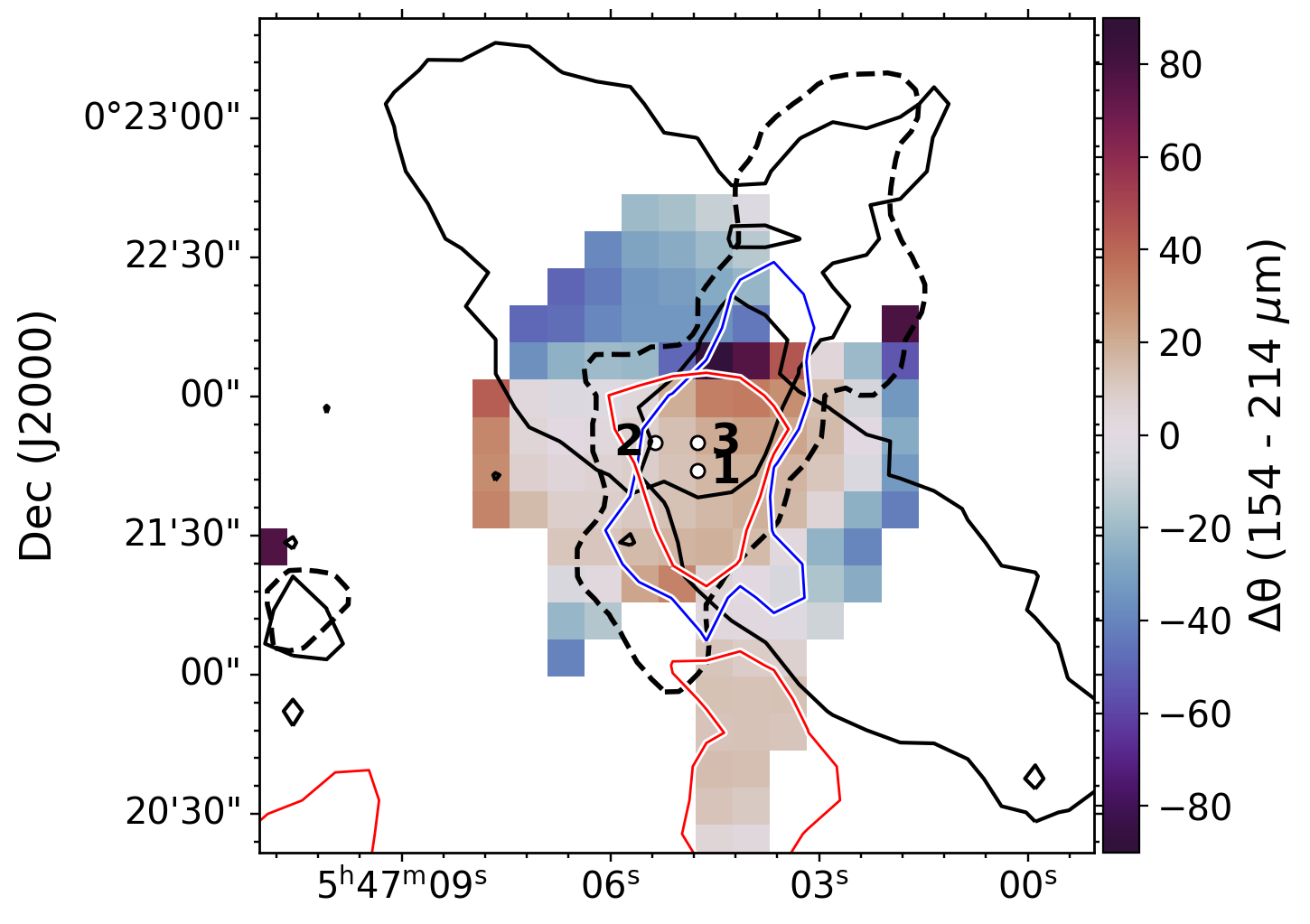}
\includegraphics[width=\columnwidth]{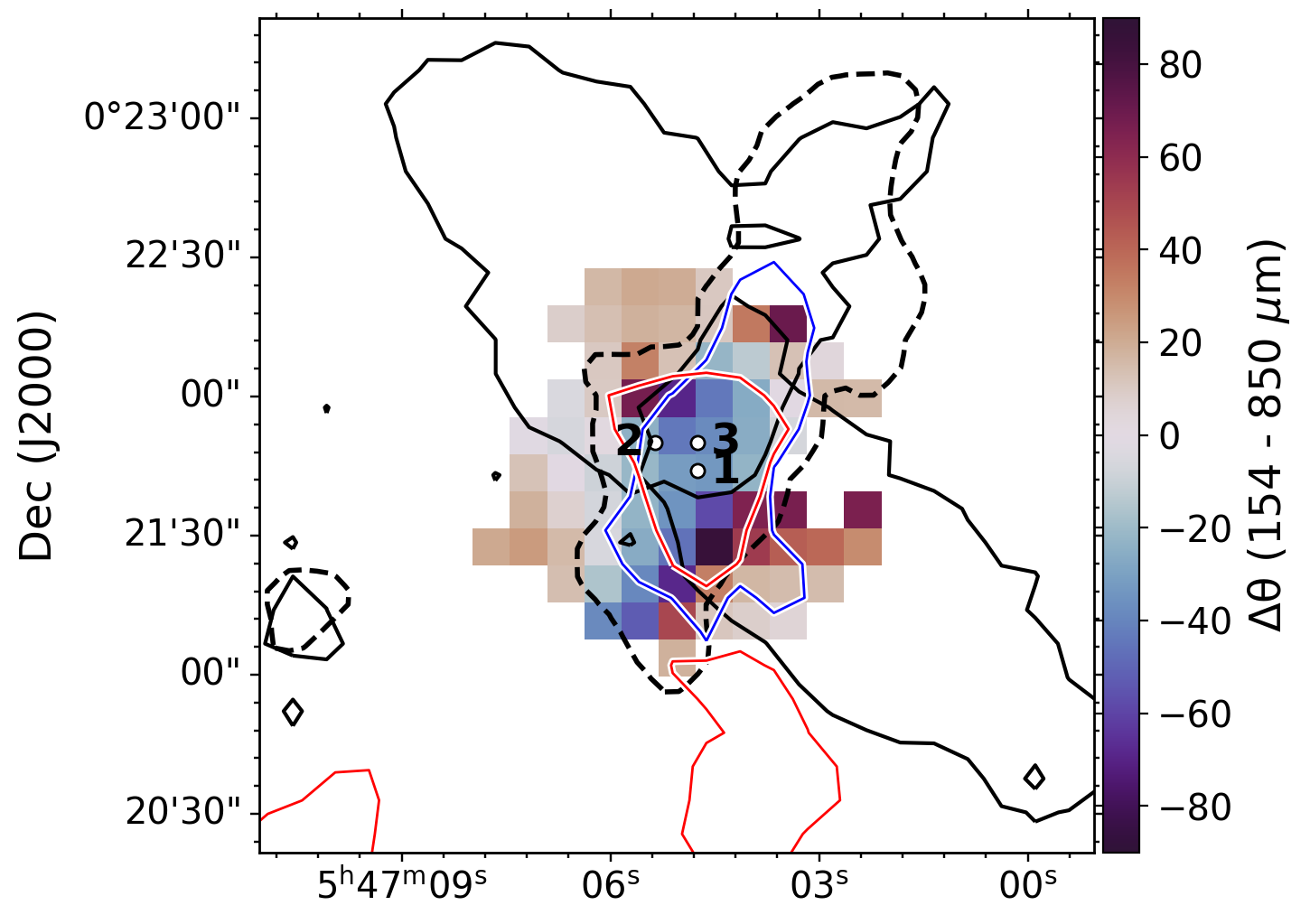}
\includegraphics[width=\columnwidth]{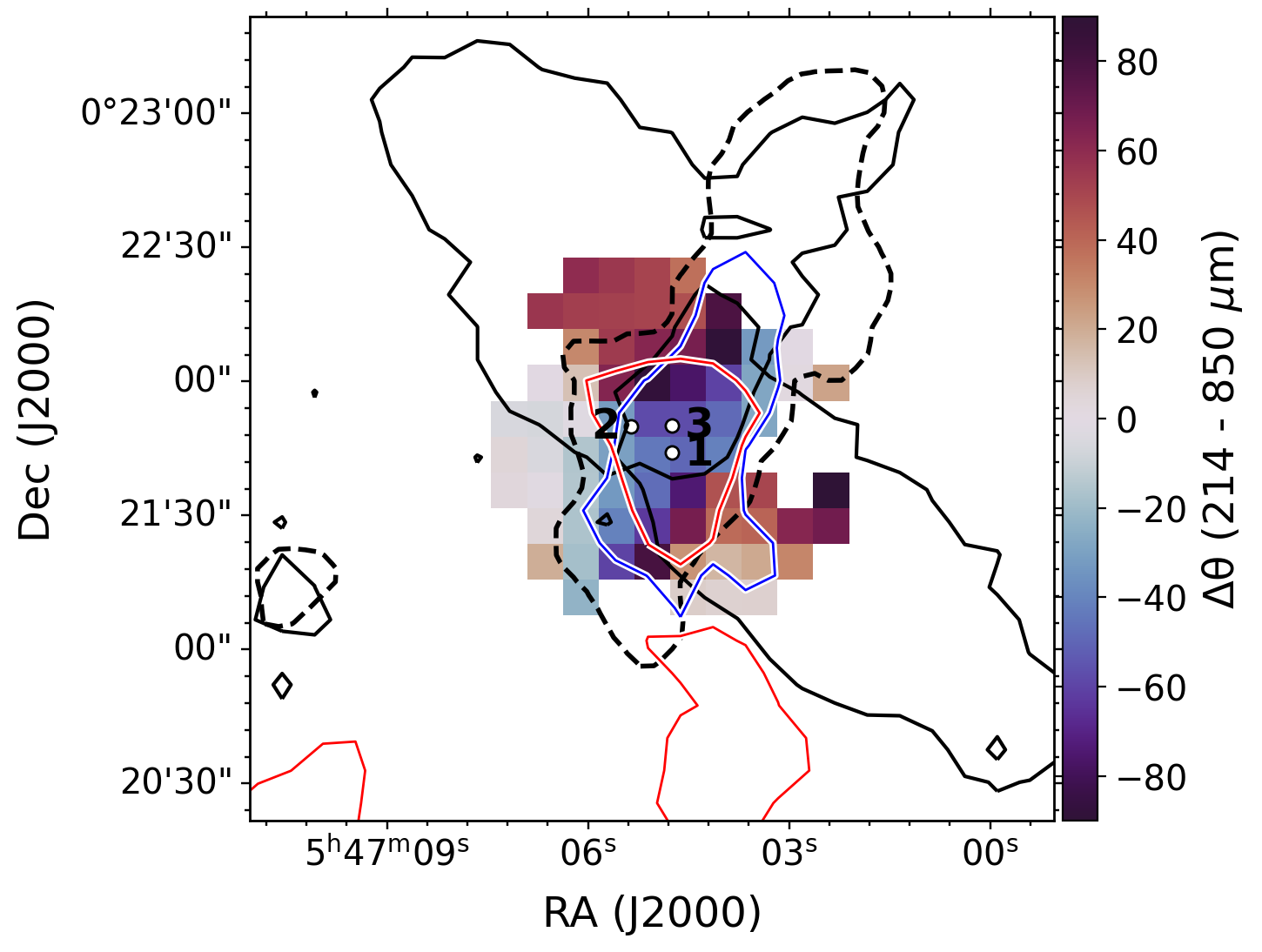}
\caption{Map of the difference in polarization angles ($\Delta \theta$) between different bands in the central 3$\arcmin$ of the maps. \textbf{Top:} $\Delta \theta_{154-214}$ (angle difference between HAWC+ bands D and E). \textbf{Middle:} $\Delta \theta_{154-850}$ (angle difference between HAWC+ D and POL-2), limited to 45$\arcsec$ from the 850~$\mu$m emission peak. \textbf{Bottom:} $\Delta \theta_{154-850}$ (angle difference between HAWC+ D and POL-2), limited to 45$\arcsec$ from the 850~$\mu$m emission peak. White dots mark the position of IRS 1, 2 and 3. The dashed contour marks $N_{\rm H_{2}} = 5 \times 10^{22}\, {\rm cm}^{-2}$ from HGBS. Contours for $^{12}$CO emission at 3.5 K km sec$^{-1}$ are shown as solid black lines, contours for redshifted (blueshifted) C$^{18}$O emission at 2.5 K km sec$^{-1}$ are shown as red (blue) lines \citep[adapted from][]{Lyo+21}.}
\label{fig:theta_diff}
\end{figure}

The angle differences in the central 3$\arcmin$ of the observed region are shown in more detail in Figure~\ref{fig:theta_diff}, which shows maps of $\Delta \theta$. The angle difference is shown for those pixels that pass the selection criteria in both bands involved (Section~\ref{sec:selection}). Superposed on the maps are contours for the $^{12}$CO and C$^{18}$O (3-2) emission observed by \citet{Lyo+21}, which trace the local outflow and possible physical structures. Another contour shows the HGBS column density at $N_{\rm H_{2}} = 5 \times 10^{22}\, {\rm cm}^{-2}$, equivalent to $\tau \sim$ 0.09 at 154 $\mu$m, $\tau \sim$ 0.05 at 214 $\mu$m or $\tau \sim$ 0.003 at 850~$\mu$m. Note that since the column density is recovered from a single-temperature modified blackbody fit (see Section~\ref{sec:ancillary}), which is likely an oversimplification, the contour should be regarded as a first-order approximation. The sign of the angle difference varies in space following an approximate NE -- SW direction, with $\Delta \theta_{154-214}$ changing from negative to positive to negative again, while $\Delta \theta_{154-850}$ and $\Delta \theta_{214-850}$ follow a positive-negative-positive distribution.

\subsection{Change of the polarization fraction $P$ with wavelength}
\label{sec:PvsP}

\begin{figure}
\includegraphics[width=\columnwidth]{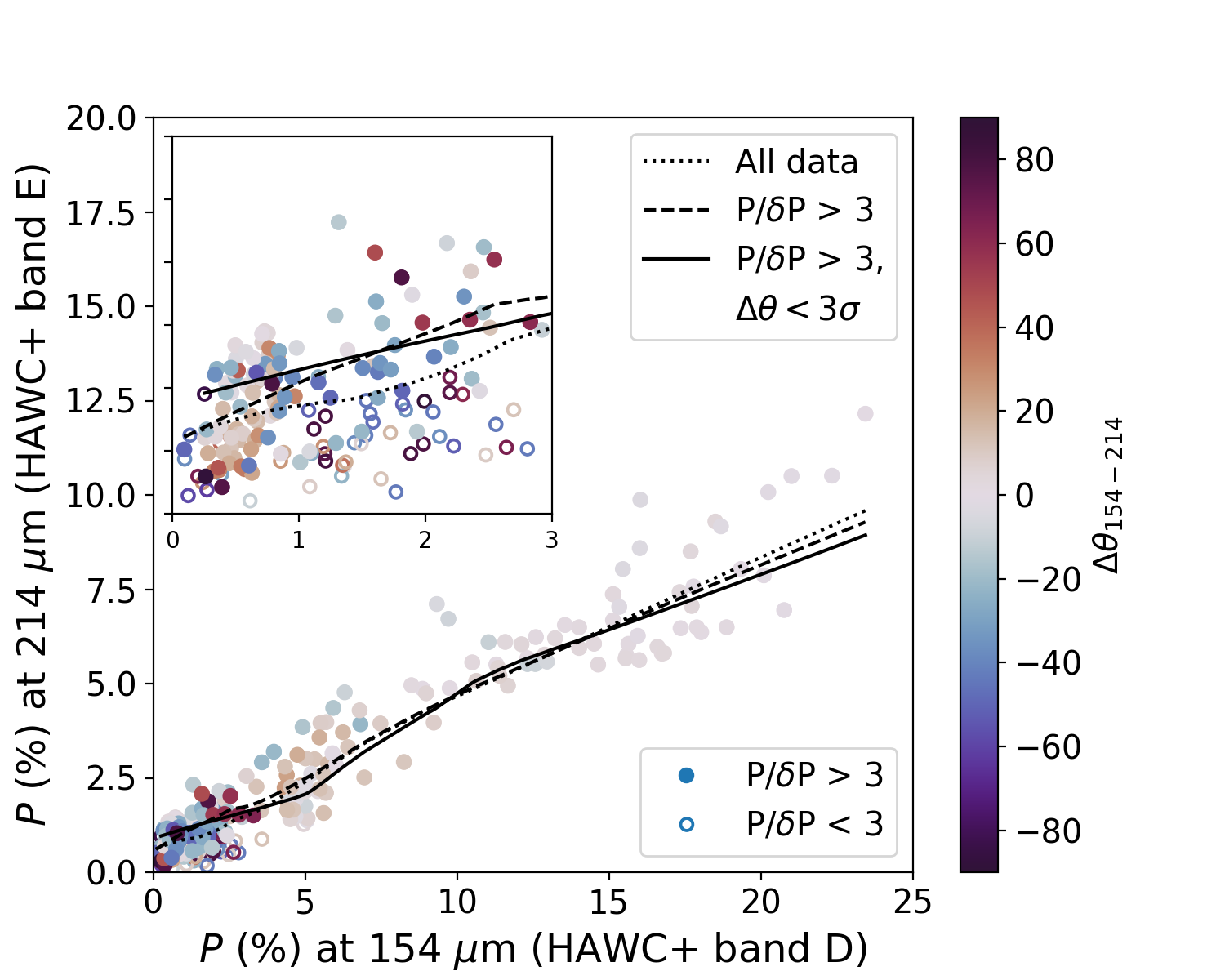}
\includegraphics[width=\columnwidth]{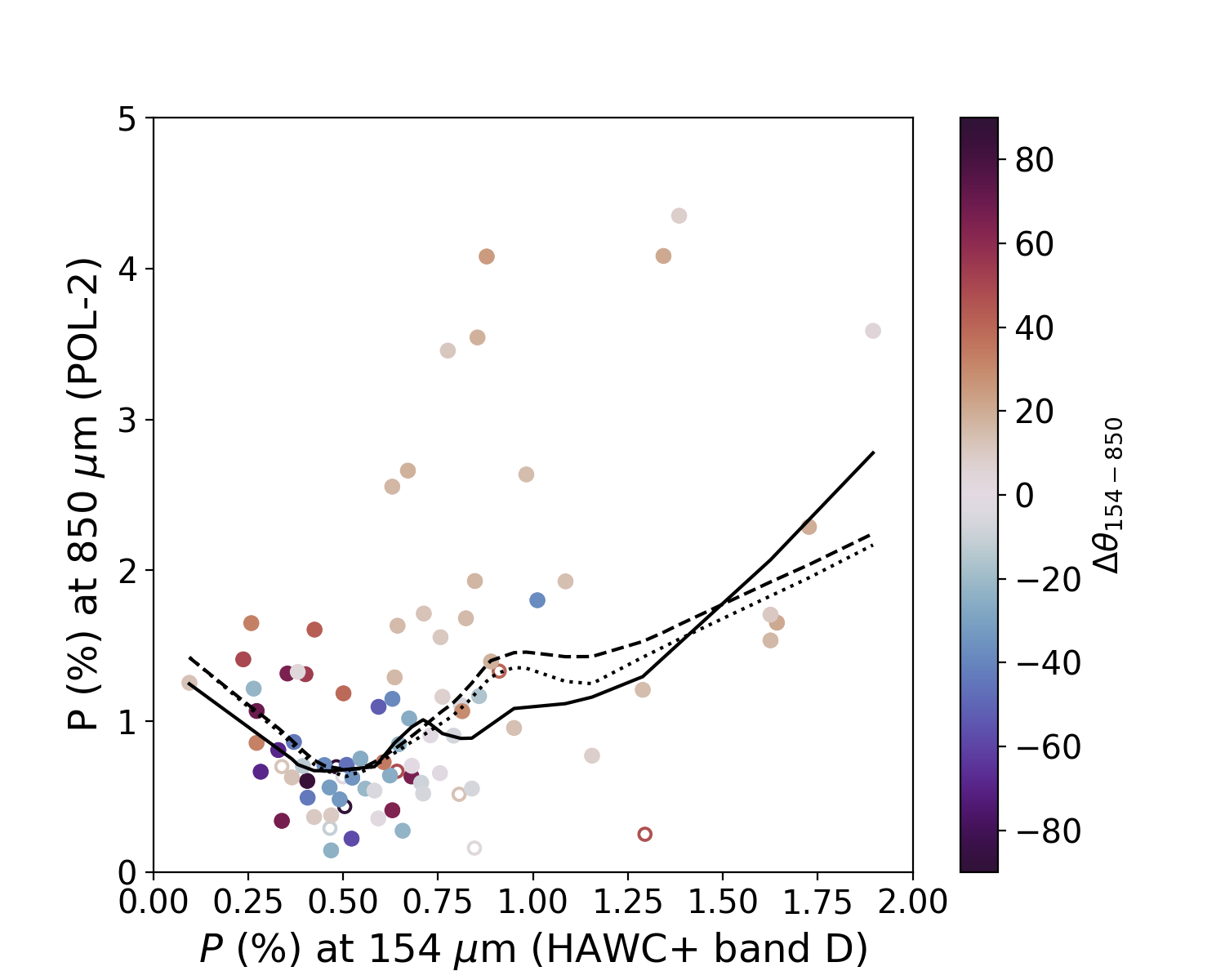}
\includegraphics[width=\columnwidth]{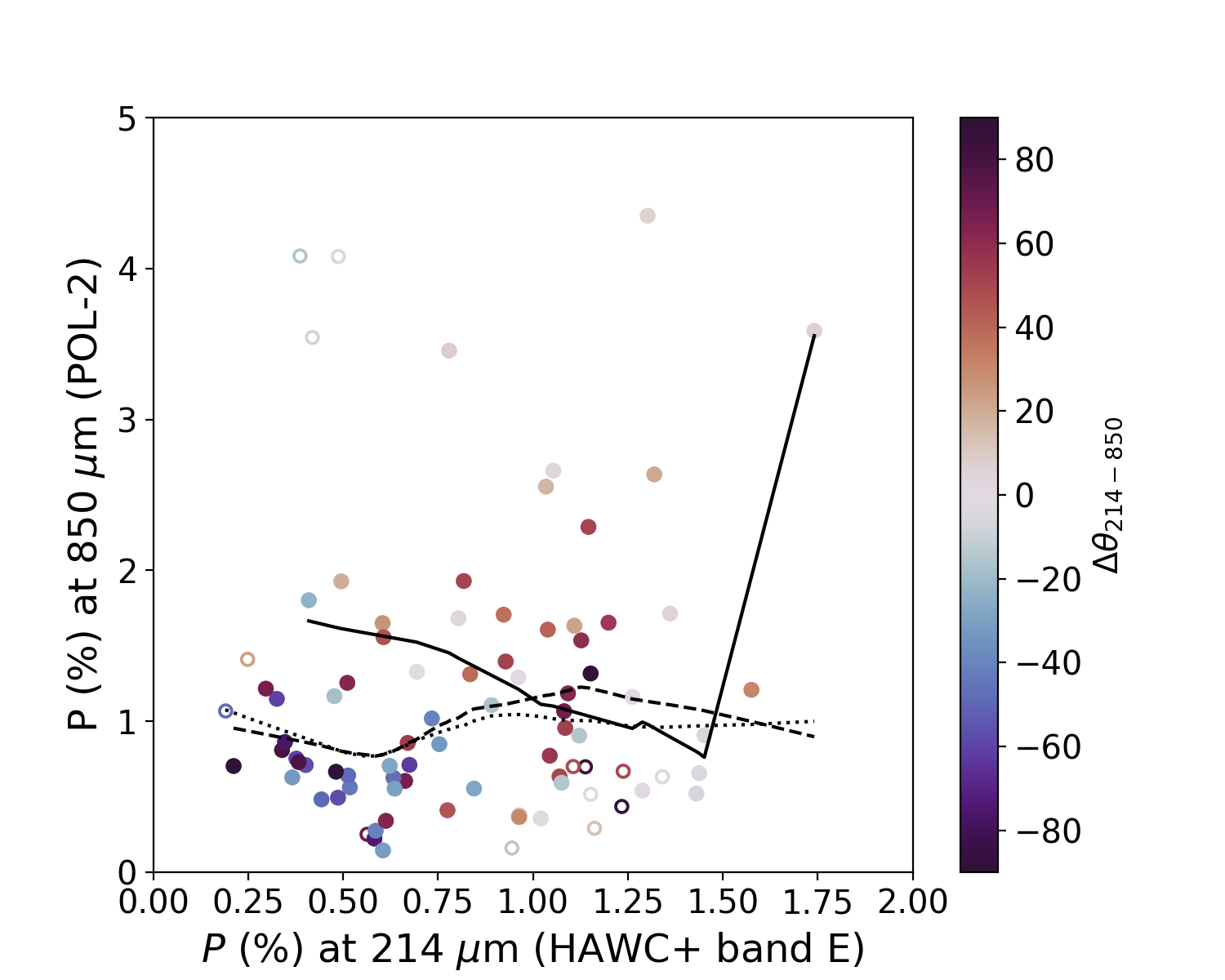}
\caption{
Comparison of the polarization fractions $P_{154}$ and $P_{214}$ (top), $P_{154}$ and $P_{850}$ (middle) and $P_{214}$ and $P_{850}$ (bottom), color coded by $\Delta \theta$ (see Section~\ref{sec:polangle}). Lines show the overall trends as traced by LOWESS-smoothed data (solid lines: all data, dashed lines: data selected by $P/\delta P > 3$, dotted lines: data selected by $P/\delta P > 3$ and $\Delta \theta < 3\sigma$). An enlargement of the bottom-left corner is shown in an inset for the upper figure. Note that the axis scales for the top plot and for the bottom two plots are not the same.
}
\label{fig:PvsP}
\end{figure}

The shape of the polarization spectrum can be studied by plotting the polarization fraction $P$ of one band versus another. We would expect the resulting plot to show a strong correlation if the observed dust is the same in both bands, with a slope providing a measure of the ratio of $P$ between the bands. A lack of correlation would imply that over at least part of N2071 a single dust model cannot explain the observations. We show these plots in Figure~\ref{fig:PvsP}, which compares $P$ between 154 and 214~$\mu$m (upper plot), between 154 and 850~$\mu$m (middle plot, showing data within 45$\arcsec$ of the peak) and between 214 and 850~$\mu$m (lower plot, showing data within 45$\arcsec$ of the peak. The points are color-coded according to $\Delta \theta_{154-214}$ and $\Delta \theta_{154-850}$ respectively (see Section~\ref{sec:polangle}). For completeness we include data with $P/\delta P < 3$, shown as empty circles to be distinguishable. To make the trends in the data more evident, we overplot a non-parametric, locally weighted smoothing \citep[LOWESS,][]{lowessref} in the form of curves.\footnote{
The smoothing was performed using the \textit{lowess} function from Python \textit{statsmodel} package, adopting a value of 0.5 for the `frac' parameter. 
}
 We computed LOWESS for three sets of data: pixels with any value of $P/\delta P$, data with $P/\delta P > 3$, and data with $P/\delta P > 3$ and the same $\theta$ (within three $\sigma$) between the two bands in exam. For the 154 vs. 214~$\mu$m plot, since most data points are clustered in the bottom left corner, an inset shows an enlargement of the $P_{154}$, $P_{214} < 3\%$ region, for additional clarity.

The polarization fractions at 154 and 214 $\mu$m (Figure~\ref{fig:PvsP}, top) show good correlation for all value ranges. This result remains qualitatively valid for all types of selections used (no selection on $P/\delta P$, selection on $P/\delta P > 3$, and selection on $P/\delta P > 3$ and no significant $\Delta \theta$). The polarization angle match is also generally good for high polarization (i.e. $P_{154}$, $P_{214}\gtrsim5\%$). We note that these high-polarization, low-$\Delta \theta$ pixels mostly come from the southward-pointing filament. This is consistent with the 154 and 214~$\mu$m being dominated by the same dust in the filament (but see Section~\ref{sec:discussion_PvsP}). It is interesting to compare our results to those of \citet{Michail+21}, who found the same values of $P_{154} \simeq P_{214}$ (within $10\%$) in OMC-1. For N2071 we find a lower $P$ at 214~$\mu$m than at 154~$\mu$m: using data with $P/\delta P > 3$ and no angle restrictions, $P_{214}/P_{154}$ has a median value of 0.63 and a median absolute deviation (MAD) of 0.23. Additionally, our data, unlike in \citet{Michail+21}, includes pixels with significant angle differences between bands, which have the highest values of $P_{214}/P_{154}$.
Adding the condition $\Delta \theta_{154-214} < 3 \sigma$, for a more meaningful comparison, we find an even lower $P_{214}/P_{154}$ ratio of $0.48 \pm 0.10$ (median $\pm$ MAD). A virtually identical result, $0.49 \pm 0.10$, is obtained when excluding the data within 45$\arcsec$ of the emission peak, where most of the significant angle differences are located.

In the case of the 154 and 850~$\mu$m bands (Figure~\ref{fig:PvsP}, middle), the values of $P$ are not as well correlated. The majority of data points, situated at $P_{154} \lesssim 2\%$ and $P_{850} \lesssim 2\%$ shows no correlation at all, as can be seen by the fact that LOWESS lines are nearly horizontal. Some data points at higher polarization -- where the LOWESS curves gain a positive slope -- suggest a positive correlation, but the trend is driven by very few data points from the southern edge of the central hub\footnote{
 Relatively high values of $P_{850}$ ($\gtrsim 3\%$) are also found at the northern edge of the hub, but they are more than 45$\arcsec$ away from the emission peak, and are therefore excluded from the analysis.
 }
and is not particularly reliable. These results are independent of whether low S/N and high $\Delta \theta$ data are kept or excluded.

Finally, polarization fractions at 214 and 850~$\mu$m (Figure~\ref{fig:PvsP}, bottom) also show no significant correlation. This is best seen by the fact that the LOWESS curves are mostly horizontal. The curve for data with $P/\delta P > 3$ and $\Delta \theta_{214-850} < 3 \sigma$ shows a sharp uptick around $P_{214} \sim 1.5 \%$, but this curve is determined with very few data points from the southern edge of the central hub and should not be considered reliable.

\subsection{Change of the $P$ vs. $I$ relation with wavelength}
\label{sec:PvsI}

\begin{figure*}
\includegraphics[width=.49\textwidth]{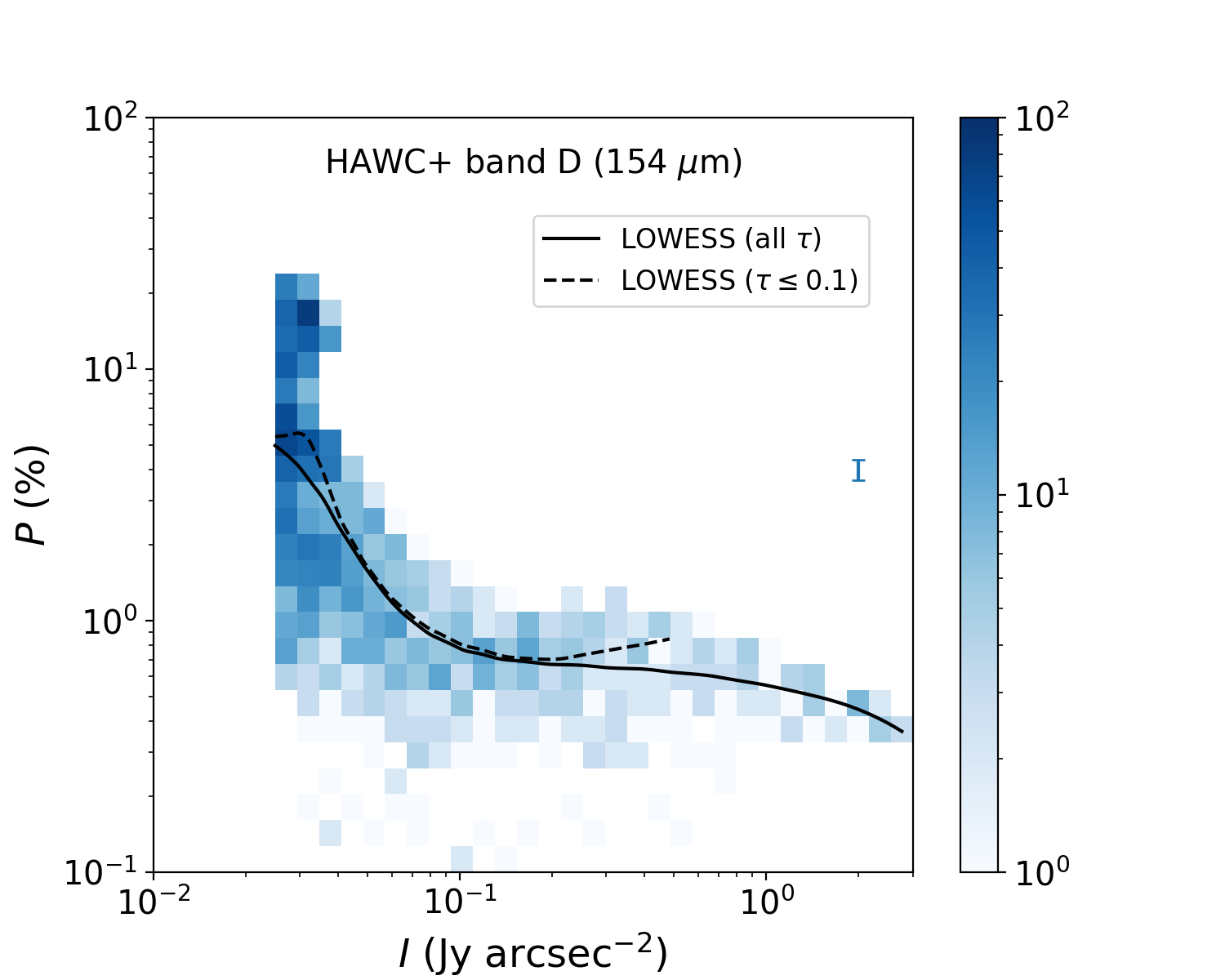}
\includegraphics[width=.49\textwidth]{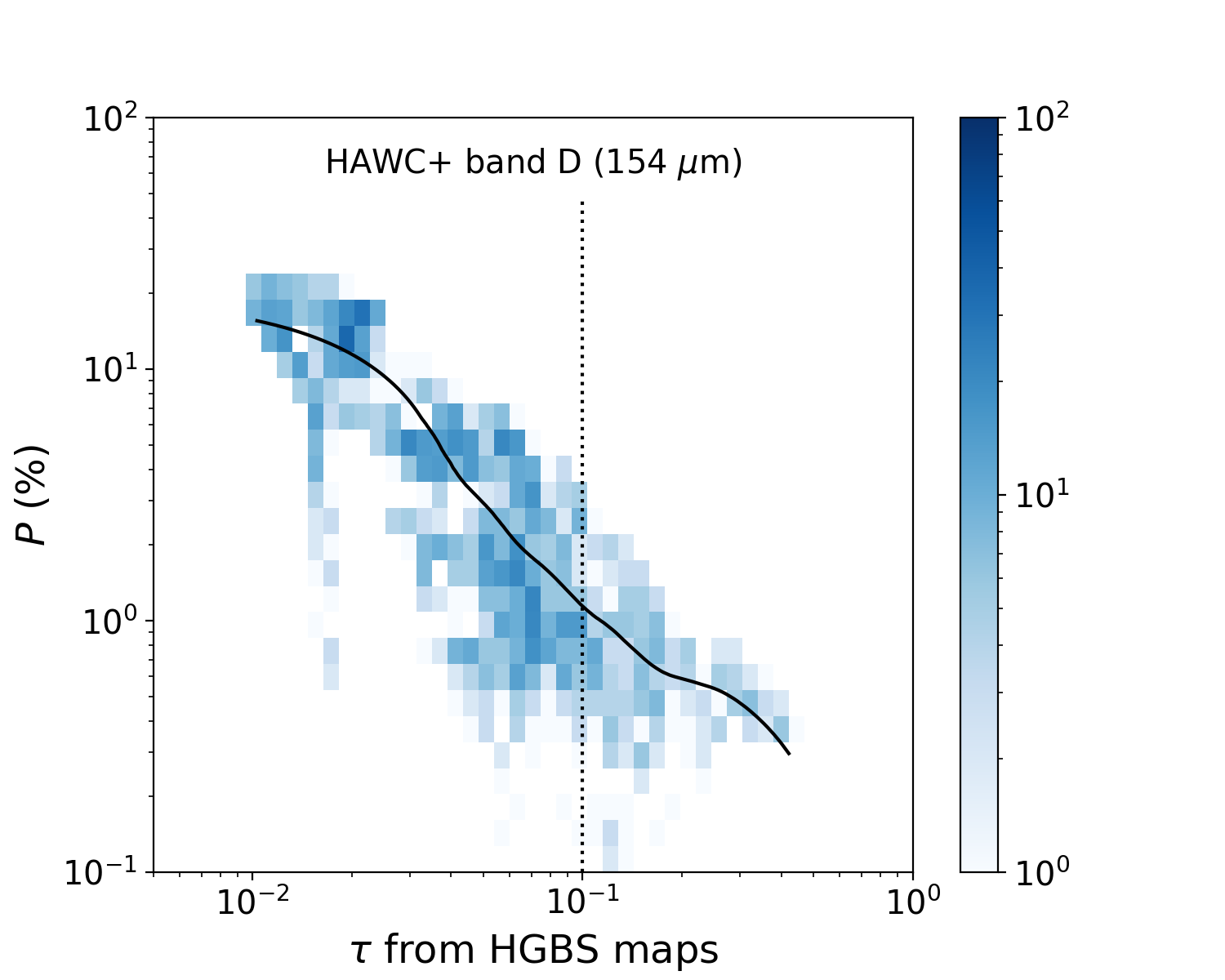}
\includegraphics[width=.49\textwidth]{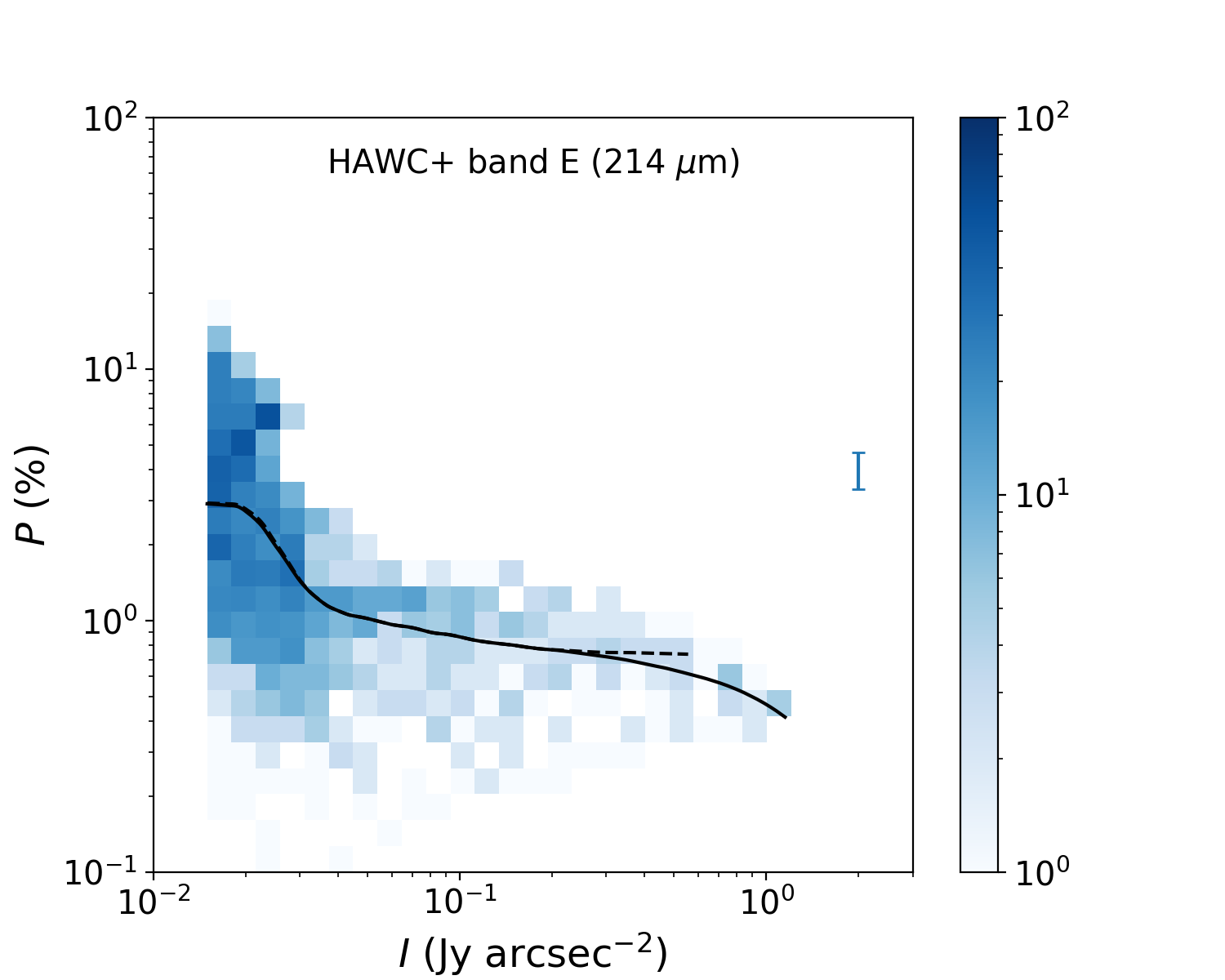}
\includegraphics[width=.49\textwidth]{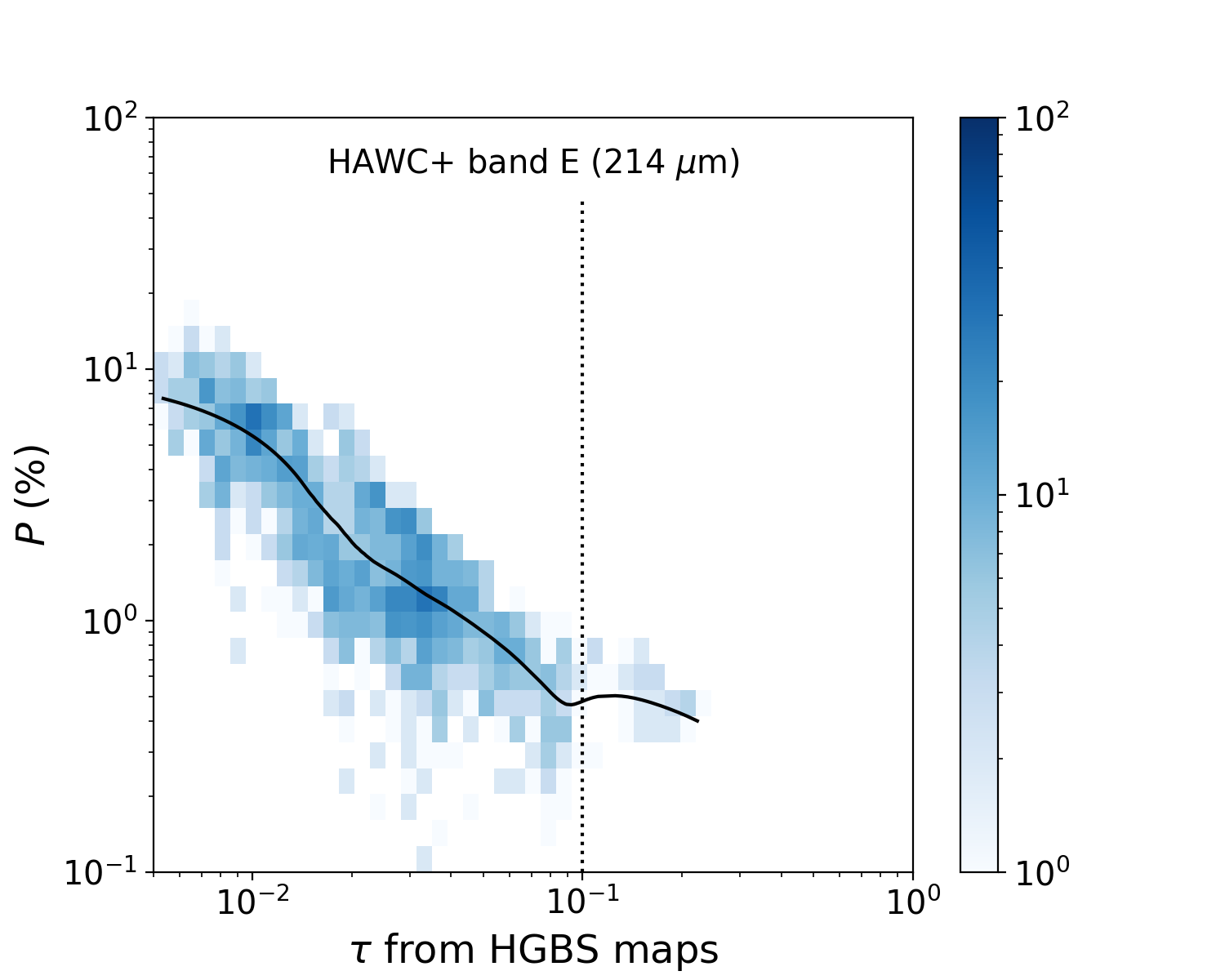}
\includegraphics[width=.49\textwidth]{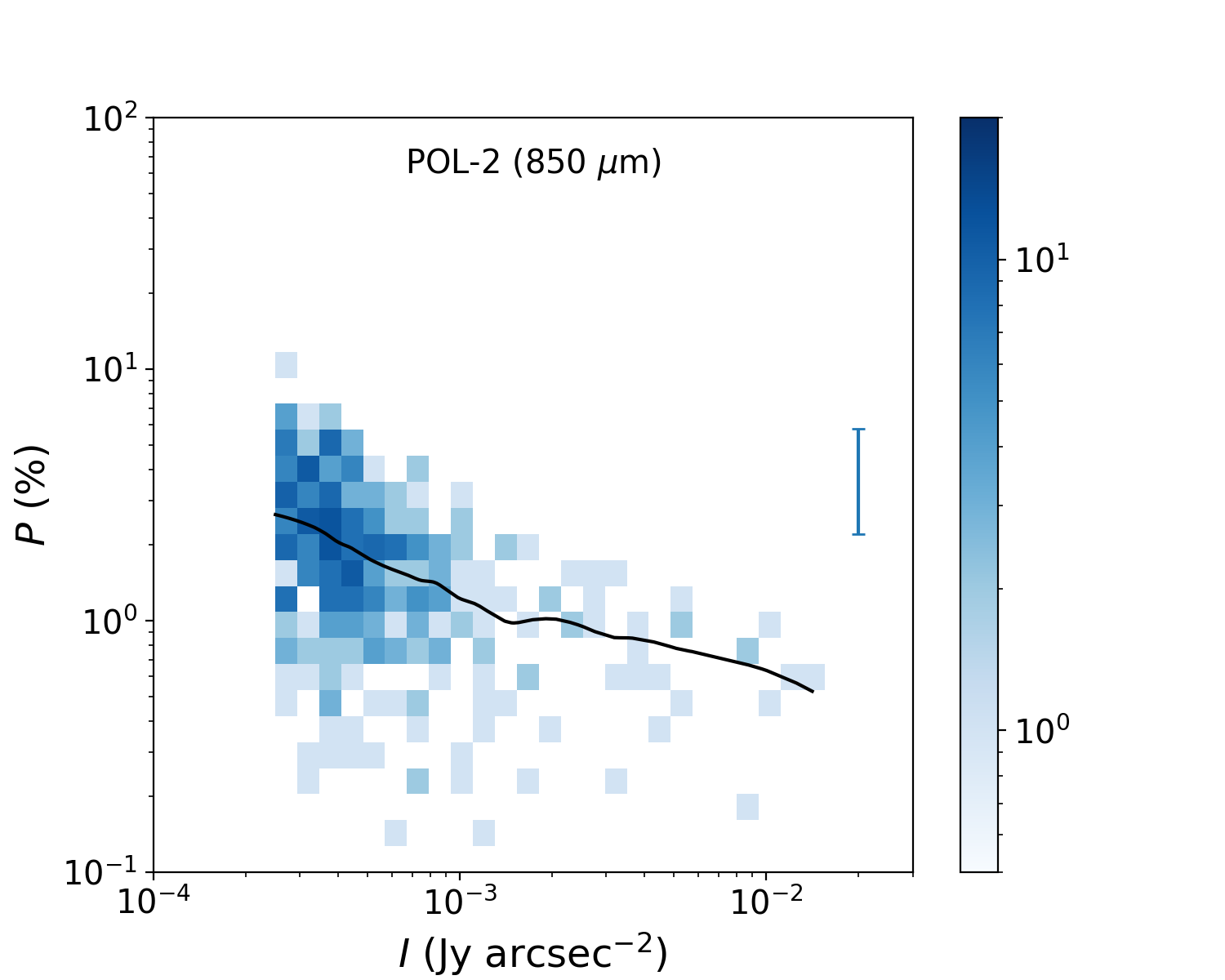}
\includegraphics[width=.49\textwidth]{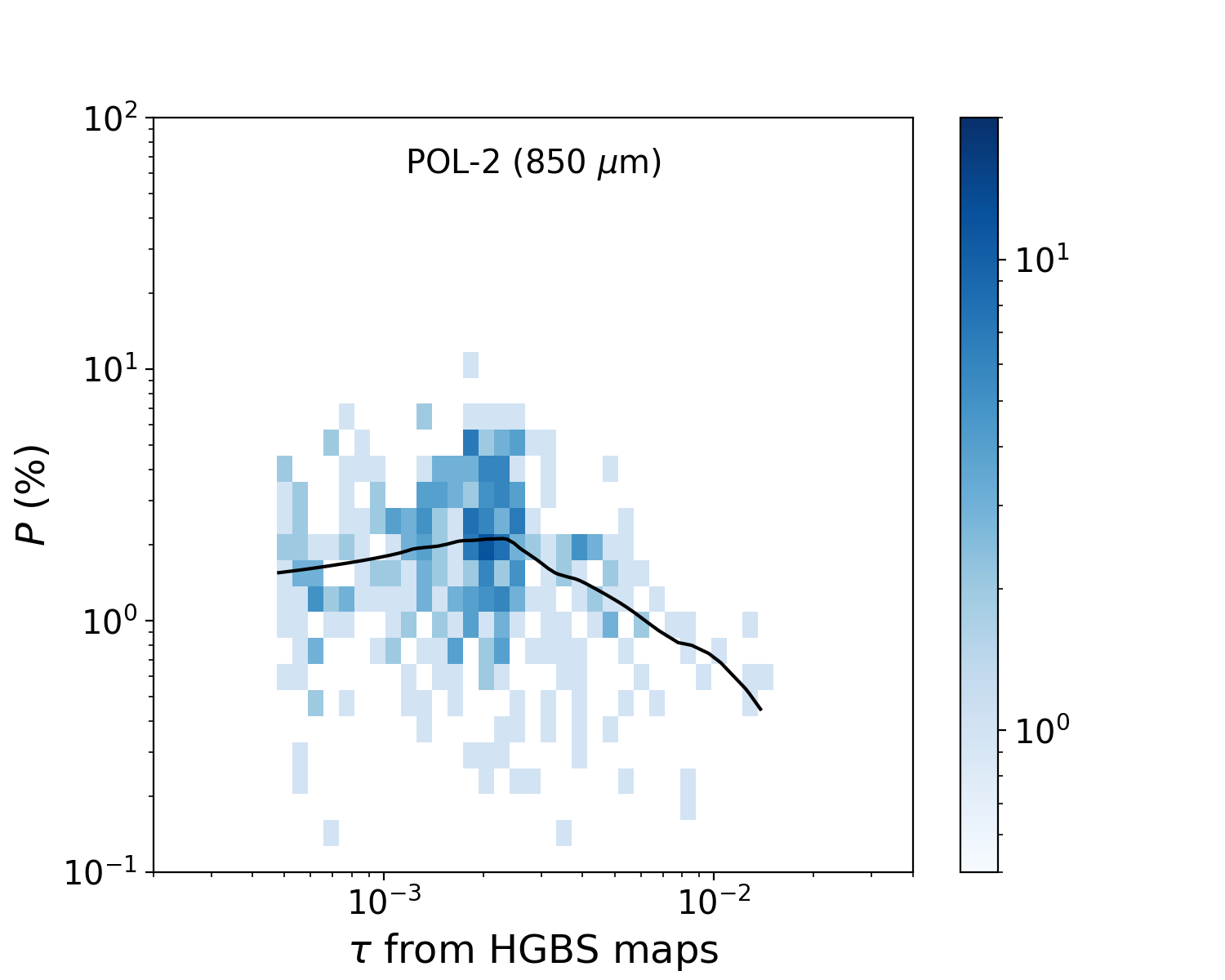}
\caption{
Plots of $P$ vs. $I$ (left column) and $P$ vs. $\tau$ (right column) in the three bands (\textbf{top:} 154 $\mu$m, \textbf{middle:} 214 $\mu$m, \textbf{bottom:} 850~$\mu$m). The shade represents the number of pixels in each histogram element. The shading on the bottom row (850~$\mu$m data) uses a different scale than the other rows since there is less data. The blue bars in the left-side plots show the typical uncertainty on $P$. The data are overplotted with the trends as traced by LOWESS smoothing (solid curves: all data; dashed curves: data with $\tau \leq 0.1$). The vertical dotted lines on the right-side plots show $\tau = 0.1$.
}
\label{fig:PvsI}
\end{figure*}

As mentioned in Section~\ref{sec:intro}, in many molecular clouds $P$ decreases with column density as traced by observational proxies. Typical proxies of column density are dust extinction $A_{\rm V}$ \citep[usually extrapolated from near-infrared observations:][]{Goodman+92, whittet+08, Jones+15}, the dust optical depth $\tau$ from an SED fit \citep[][]{Hildebrand+99, Juvela+18}, and the dust continuum intensity $I$ in the submillimeter \citep[][]{W-T+00, Koch+18, Pattle+19}. Our data gives us the opportunity to compare $P$ vs. column density plots at three different submillimeter wavelengths, and using two different proxies: the intensity $I$ and the modified-blackbody optical depth $\tau$ from the HGBS (see Section~\ref{sec:ancillary}). The plots are shown in Figure~\ref{fig:PvsI} in the form of two-dimensional histograms. Since in this case we are not making a pixel-by-pixel comparison of the maps, we use data that has been resampled to 8$\arcsec$, but not smoothed. Additionally, we do not apply the selection on $P/\delta P$ in this case, to avoid selection biases. To emphasize the main trend in each of the plots, we added the LOWESS-smoothed data (see Section~\ref{sec:PvsP}). Note that the values of $\tau$ are from a fit to \textit{Herschel} $\lambda~\leq~500\,\mu$m data. The value at 850~$\mu$m is therefore an extrapolation beyond the original wavelength range. The plot on the bottom right of Figure~\ref{fig:PvsI} should be regarded as an approximation.

All plots show a similar qualitative trend of $P$ decreasing when the proxy for column density increases, as expected. However, individual plots do show different behavior. Most notably, when using $I$ as a proxy for column density, polarization at short wavelengths (154 and 214 $\mu$m) seems to reach a plateau of $\sim 1 \%$ for $I \gtrsim 100$ mJy arcsec$^{-2}$. This trend is absent, or much smaller, at 850~$\mu$m wavelength and when $\tau$ is used as a proxy. Therefore, while the decrease of $P$ with column density is a robust finding, finer results such as variations of the slope may be influenced by the type and preparation of the data.

\section{Discussion}
\label{sec:discussion}

\subsection{polarization angle}
\label{sec:discussion_polangle}

As mentioned in Section~\ref{sec:polangle}, polarization angles in some regions change with wavelength. This is qualitatively consistent with a variation of dust temperature along the line of sight -- one of the ingredients of the ``heterogeneous cloud model'' discussed in  Section~\ref{sec:intro} -- if variations in the average magnetic field orientation are also present. We also explore a few alternative possibilities: (1) Alternative alignments scenarios within the RATs paradigm (k-RATs and ``wrong internal alignment'' -- see below), (2) polarized thermal emission produced by anisotropic radiation \citep[PTEAR,][]{Onaka+95, Onaka+00} and (3) grain growth. These scenarios are not necessarily mutually exclusive.

A significant gradient of dust temperature along the line of sight is expected to occur in molecular clouds such as N2071, and its effects to be especially visible at high column density. The expected trend for molecular clouds is a decrease of dust temperature with optical depth, although in the proximity of embedded sources temperature might increase with optical depth instead. Since different bands will have different sensitivity to  dust at different temperatures, they may be dominated by emission from different points on the line of sight. In this scenario the different polarization patterns seen in the different bands would trace magnetic fields at different depths in the cloud. For example, the HAWC+ 154~$\mu$m band would be more likely to trace magnetic fields in the warmer envelopes, whereas POL-2 at 850~$\mu$m would probe the colder, denser material within those envelopes, and the 214~$\mu$m band would probe intermediate regions. If the IR sources are able to efficiently heat their surroundings, on the other hand, it's possible that the HAWC+ bands could trace the warm, high density material surrounding the protostars, but this possibility would require testing with models. Such tests are beyond the scope of this paper. Another effect is that dust opacity decreases with wavelength, so that optical depth will be largest at 154~$\mu$m and smallest at 850~$\mu$m. However, we do not expect the change of optical depth with wavelength to play a large role, since N2071 is optically thin in all bands at the spatial resolution we employ. The maximum value of $\tau$ at 154~$\mu$m is $\sim$0.4 at a resolution of 18\hbox{$\,.\!\!^{\prime\prime}$}2, and it decreases rapidly with distance from the central emission peak. As Figure~\ref{fig:theta_diff} shows, high $\Delta \theta$ values persist even in the area where $\tau\, (154\, \mu{\rm m}) < 0.09$. 

Another possibility is that the alignment mechanism of grains, rather than in the magnetic field orientation, changes on the line of sight. RAT theory predicts that in a strong, anisotropic radiation field grains will align with the radiation anisotropy direction rather than with the magnetic field \citep[k-RATs alignment;][]{Lazarian+Hoang07, Tazaki+17}. If the embedded IR sources are intense enough to cause k-RATs alignment, we would observe a change of polarization angle in their vicinity. This change would have a non-trivial wavelength dependence: on one hand, k-RATs mainly affects larger grains, which tend to be cooler and therefore would be better observed at 850~$\mu$m. On the other hand, since k-RATs need intense radiation field, they may influence mainly the warmer part of the cloud near the embedded IR sources. However, k-RATS alignment predicts an azimuthal pattern in polarization angles, which is not evident in our maps (see Figure~\ref{fig:map_all}). Therefore, it seems unlikely that this mechanism be the main reason for the change in polarization angle with wavelength.
An alternative explanation lies in the ``wrong internal alignment'' described in \citet{Hoang+Lazarian09} and \citet{Hoang21}: under conditions of large grains and/or high gas density, internal alignment (see Section~\ref{sec:intro}) results in the longest axis, rather than the shortest, to become parallel to the rotation axis. External alignment also becomes less efficient in the same conditions. This results in the the polarization angle rotating by 90$^\circ$ or becoming erratic. For the density in the central hub of N2071, estimated by \citet{Lyo+21} at $n$(H$_2$) = $3.7 (\pm 2.1) \times 10^{5}$ cm$^{-3}$ or $n_{\rm H} \sim 7 \times 10^{5}$ cm$^{-3}$ in the central 60$\arcsec$, the size threshold for ``wrong'' alignment is 0.15~--~0.5~$\mu$m, depending on grain composition \citep[see][]{Hoang21}, so we can expect it to affect the larger grains there. Since larger grains are cooler and better observed at longer wavelengths, this scenario could explain why the polarization angle is wavelength-dependent, as well as why the central hub has low polarization fractions.

An alternative involves a polarization mechanism that does not rely on grain alignment, PTEAR, which allows non-spherical grains in anisotropic radiation fields to emit polarized radiation as described by \citet{Onaka+95}. Grains with their largest surface facing the heating source will have higher temperatures and emit more than grains with different orientations. Since a grain's thermal emission is polarized in the direction of its longer axis, the PTEAR mechanism tends to produce polarization orthogonal to the direction of the radiation anisotropy.\footnote{
For this polarization to be observed, the direction of the anisotropy must not be parallel to the line of sight.}
This effect does not require grain alignment, although it can be present even when grains are aligned with a magnetic field \citep{Onaka+00}.The PTEAR mechanism is most efficient at short wavelengths, and is therefore best suited to explain the angle difference between the 154 and 214~$\mu$m bands. However, PTEAR effects are generally small compared to those of dust alignment, and they rapidly decrease with wavelength for $\lambda > 100\, \mu$m. We conclude that, while PTEAR may play a role in the origin of $\Delta \theta$, especially at shorter wavelengths, it is unlikely to be the main reason for it.

A final possibility is self-scattering by very large ($\gtrsim 30\, \mu$m) grains. For grain sizes of the order of $\sim \lambda / 2\pi$, where $\lambda$ is the observed wavelength, polarization can arise from self-scattering processes \citep[][]{Kataoka+15, Sadavoy+19, Kirch+Bert20}. One drawback in this scenario is that, while grain growth is expected to take place in dense environments, the sizes required for this effect ($\sim 30\, \mu$m for the HAWC+ bands) are much larger than the $\sim \mu$m grains expected in regions with $N_{\rm H} \sim 10^5 \, {\rm cm}^-3$ \citep[e.g.][]{Hirashita+Li13}, even though some models suggest that outflows can transport large grains out to lower density regions \citep{Wong+16}.

\subsection{P vs. P plots and dust models}
\label{sec:discussion_PvsP}

In this section, we compare the $P$ vs. $P$ plots from Section~\ref{sec:PvsP} to dust polarization models from \citet{Guillet+18}. 
Note that, as discussed in Section~\ref{sec:discussion_polangle}, the change in the polarization angle with wavelength is sign that different bands may be observing different dust populations. Therefore, while a single dust model may reproduce $P$ for the two wavelengths with the best angle match -- 154 and 214~$\mu$m -- we expect that $P$ correlations including the 850~$\mu$m band will be harder to reproduce.

We selected two models from \citet{Guillet+18}, named A and D. Both models include a population of polycyclic aromatic hydrocarbons, one of large amorphous carbon grains \citep{Zubko+96, Compiegne+11} and one of large astronomical silicate grains \citep{Wein+Draine01}. The main differences between the models are that (1) amorphous carbon grains are unaligned in model A and aligned in model D, (2) silicate grains in model D have amorphous carbon inclusions (6\% in volume), and (3) only large grains are aligned in either model, but the size threshold for alignment is 108 nm in model A and 89 nm in model D. We chose these two specific models from \citet{Guillet+18} because they show the largest difference in $P$ for $\lambda \lesssim 200\, \mu$m, thus providing the widest possible fork in predicted values for HAWC+ bands. We calculated the model output using the {\tt DustEM} tool\footnote{\url{https://www.ias.u-psud.fr/DUSTEM/}} \citep[][]{Compiegne+11, Guillet+18}, which is also the one that was used to develop them in the first place.

\begin{figure}
\includegraphics[width=\columnwidth]{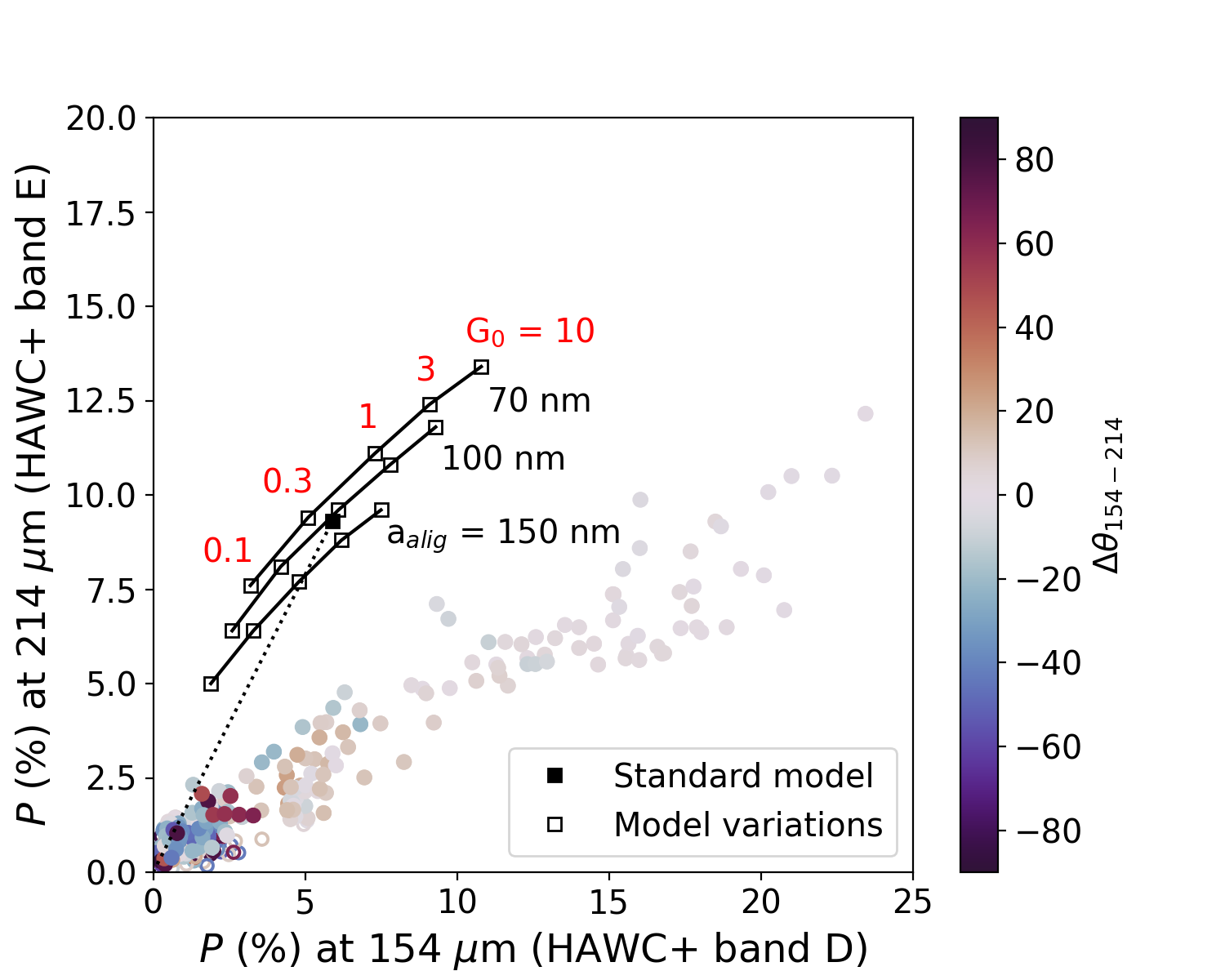}
\includegraphics[width=\columnwidth]{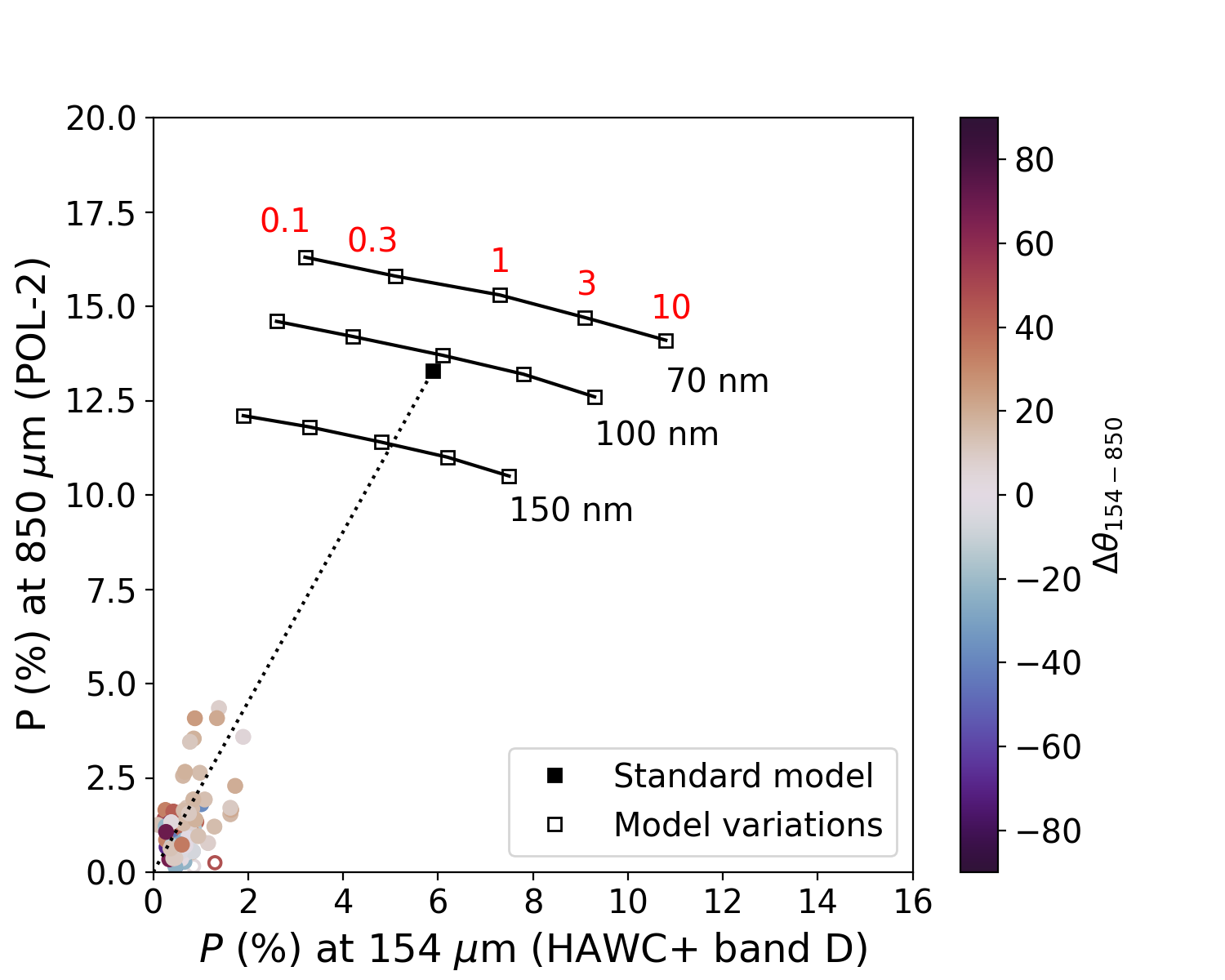}
\includegraphics[width=\columnwidth]{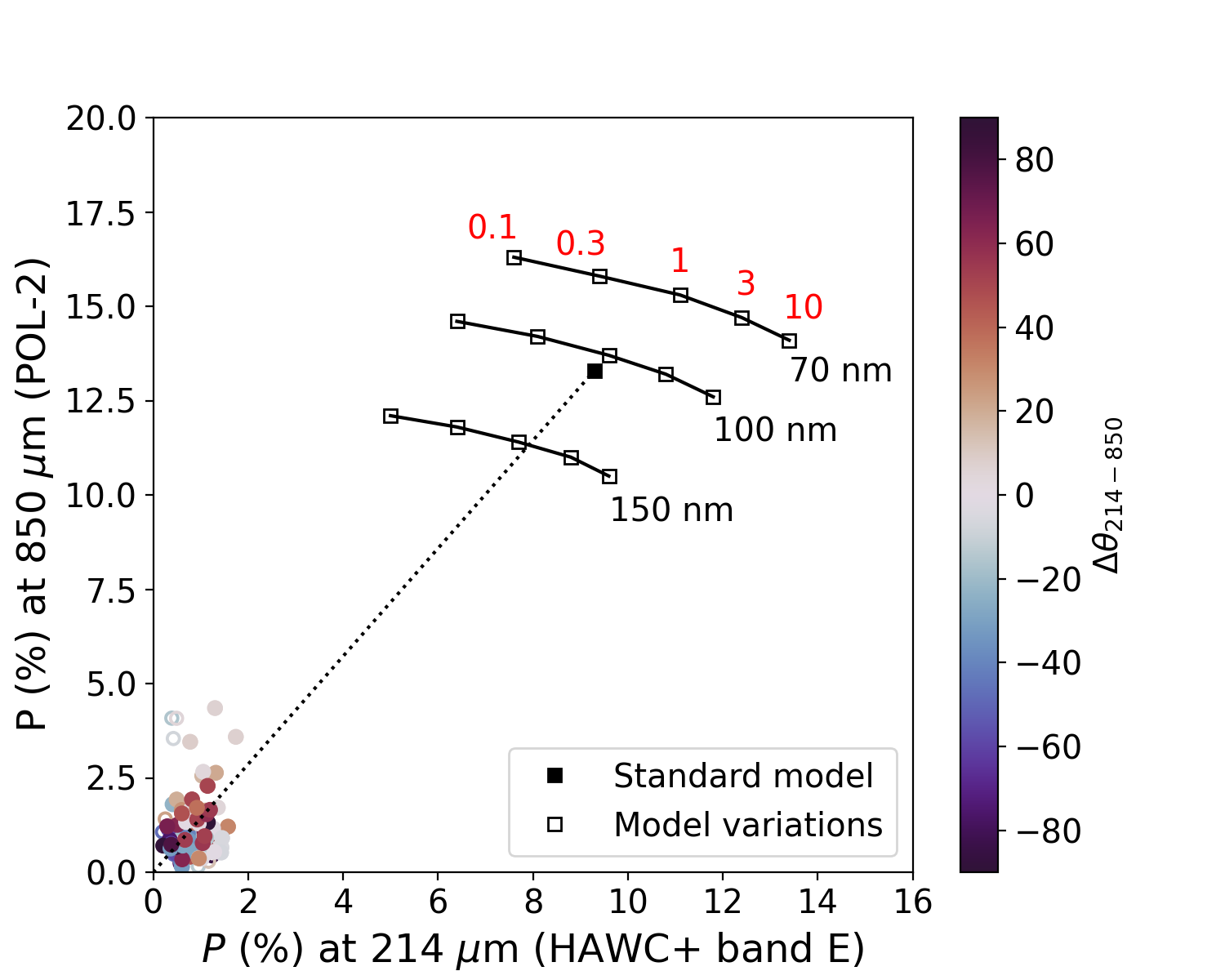}
\caption{
Polarization fraction, compared to a model grid for model A from \citet{Guillet+18} with variable $G_0$ and $a_{\rm alig}$. \textbf{Top:} $P_{214}$ vs. $P_{154}$. \textbf{Middle:} $P_{850}$ vs. $P_{154}$ (for data within 45$\arcsec$ of emission peak). \textbf{Bottom:} $P_{850}$ vs. $P_{214}$ (for data within 45$\arcsec$ of emission peak). The black square marker shows the results for the \citet{Guillet+18} original model, white square markers show the effect of varying the interstellar radiation field intensity and the alignment size threshold. Note that the horizontal axis scale for the top plot and for the bottom two plots is not the same. Color coded by $\Delta \theta$ (see Section~\ref{sec:polangle}).
}
\label{fig:PvsP_grids_A}
\end{figure}

\begin{figure}
\includegraphics[width=\columnwidth]{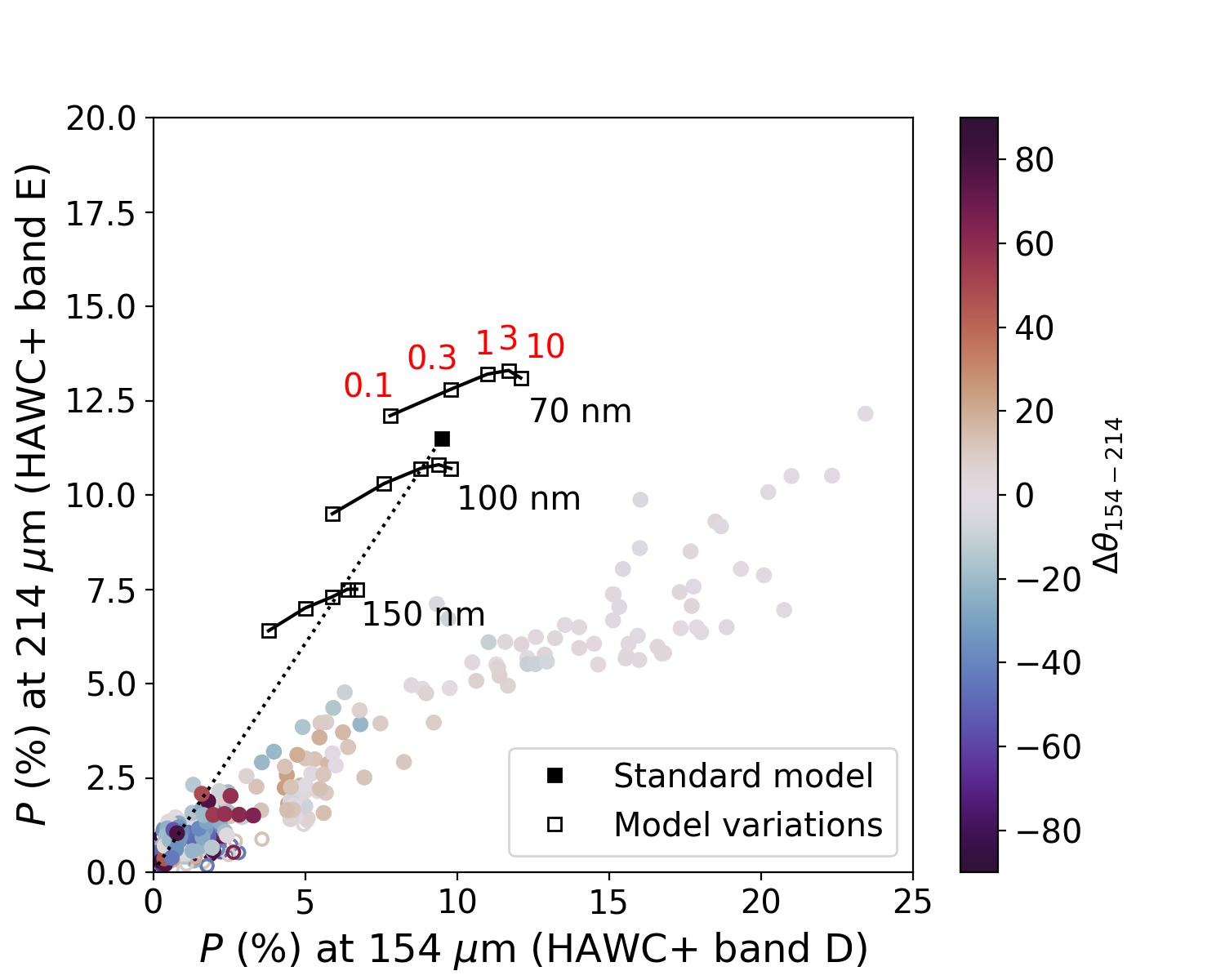}
\includegraphics[width=\columnwidth]{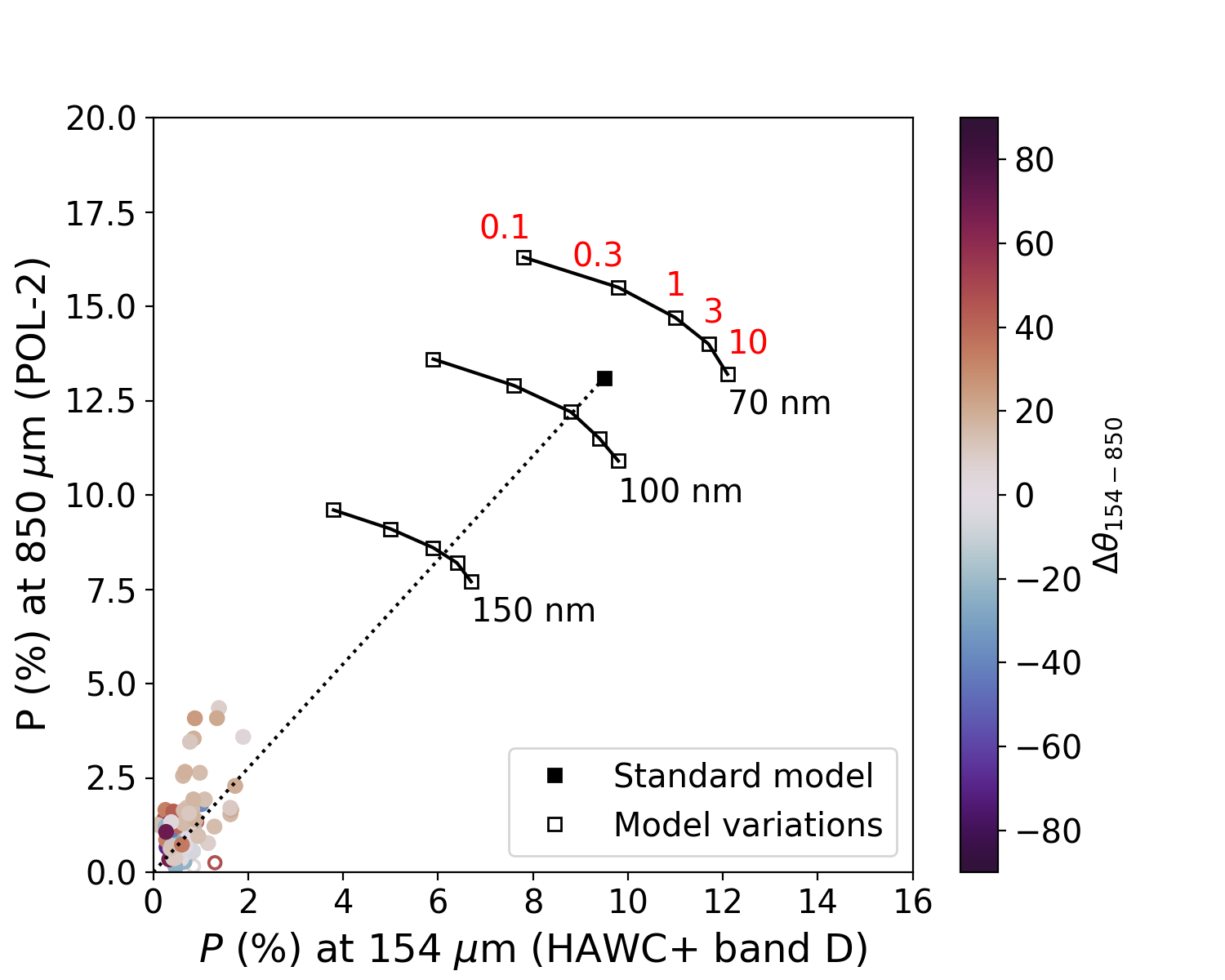}
\includegraphics[width=\columnwidth]{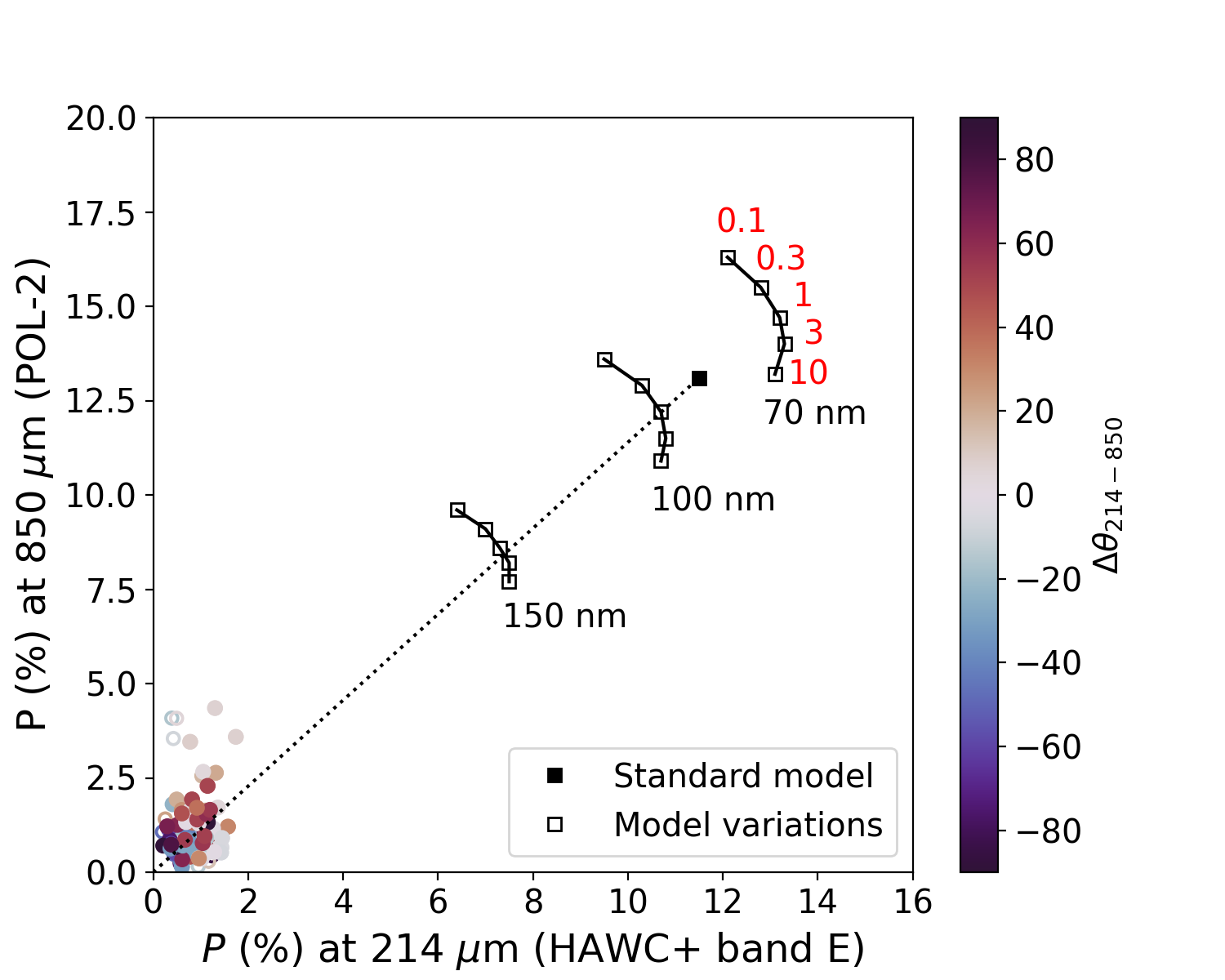}
\caption{
Polarization fraction, compared to a model grid for model D from \citet{Guillet+18} with variable $G_0$ and $a_{\rm alig}$. \textbf{Top:} $P_{214}$ vs. $P_{154}$. \textbf{Middle:} $P_{850}$ vs. $P_{154}$ (for data within 45$\arcsec$ of emission peak). \textbf{Bottom:} $P_{850}$ vs. $P_{214}$ (for data within 45$\arcsec$ of emission peak). The black square marker shows the results for the \citet{Guillet+18} original model, white square markers show the effect of varying the interstellar radiation field intensity and the alignment size threshold. Note that the horizontal axis scale for the top plot and for the bottom two plots is not the same. Color coded by $\Delta \theta$ (see Section~\ref{sec:polangle}).
}
\label{fig:PvsP_grids_D}
\end{figure}

The observational $P$ vs. $P$ plots are compared to these models in Figure~\ref{fig:PvsP_grids_A} (model A) and \ref{fig:PvsP_grids_D} (model D). The polarized fraction predicted by the models has been obtained by interpolation of their output $P(\lambda)$ at 154, 214 and 850~$\mu$m. Square markers show model polarization in the ideal case where the magnetic field is uniform, constant and parallel to the plane of the sky, ($F_{\rm dis}~=~1$, $\gamma~=~0$ in the terminology of equation~\ref{eq:P_decomposition}). Adding the effects of magnetic field orientation and structure back into the picture reduces the model $P$ in all bands. Under our assumption that the magnetic field reduction factors are independent of wavelength, the model result moves along the segment joining the marker (representing the ``maximum polarization'' case) to the origin of the axes. We only show this segment for a single marker in Figures \ref{fig:PvsP_grids_A} and \ref{fig:PvsP_grids_D}, to maintain visibility.
Black markers show results for models as they are in \citet{Guillet+18}. White markers show how the results change when we vary two parameters: radiation field intensity $G_0$ and grain alignment threshold $a_{\rm alig}$. The parameter $G_0$ is a dimensionless multiplicative factor for the interstellar radiation field by \citet{Mathis+83}, while $a_{\rm alig}$ is the threshold size above which most grains are aligned. It is important to bear in mind that the parameter space grids represented here are unlikely to be uniformly populated. For instance, according to the RATs model grain alignment is more efficient in strong radiation fields \citep{Lazarian+Hoang07, Andersson+15, Hoang+Lazarian16}. Within the RAT paradigm we therefore expect a negative correlation between $G_0$ and $a_{\rm alig}$.

The models are unable to reproduce the shape of the $P$ vs. $P$ plot at 154~$\mu$m and 214~$\mu$m, as can be seen in the upper panels of Figures \ref{fig:PvsP_grids_A} and \ref{fig:PvsP_grids_D}. Most notably, the model grid cannot reproduce the low value for the observed $P_{214}/P_{154}$ ratio. This discrepancy could in principle be alleviated by a heterogeneous cloud model (see Section~\ref{sec:intro}). A cold, unaligned or poorly-aligned dust component would ``dilute'' the polarization fraction more efficiently at longer wavelengths, thus lowering the value of the model's $P_{214}/P_{154}$. This is in apparent contradiction with the conclusion in Section~\ref{sec:PvsP}, where the good correlation between $P_{154}$ and $P_{214}$ and the low values of $\Delta \theta_{154-214}$ suggested that, in the southward filament, both bands are seeing the same dust. A possible solution is that, even if there are two or more dust populations in the filament, they trace the same magnetic field and their relative abundance and temperatures are constant along the filament, so that depolarization and wavelength-dependent ``dilution'' are fixed. Another possibility is that the dust properties used in models A and D may not be ideal for molecular cloud dust, since they have been developed to match \textit{Planck} polarization at $\lambda \geq 850\, \mu$m in regions of moderate column density. This would point to the necessity of developing polarized dust models specifically for dense environments.

In the $P_{154}$ vs. $P_{850}$ and $P_{214}$ vs. $P_{850}$ cases, the comparison with a single dust model is not as informative as for $P_{154}$ vs. $P_{214}$, as predicted at the beginning of this section. While the model grids produced can recover the observed $P$ vs. $P$ relations as long as one includes significant depolarization (in the language of equation~\ref{eq:P_decomposition}, for the choice of proper values of $F_{\rm dis}$ and $\gamma$), the majority of observational points have very low polarization and do not show a strong correlation.

\subsection{$P$ vs. $I$ plots}
\label{sec:discussion_PvsI}

We saw in Section~\ref{sec:PvsI} that, while $P$ decreases with column density at all available wavelengths and for both proxies used ($I$ and $\tau$), some of the observed results are not robust. Most notably, the $P$ vs. $I$ relation seems to plateau for high values of $I$ at short wavelengths ($\leq 214 \, \mu$m). However, the flattening is absent, or much less marked, at longer wavelengths (850~$\mu$m) and in the $P$ vs. $\tau$ relation. We interpret this as a sign that, for $\lambda \lesssim 200\, \mu$m, $I$ is strongly dependent on dust temperature and therefore does not make a good proxy for column density. This is relevant to studies that focus on the change in $P$-vs-column density slope, such as \citet{Pattle+19}, who interpret a break in the slope of the $P$ vs. $I$ in $\rho$ Oph as an effect of noise in the low-$I$ section of the plot. It should be noted that studies using $I$ as a proxy for column density are generally conducted at long wavelengths ($\lambda \geq 850\, \mu$m), which are less sensitive to temperature variations than the shorter wavelengths observed by HAWC+ \citep[See also][Section~5.8, for a discussion of the assumptions in using $I$ as a column density tracer]{Pattle+19}. Thus, our results suggest that $I$ as a tracer of column density should be employed with caution. If there is no column density information, long wavelength ($\lambda \gtrsim 850\, \mu$m) intensity maps may be a reasonable proxy for column density, but shorter wavelength data such as those from HAWC+ will be unreliable.

We point out that, while $\tau$ from the HGBS is likely a more robust column density estimator than $I$, it is derived from a single-temperature modified blackbody fit, which is bound to be an oversimplification in the central regions of N2071. Therefore, it should be regarded as one possible estimator of column density, rather than the ``correct'' value for it. This caveat is especially important for the 850~$\mu$m band, as mentioned in Section~\ref{sec:PvsI}.

Another caveat is that, compared to the short-wavelength HAWC+ data, the 850~$\mu$m POL-2 data suffers from loss of large-scale, low-intensity flux. Therefore, the direct comparison of $P$ vs. $I$ curves between HAWC+ and POL-2 may be misleading at low intensities.

\section{Conclusions}
\label{sec:conclusions}

We made a multi-wavelength study of FIR/submm polarization in NGC 2071. By combining 850~$\mu$m data from the BISTRO (JCMT SCUBA-2/POL-2) survey with SOFIA/HAWC+ observations in bands D (154~$\mu$m) and E (214~$\mu$m) we managed to examine a wider wavelength range than previous single-instrument studies \citep[e.g.][]{Santos+19, Michail+21}; furthermore, we analyse the change of polarization angle with wavelength rather than limiting the analysis to areas where the angle is constant. The areas we selected in N2071 can be described as a central hub and a southward-pointing filament. More filaments radiating from the hub can be seen at 850~$\mu$m, but are not as evident at short wavelengths. The comparison of polarization angles $\theta$ and polarization fraction $P$ between different bands yielded the following results:

\begin{enumerate}
    \item The distribution of polarization angle difference $\Delta \theta_{154-214}$ between 154~$\mu$m and 214~$\mu$m peaks at $\sim0^\circ$; however, it has wide wings indicating that the angle differs significantly over part of the area studied (Section~\ref{sec:polangle}). The angle difference $\Delta \theta_{154-850}$ between 154~$\mu$m and 850~$\mu$m is wider and has a less marked central peak, while the distribution for $\Delta \theta_{214-850}$ shows no main peak at all. This indicates that the angle differences between 154~$\mu$m and 850~$\mu$m are greater than those between 154~$\mu$m and 214~$\mu$m, and those between 214~$\mu$m and 850~$\mu$m even more so.
    \item One consequence is that where $\Delta \theta$ is significant one cannot obtain a magnetic field direction simply by rotating polarization angles by 90$^\circ$. The relation between magnetic field and polarization angles will depend on the physical origin of $\Delta \theta$ and on the wavelength observed.
    \item Lines of sight with significant $\Delta \theta_{154-214}$ are concentrated in the central hub. The central hub area, which contains an outflow and embedded infrared sources, has a complex structure and temperature distribution, lending credence to the explanation that the $\Delta \theta$ can be explained by changes in the magnetic field morphology and dust temperature along the line of sight. 
    \item Alternative explanations for the observed $\Delta \theta$ (Section~\ref{sec:discussion_polangle}) include the loss of internal alignment for grains in dense environments and polarization produced by anisotropic radiation (PTEAR). Other phenomena that we considered, such as the alignment of grains on the direction of radiation field anisotropy (k-RATs) and scattering by very large grains, are at odds with either the observed $\theta$ patterns or grain growth models and seem less plausible. Further modelling work is needed to determine if these phenomena can account for the observed behavior of $\Delta \theta$.
    \item We compared the polarization observed in our three bands to predictions from dust models (Section~\ref{sec:discussion_PvsP}). We found that the models failed to reproduce the observations, even when we included parameter variations. 
    This seems to support the idea that a heterogeneous cloud model with more than one dust population may be needed to reproduce observations \citep{Hildebrand+99, Michail+21}. Dust models specifically developed for dense clouds may also be necessary. Most notably, the models predict a higher ratio of $P_{214}$ to $P_{154}$ than the observed value of $\sim0.5$. Note that our value for $P_{214} / P_{154}$ is lower than the $\sim 1$ obtained by \citet{Michail+21} in OMC-1, suggesting strong environmental dependence.
    \item We also compared $P$ versus column density, using as proxy for column density the intensity $I$ and the optical depth $\tau$ from \textit{Herschel} in all three bands (Section~\ref{sec:PvsI}). In general, we found $P$ decreases with with column density as expected. However, the trends with $I$ showed a plateau at the highest intensities at 154~$\mu$m and 214~$\mu$m whereas the trends with $\tau$ did not (Section~\ref{sec:discussion_PvsI}). We recommend caution when using $I$ as a tracer of optical depth, especially at short wavelengths (FIR as opposed to submm/mm).
\end{enumerate}

Our overall conclusion is that although we were unable to fit a single dust model to our data, multi-wavelength polarimetry revealed a number of interesting variations that would otherwise have gone unnoticed in a single-band study. The combination of ``short-wavelength'' HAWC+ and ``long-wavelength'' POL-2 data was crucial in obtaining these results. Future work will include modelling the temperature gradient and dust size distribution in the cloud core, as well as extending this type of analysis to other Gould Belt molecular clouds such as Serpens Main, $\rho$ Ophiuchi and Orion A, to determine whether or not the same behavior is observed.

\textit{Facilities:} JCMT (SCUBA-2, POL-2), SOFIA (HAWC+), Herschel (SPIRE, PACS).

\textit{Software:} Starlink, Numpy, Matplotlib, APLpy, {\tt DustEM} 

\bigskip

\textit{This is a pre-copyedited, author-produced PDF of an article accepted for publication} in Monthly Notices of the Royal Astronomical Society \textit{following peer review. The version of record (}MNRAS\textit{, Volume 512, Issue 2, May 2022, Pages 1985–2002) is available online at: \url{https://academic.oup.com/mnras/article-abstract/512/2/1985/6544656}}

\section*{Acknowledgements}

We thank Jonathan Marshall for his help in testing the online scripts provided with the article, and Daniel Cotton for his help with converting the subtraction of polarization angles to code.

The James Clerk Maxwell Telescope is operated by the East Asian Observatory on behalf of The National Astronomical Observatory of Japan; Academia Sinica Institute of Astronomy and Astrophysics; the Korea Astronomy and Space Science Institute; Center for Astronomical Mega-Science (as well as the National Key R\&D Program of China with No. 2017YFA0402700). Additional funding support is provided by the Science and Technology Facilities Council of the United Kingdom and participating universities and organizations in the United Kingdom, Canada and Ireland. Additional funds for the construction of SCUBA-2 were provided by the Canada Foundation for Innovation. The authors wish to recognize and acknowledge the very significant cultural role and reverence that the summit of Maunakea has always had within the indigenous Hawaiian community.  We are most fortunate to have the opportunity to conduct observations from this mountain.

JCMT data for this program were obtained under M16AL004.

This research was conducted in part at the SOFIA Science Center, which is operated by the Universities Space Research Association under contract NNA17BF53C with the National Aeronautics and Space Administration.

This research has made use of the NASA/IPAC Infrared Science Archive, which is funded by the National Aeronautics and Space Administration and operated by the California Institute of Technology.

This research has made use of data from the Herschel Gould Belt survey (HGBS) project (http://gouldbelt-herschel.cea.fr). The HGBS is a Herschel Key Programme jointly carried out by SPIRE Specialist Astronomy Group 3 (SAG 3), scientists of several institutes in the PACS Consortium (CEA Saclay, INAF-IFSI Rome and INAF-Arcetri, KU Leuven, MPIA Heidelberg), and scientists of the Herschel Science Center (HSC).

This research used the facilities of the Canadian Astronomy Data Centre operated by the National Research Council of Canada with the support of the Canadian Space Agency. 

L.~F. and F.~K. acknowledge support from the Ministry of Science and Technology of Taiwan, under grant MoST107-2119-M-001-031-MY3 and from Academia Sinica under grant AS-IA-106-M03. 

M.~T. is supported by JSPS KAKENHI grant Nos. 18H05442, 15H02063, and 22000005. J.~K. is supported JSPS KAKENHI grant No.19K14775.

C.~L.~H.~H. acknowledges the support of the NAOJ Fellowship and JSPS KAKENHI grants 18K13586 and 20K14527.

S.~P.~L.  acknowledges grants from the Ministry of Science and Technology of Taiwan 106-2119-M-007 -021 -MY3 and 109-2112-M-007 -010 -MY3).

E.~J.~C. was supported by the National Research Foundation of Korea(NRF) grant funded by the Korea government(MSIT) (No. NRF-2019R1I1A1A01042480).

D.~J. is supported by the National Research Council of Canada and by an NSERC Discovery Grant.

K.~H.~Kim is supported by the center for Women In Science, Engineering and Technology (WISET) grant funded by the Ministry of Science and ICT (MSIT) under the program for returners into R\&D (WISET-2019-288; WISET-2020-247).

W.~K. was supported by the National Research Foundation of Korea (NRF) grant funded by the Korea government (MSIT) (NRF-2021R1F1A1061794).

T.~H. acknowledges the support by the National Research Foundation of Korea (NRF) grant funded by the Korea government (MSIT) through the Mid-career Research Program (2019R1A2C1087045).

C.~W.~L. is supported by the Basic Science Research Program through the NRF funded by the Ministry of Education, Science and Technology (NRF-2019R1A2C1010851).

P.~M.~K is supported by the Ministry of Science and Technology (MoST) through grants MoST 109-2112-M-001-022 and MoST 110-2112-M-001-057.

A.~S. was supported by National Science Foundation under grant AST-1715876.

C.~E. acknowledges the financial support from grant RJF/2020/000071 as a part of Ramanujan Fellowship awarded by Science and Engineering Research Board (SERB), Department of Science and Technology (DST), Govt. of India.

I.~W.~S. acknowledges financial support for this work was provided by NASA through award \#06\_0119 issued by USRA.

\section*{Data Availability}

The scripts used in this paper can be found at the following Github: \url{https://github.com/lfanciullo/Fanciullo_etal_2022_BISTRO-HAWCplus_N2071_polarimetry}

The BISTRO data for N2071 can be downloaded from \url{https://www.eaobservatory.org/jcmt/data-access/}

The HAWC+ data for N2071 can be found under the target name `Orion B N2071' on the SOFIA IRSA database at \url{https://irsa.ipac.caltech.edu/applications/sofia/}

The HGBS files for Orion B can be downloaded from \url{ http://gouldbelt-herschel.cea.fr/archives}




\bibliographystyle{mnras}
\bibliography{biblio}



\appendix

\section{Three-band comparison of polarization angles}
\label{sec:3bdpol}

\begin{figure}
    \centering
    \includegraphics[width=\columnwidth]{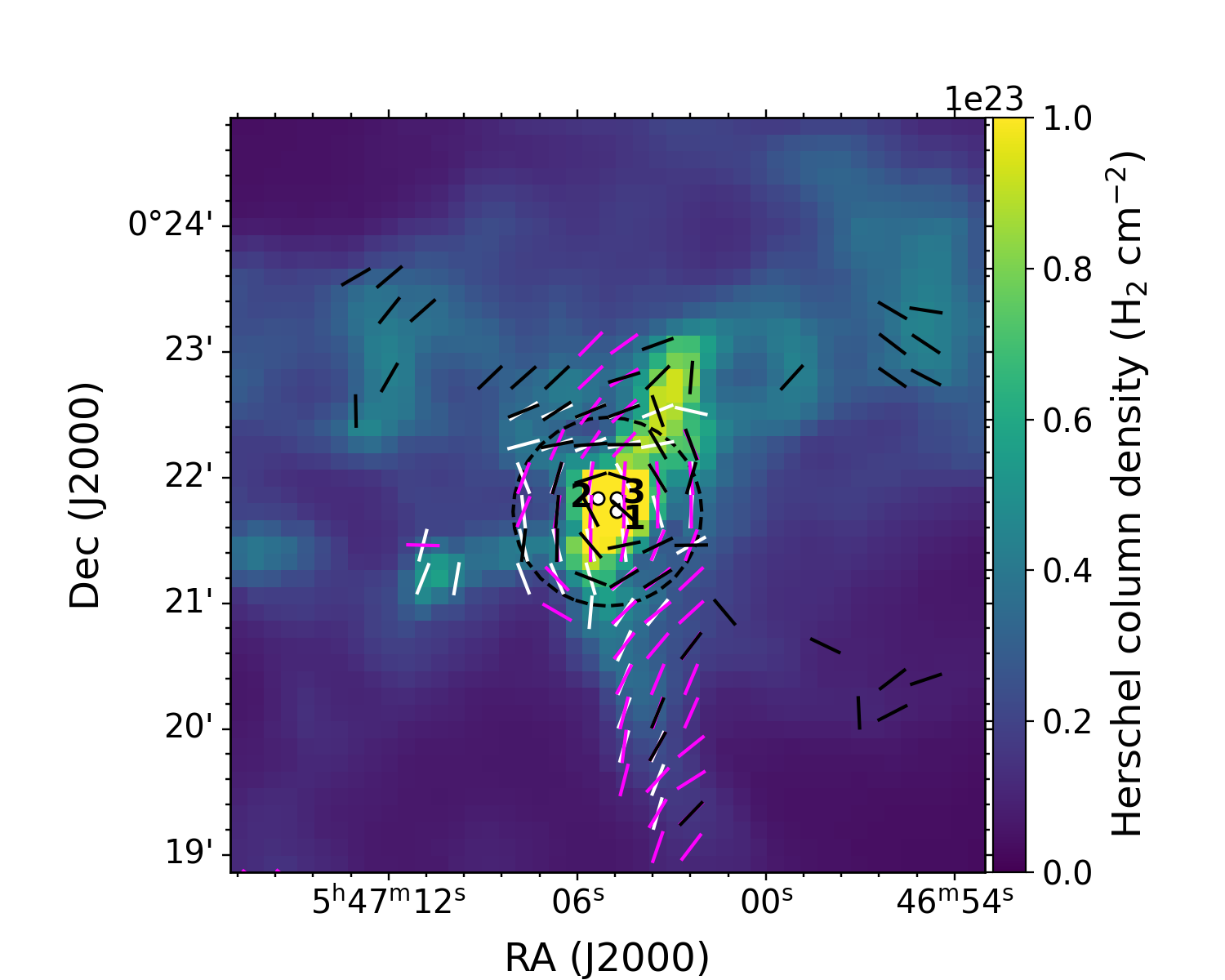}
    \caption{
    Polarization vector directions for all three bands, superposed on the \textit{Herschel} $N_{\rm H_{2}}$ column density map. Vectors for the 154~$\mu$m band are in white, vectors for the 214~$\mu$m band are in magenta and vectors for the 850~$\mu$m band are in black. The dashed circle shows a 90$\arcsec$ diameter centered on the emission peak. The side of the map measures 6$\arcmin$. Compare Figure~\ref{fig:map_all}.
    }
    \label{fig:3band_theta_quiverplot}
\end{figure}

Figure~\ref{fig:3band_theta_quiverplot} shows a quiver plot of the polarization angles in all three bands for comparison purposes. The angles are superposed on the central 6$\arcmin$ of the \textit{Herschel} $N_{\rm H_{2}}$ column density map. Only vectors with $P/\delta P > 3$ are shown. As in Section~\ref{sec:maps}, we show every other polarization vector. The distance between polarization vectors is therefore 16$\arcsec$.

\section{Preliminary comparison of selection and filtered maps}
\label{sec:filtered}

\begin{figure*}
\includegraphics[width=.49\textwidth]{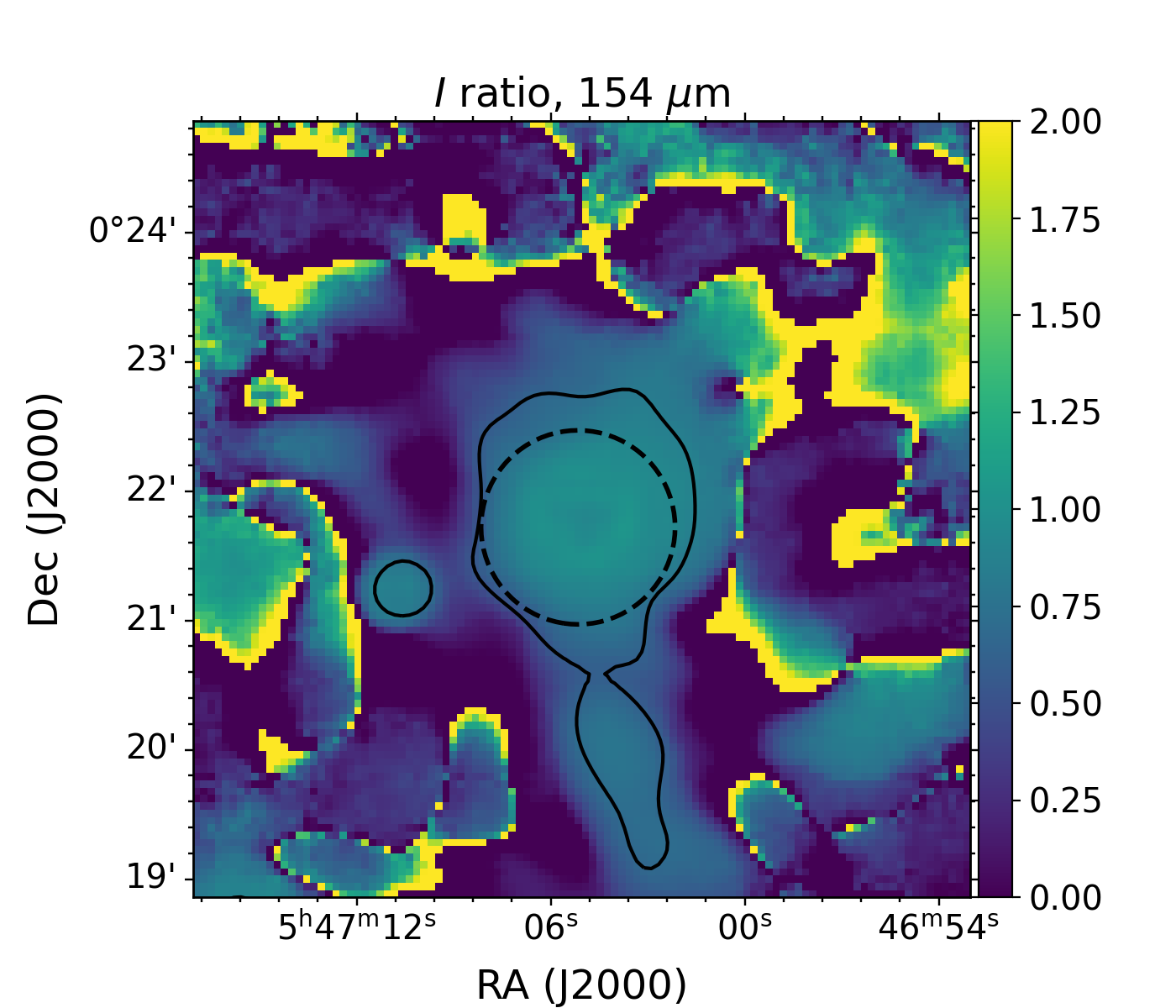}
\includegraphics[width=.49\textwidth]{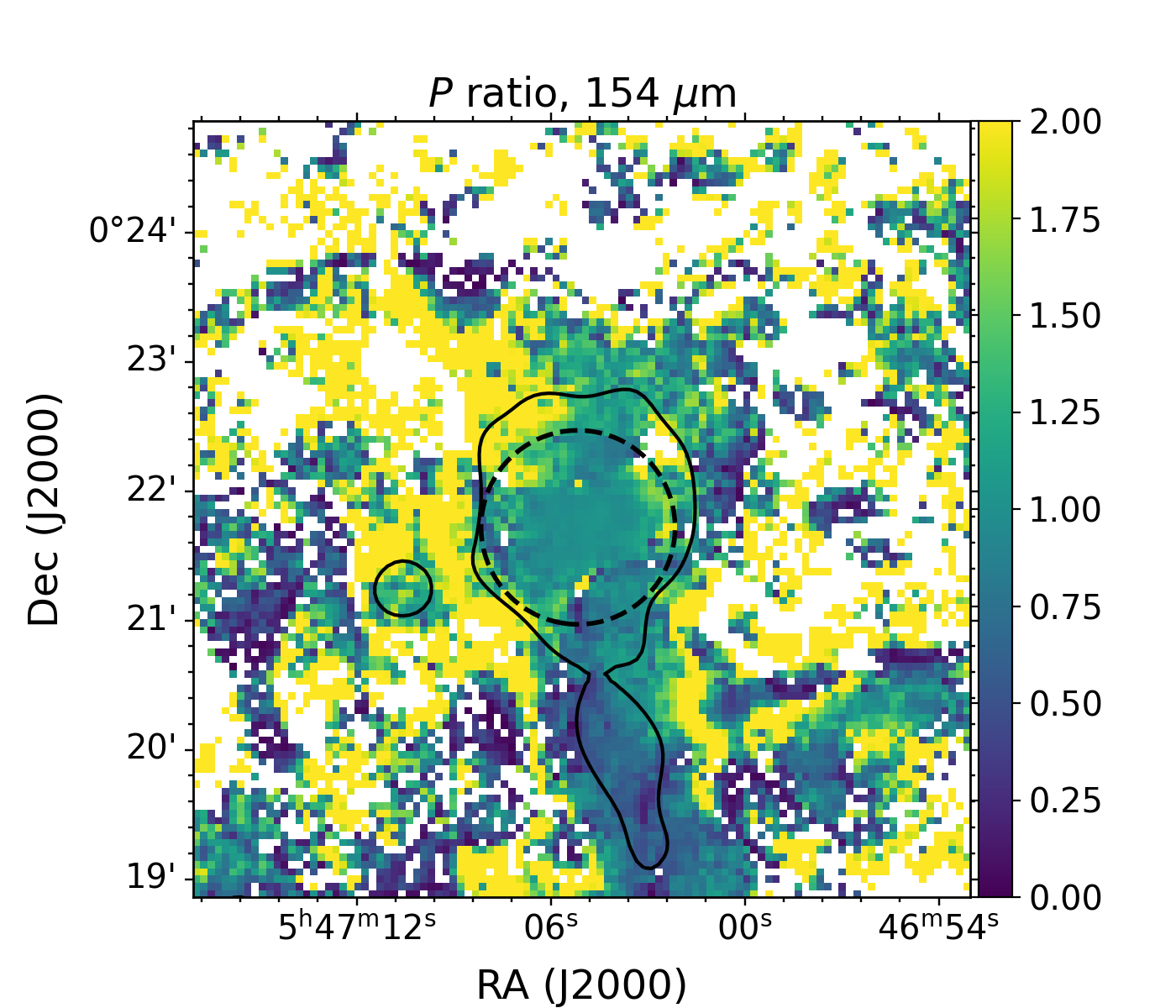}
\includegraphics[width=.49\textwidth]{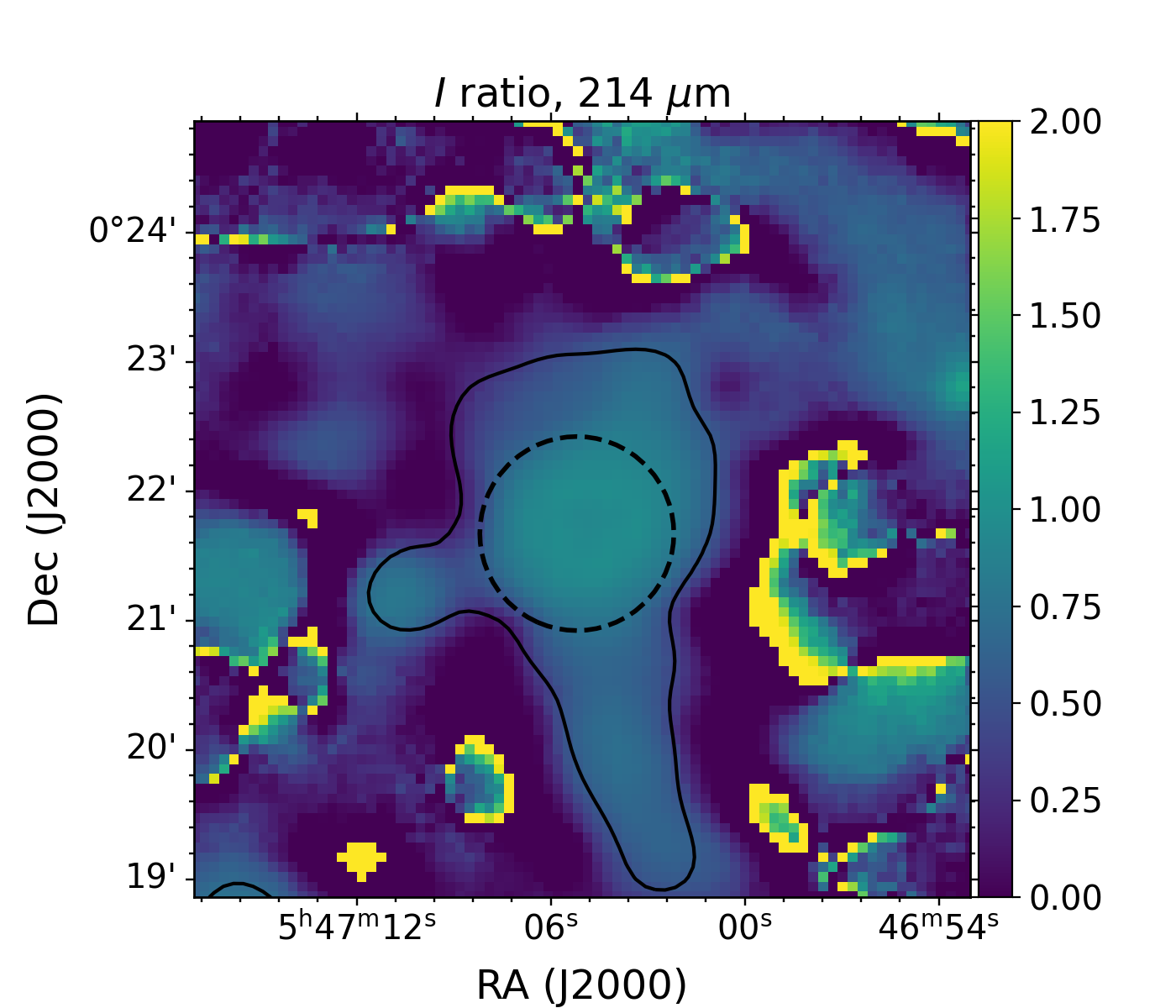}
\includegraphics[width=.49\textwidth]{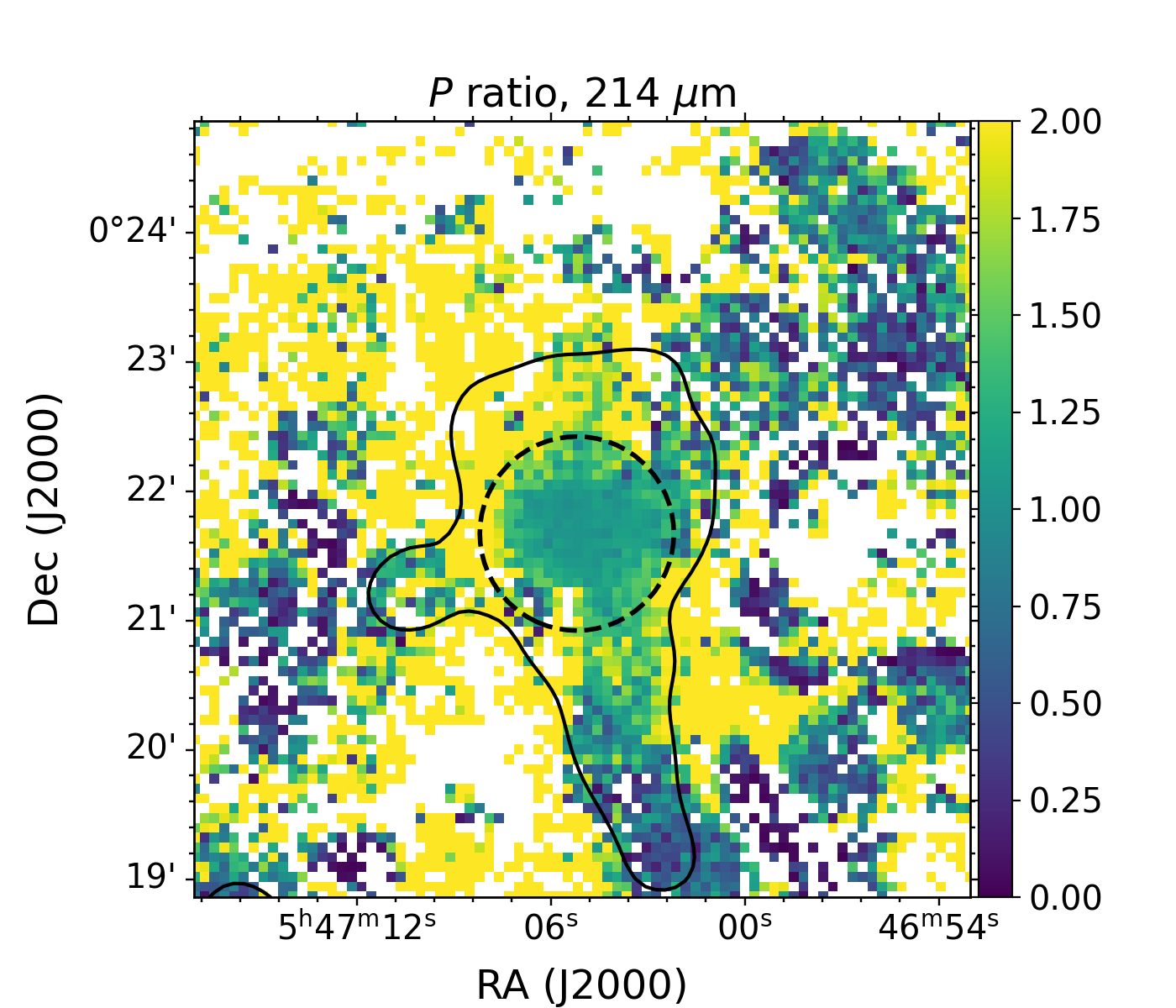}
\caption{
Effects of POL-2-like filtering on $I$ and $P$ HAWC+ maps. \textbf{Top left:} Ratio of filtered to unfiltered $I$ at 154~$\mu$m. \textbf{Top right:} Ratio of filtered to unfiltered $P$ at 154~$\mu$m. \textbf{Bottom left:} Ratio of filtered to unfiltered $I$ at 214~$\mu$m. \textbf{Bottom right:} Ratio of filtered to unfiltered $P$ at 214~$\mu$m. The dashed contours indicate 45$\arcsec$ from the 850~$\mu$m emission peak; solid contours show flux thresholds of 25 mJy arcsec$^{-2}$ (154~$\mu$m) and 15 mJy arcsec$^{-2}$ (214~$\mu$m).
}
\label{fig:filtering}
\end{figure*}

SCUBA-2 is insensitive to large-scale structure due to atmospheric filtering effects \citep{Chapin+13}, and the low scanning speed of POL-2 exacerbates this effect \citep{Friberg+16}. In order to test where the size scales detected in the POL-2 and HAWC+ data are comparable, we filtered large-scale emission from the HAWC+ maps in order to match the spatial scales in the POL-2 data.  To do so, the POL-2 data reduction process was applied to the SOFIA observations, following the method developed by the JCMT Gould Belt Survey for filtering Herschel Space Telescope data to match SCUBA-2 spatial scales, as described in detail by \citet{Sadavoy+13} and \citet{Pattle+15}. We first recentred the SOFIA maps to match the central coordinates of a POL-2 field without significant emission in the centre of the map. The SOFIA $I$, $Q$ and $U$ data were added to the POL-2 $I$, $Q$ and $U$ bolometer time series for this field, and the second stage of the reduction process, as described by \citet{Lyo+21}, was repeated, including the application of the fixed mask for NGC 2071 to the SOFIA data.  We scaled the SOFIA fluxes by a factor of 0.1 before calibrating them from their native units into pW of artificial POL-2 emission, adding them to the POL-2 timestream.  In this way, the significantly higher surface brightness of dust emission in the HAWC+ wavebands compared to that at 850~$\mu$m does not prevent the POL-2 map-making process from converging.  This scaling was reversed at the end of the filtering process, and the central coordinates of the map were returned to their original values.

The results of the filtering process are shown in Figure~\ref{fig:filtering}. The figure consists of the maps of the ratio of filtered $I$ to unfiltered $I$ and that of filtered $P$ to unfiltered $P$, for both HAWC+ bands (154 and 214~$\mu$m). Within 45$\arcsec$ of the POL-2 emission peak -- as shown by a dashed circle -- most pixels show ratios close to 1, although $P$ shows more variability. These maps are a preliminary result, sampled at the same pixel size as the orignal HAWC+ observations (3\hbox{$\,.\!\!^{\prime\prime}$}40 for band D and 4\hbox{$\,.\!\!^{\prime\prime}$}55 for band E). A fuller analysis of filtering, including the effects of smoothing and resampling, will be the subject of a follow-up paper.

\section{Author affiliations}
\label{affiliations}

$^{1}$Academia Sinica Institute of Astronomy and Astrophysics, 11F of AS/NTU Astronomy-Mathematics Building, No.1, Sec. 4, Roosevelt Rd, Taipei 10617, Taiwan, R.O.C.\\
$^{2}$European Southern Observatory, Karl-Schwarzschild-Str.~2, 85748 Garching, Germany\\
$^{3}$Centre for Astronomy, School of Physics, National University of Ireland Galway, University Road, Galway H91 TK33, Ireland\\
$^{4}$Department for Physics, Engineering Physics and Astrophysics, Queen's University, Kingston, ON, K7L 3N6, Canada\\
$^{5}$SOFIA Science Center, Universities Space Research Association, NASA Ames Research Center, Moffett Field, California 94035, USA\\
$^{6}$Korea Astronomy and Space Science Institute, 776 Daedeokdae-ro, Yuseong-gu, Daejeon 34055, Republic of Korea\\
$^{7}$Department of Physics, Faculty of Science and Engineering, Meisei University, 2-1-1 Hodokubo, Hino, Tokyo 191-8506, Japan\\
$^{8}$Department of Astronomy, Graduate School of Science, The University of Tokyo, 7-3-1 Hongo, Bunkyo-ku, Tokyo 113-0033, Japan\\
$^{9}$Universit\'e Paris-Saclay, CNRS, CEA, Astrophysique, Instrumentation et Mod\'elisation de Paris-Saclay, 91191 Gif-sur-Yvette, France\\
$^{10}$Aix Marseille Univ, CNRS, CNES, LAM, Marseille, France\\
$^{11}$East Asian Observatory, 660 N. A'oh\={o}k\={u} Place, University Park, Hilo, HI 96720, USA\\
$^{12}$CAS Key Laboratory of FAST, National Astronomical Observatories, Chinese Academy of Sciences, People's Republic of China\\
$^{13}$Department of Astronomy and Space Science, Chungnam National University, 99 Daehak-ro, Yuseong-gu, Daejeon 34134, Republic of Korea\\
$^{14}$Tokushima University, Minami Jousanajima-machi 1-1, Tokushima 770-8502, Japan\\
$^{15}$National Astronomical Observatory of Japan, Alonso de C\'ordova 3788, Office 61B, Vitacura, Santiago, Chile\\
$^{16}$Joint ALMA Observatory, Alonso de C\'ordova 3107, Vitacura, Santiago, Chile\\
$^{17}$NAOJ Fellow\\
$^{18}$University of Science and Technology, Korea (UST), 217 Gajeong-ro, Yuseong-gu, Daejeon 34113\\
$^{19}$NRC Herzberg Astronomy and Astrophysics, 5071 West Saah Road, Victoria, BC, V9E 2E7, Canada\\
$^{20}$Department of Physics and Astronomy, University of Victoria, 3800 Finnerty Road, Elliot Building, Victoria, BC, V8P 5C2, Canada\\
$^{21}$Department of Physics and Astronomy, University College London, WC1E 6BT London, UK\\
$^{22}$Jeremiah Horrocks Institute, University of Central Lancashire, Preston PR1 2HE, UK\\
$^{23}$Department of Earth Science Education, Seoul National University, 1 Gwanak-ro, Gwanak-gu, Seoul 08826, Republic of Korea\\
$^{24}$SNU Astronomy Research Center, Seoul National University, 1 Gwanak-ro, Gwanak-gu, Seoul 08826, Republic of Korea\\
$^{25}$Institute of Astronomy and Department of Physics, National Tsing Hua University, Hsinchu 30013, Taiwan\\
$^{26}$Key Laboratory for Research in Galaxies and Cosmology, Shanghai Astronomical Observatory, Chinese Academy of Sciences, 80 Nandan Road, Shanghai 200030, People’s Republic of China\\
$^{27}$National Astronomical Observatory of Japan, National Institutes of Natural Sciences, Osawa, Mitaka, Tokyo 181-8588, Japan\\
$^{28}$Xinjiang Astronomical Observatory, Chinese Academy of Sciences, 830011 Urumqi, People's Republic of China\\
$^{29}$School of Physics and Astronomy, Cardiff University, The Parade, Cardiff, CF24 3AA, UK\\
$^{30}$Astrobiology Center, Osawa, Mitaka, Tokyo 181-8588, Japan\\
$^{31}$Department of Physics and Astronomy, Graduate School of Science and Engineering, Kagoshima University, 1-21-35 Korimoto, Kagoshima, Kagoshima 890-0065, Japan\\
$^{32}$Indian Institute of Astrophysics, II Block, Koramangala, Bengaluru 560034, India\\
$^{33}$Indian Institute of Science Education and Research (IISER) Tirupati, Rami Reddy Nagar, Karakambadi Road, Mangalam (P.O.), Tirupati 517 507, India\\
$^{34}$Department of Earth, Environment, and Physics, Worcester State University, Worcester, MA 01602, USA\\
$^{35}$Center for Astrophysics | Harvard \& Smithsonian, 60 Garden Street, Cambridge, MA 02138, USA\\


\bsp	
\label{lastpage}
\end{document}


\section*{Analysis using absolute $\Delta \theta$ values}
\label{sec:Dtheta0-90}

\begin{figure}
    \centering
    \includegraphics[width=\columnwidth]{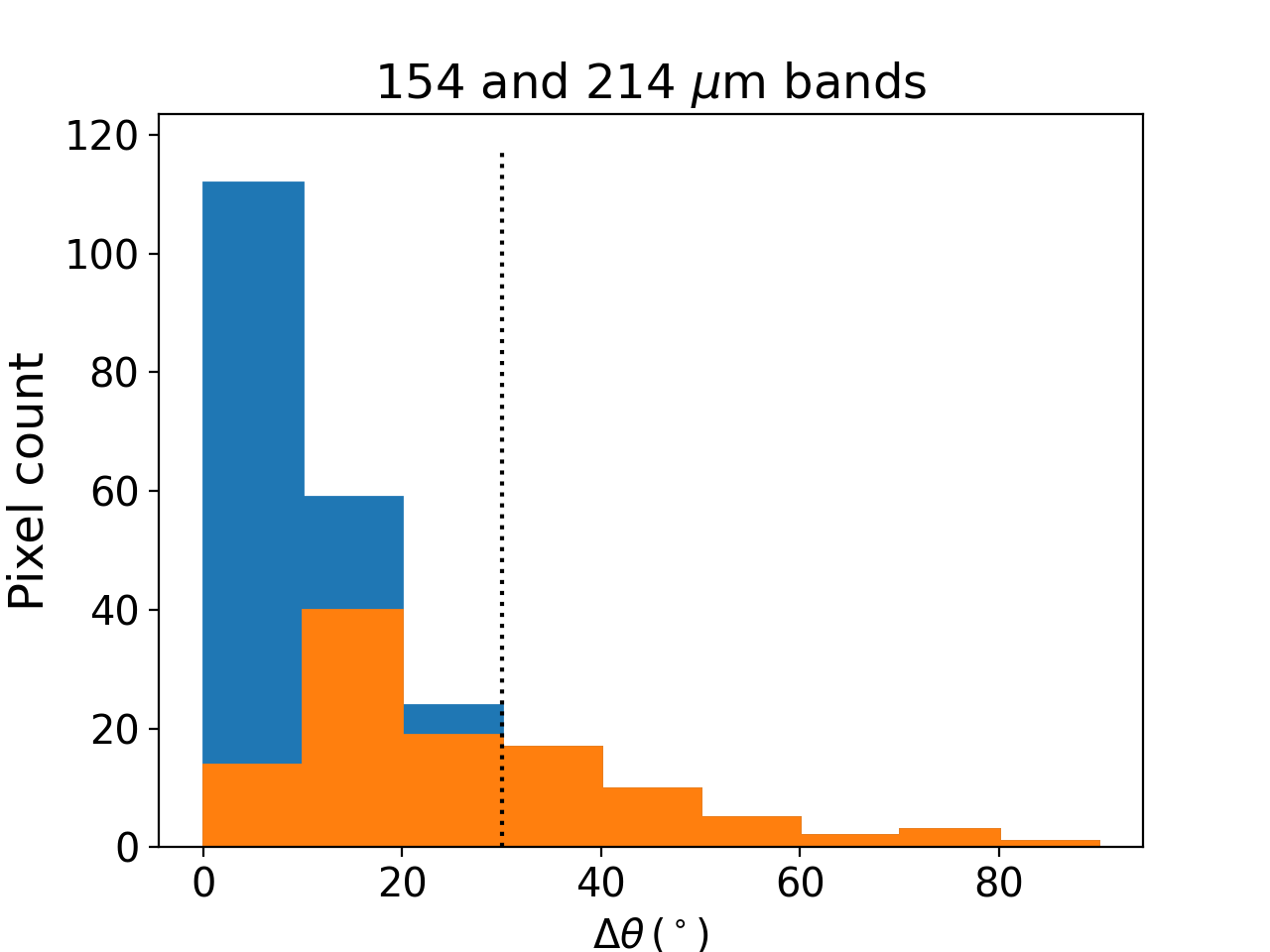}
    \includegraphics[width=\columnwidth]{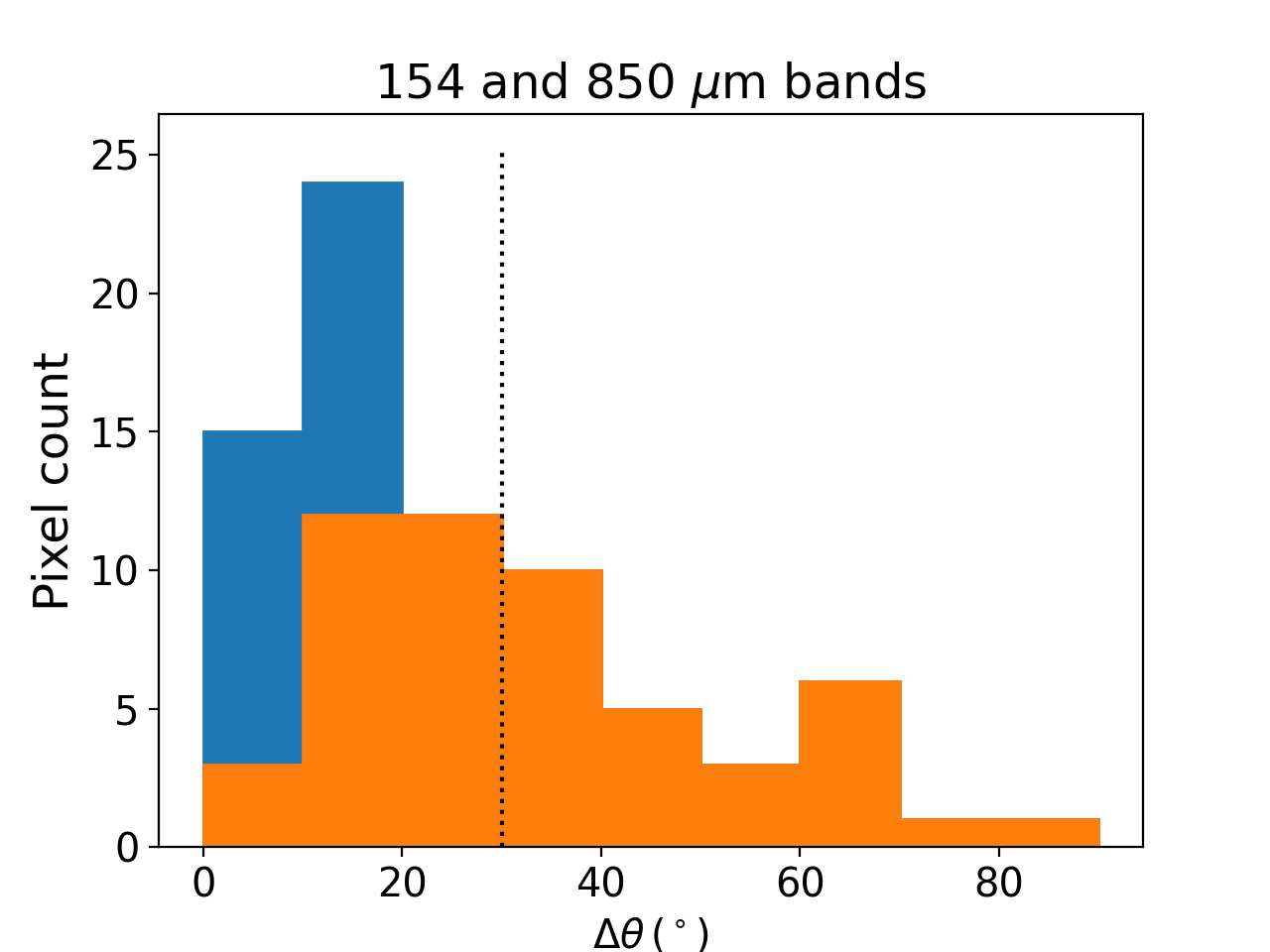}
    \includegraphics[width=\columnwidth]{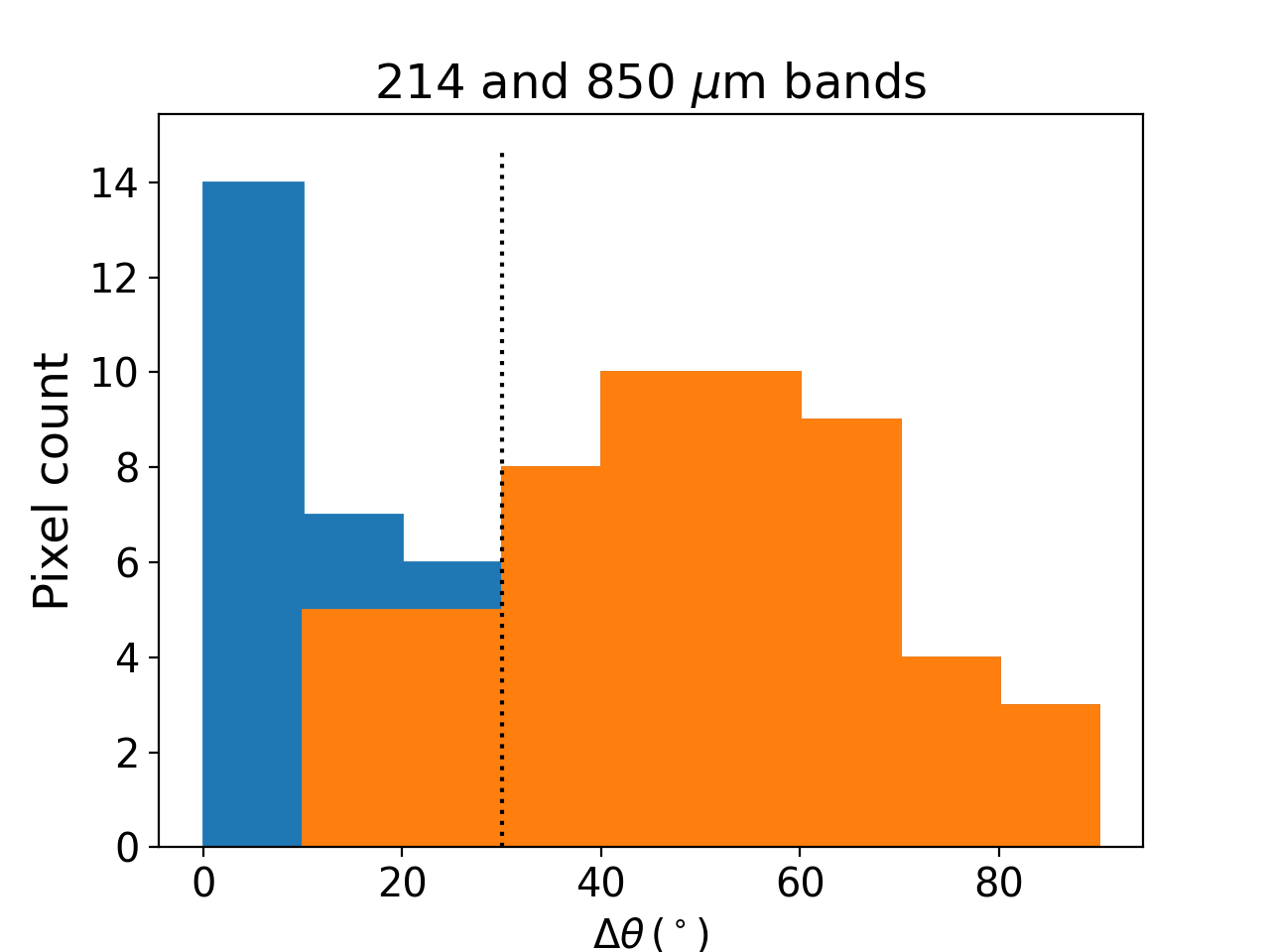}
    \caption{
    Histogram of the difference in polarization angles (absolute value) between 154 and 214 $\mu$m (top), between 154 and 850~$\mu$m (middle) and between 214 and 850~$\mu$m (bottom). Blue bars show data where the angles agree within $3\sigma$ ($\Delta \theta / \delta \Delta \theta < 3$), orange bars show data where angles differ by $3\sigma$ or more. The vertical dotted line shows the $30^\circ$ limit. Compare Figure 3 in the main paper.
    }
    \label{fig:theta-abs_compare}
\end{figure}

\begin{figure}
\includegraphics[width=.95\columnwidth]{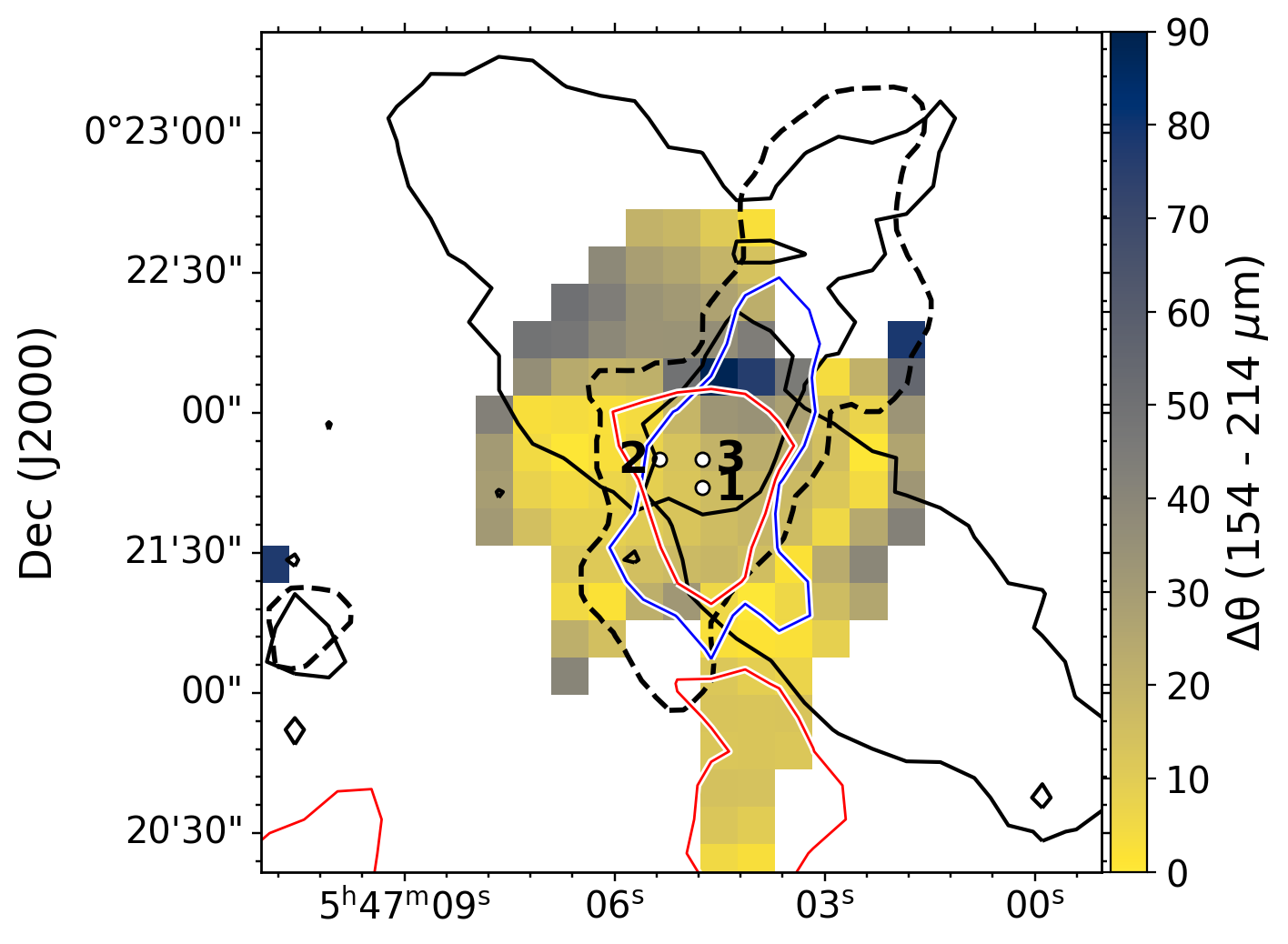}
\includegraphics[width=.95\columnwidth]{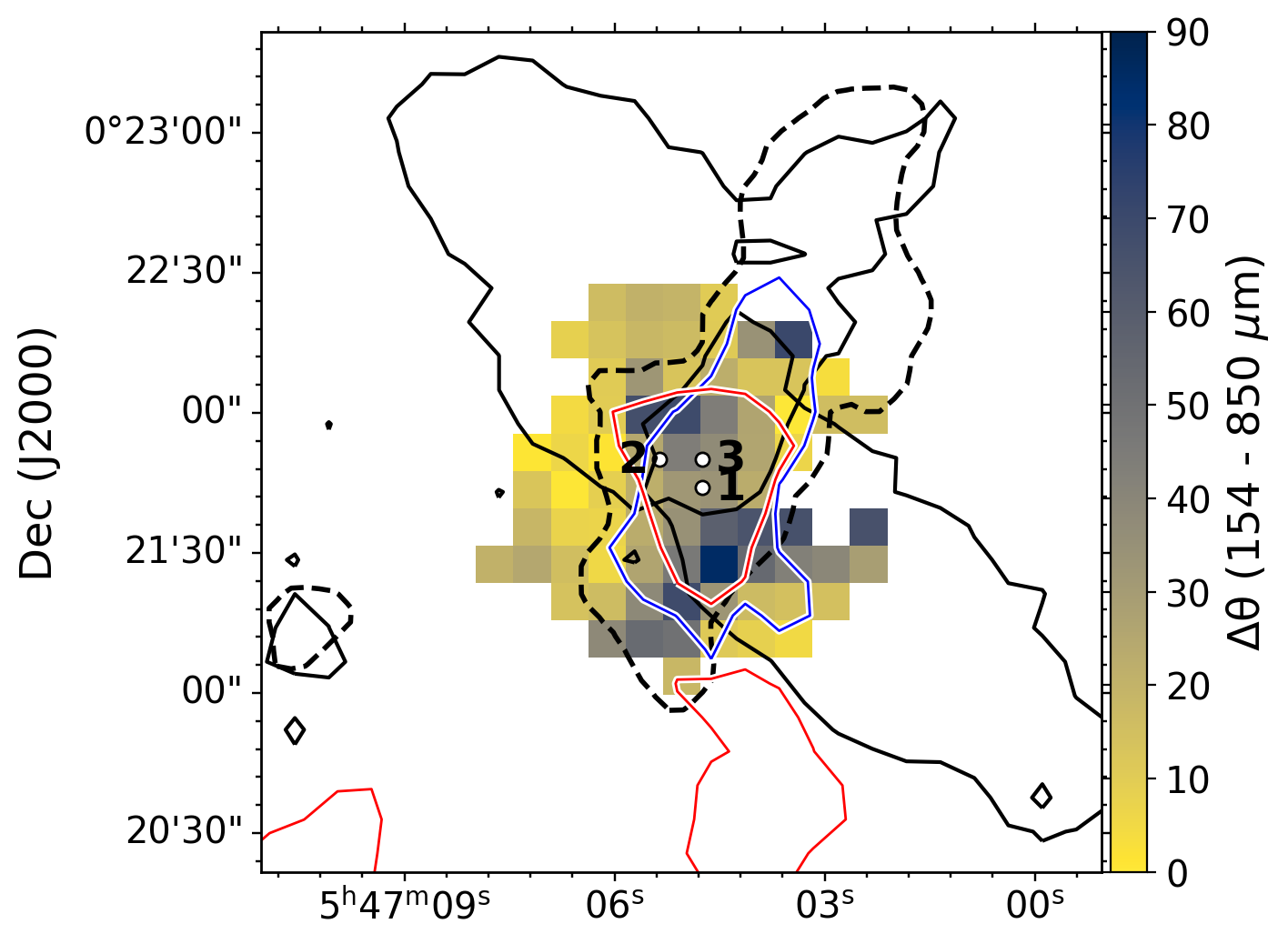}
\includegraphics[width=.95\columnwidth]{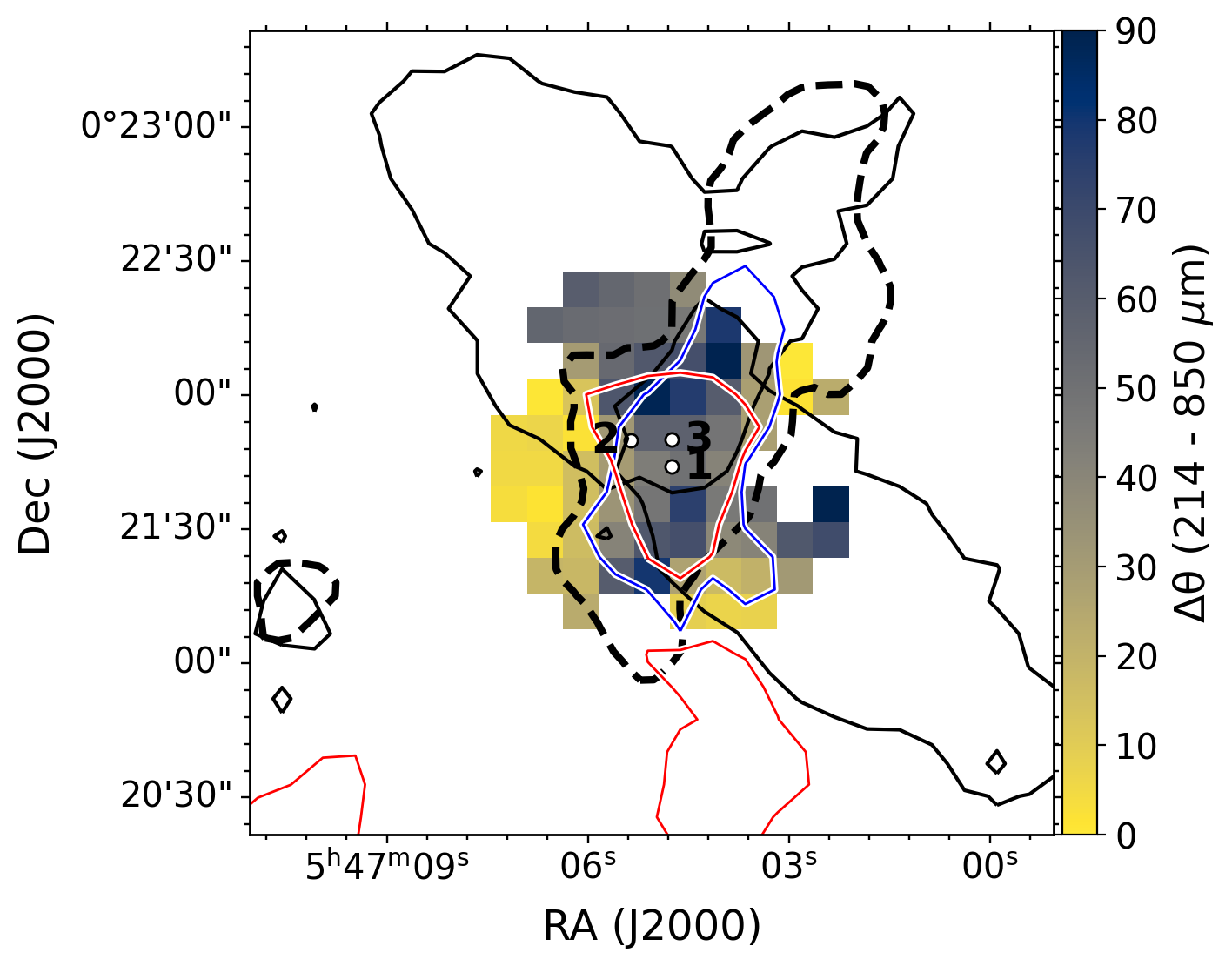}
\caption{
Map of the differences in polarization angles ($\Delta \theta$) between different bands, in absolute value. \textbf{Top:} absolute value of $\Delta \theta_{154-214}$; \textbf{center:} absolute value of $\Delta \theta_{154-850}$, \textbf{bottom:} absolute value of $\Delta \theta_{214-850}$. The dashed lines show the contour for $N_{\rm H_{2}} = 5 \times 10^{22}\, {\rm cm}^{-2}$ from HGBS; solid black contours show $^{12}$CO emission; red (blue) contours show redshifted (blueshifted) C$^{18}$O emission (adapted from Lyo et al. 2021); 
white dots mark the position of IRS 1, 2 and 3. Compare Figure 5 in the main paper.
}
\label{fig:theta-abs_diff}
\end{figure}

\begin{figure}
\includegraphics[width=\columnwidth]{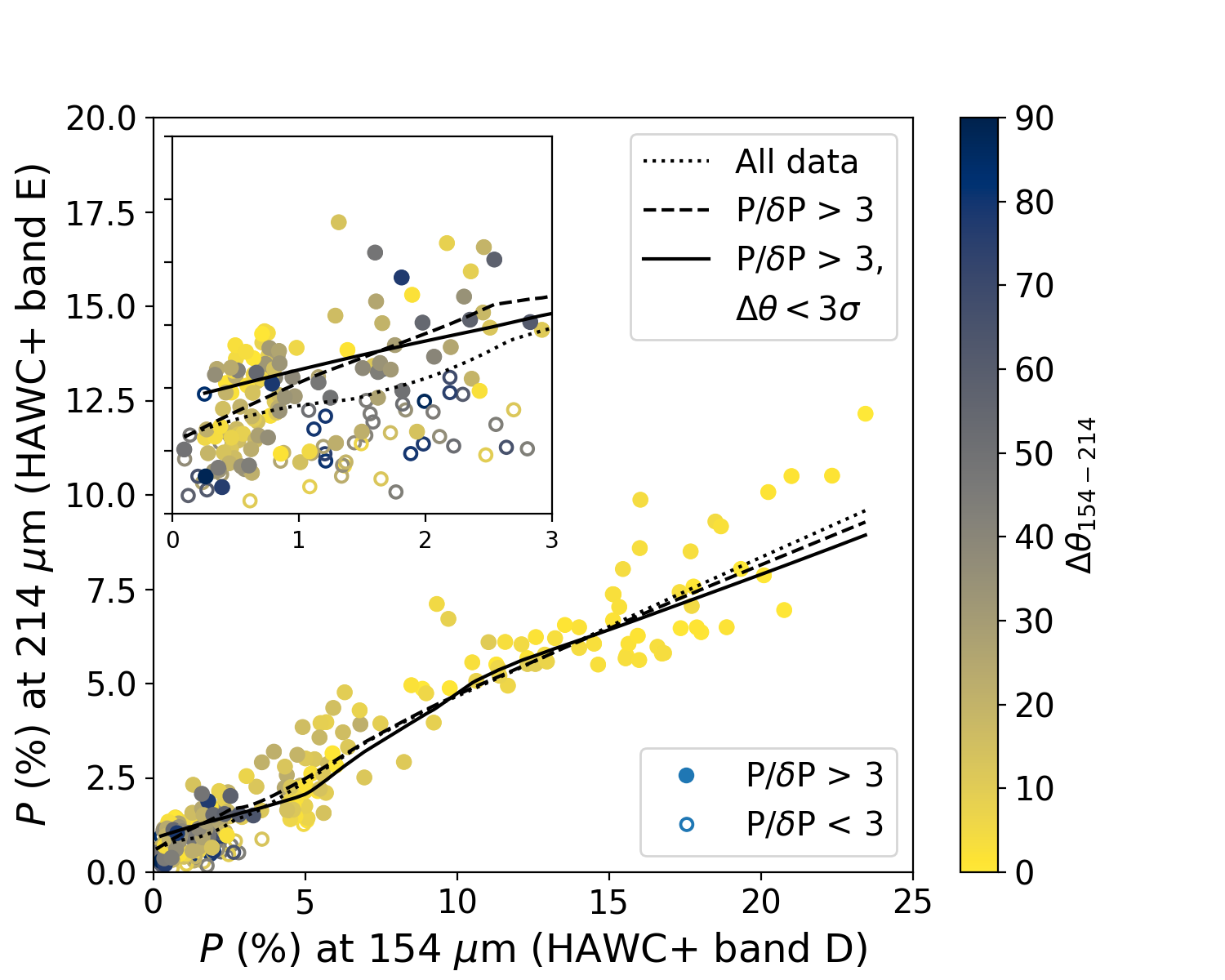}
\includegraphics[width=\columnwidth]{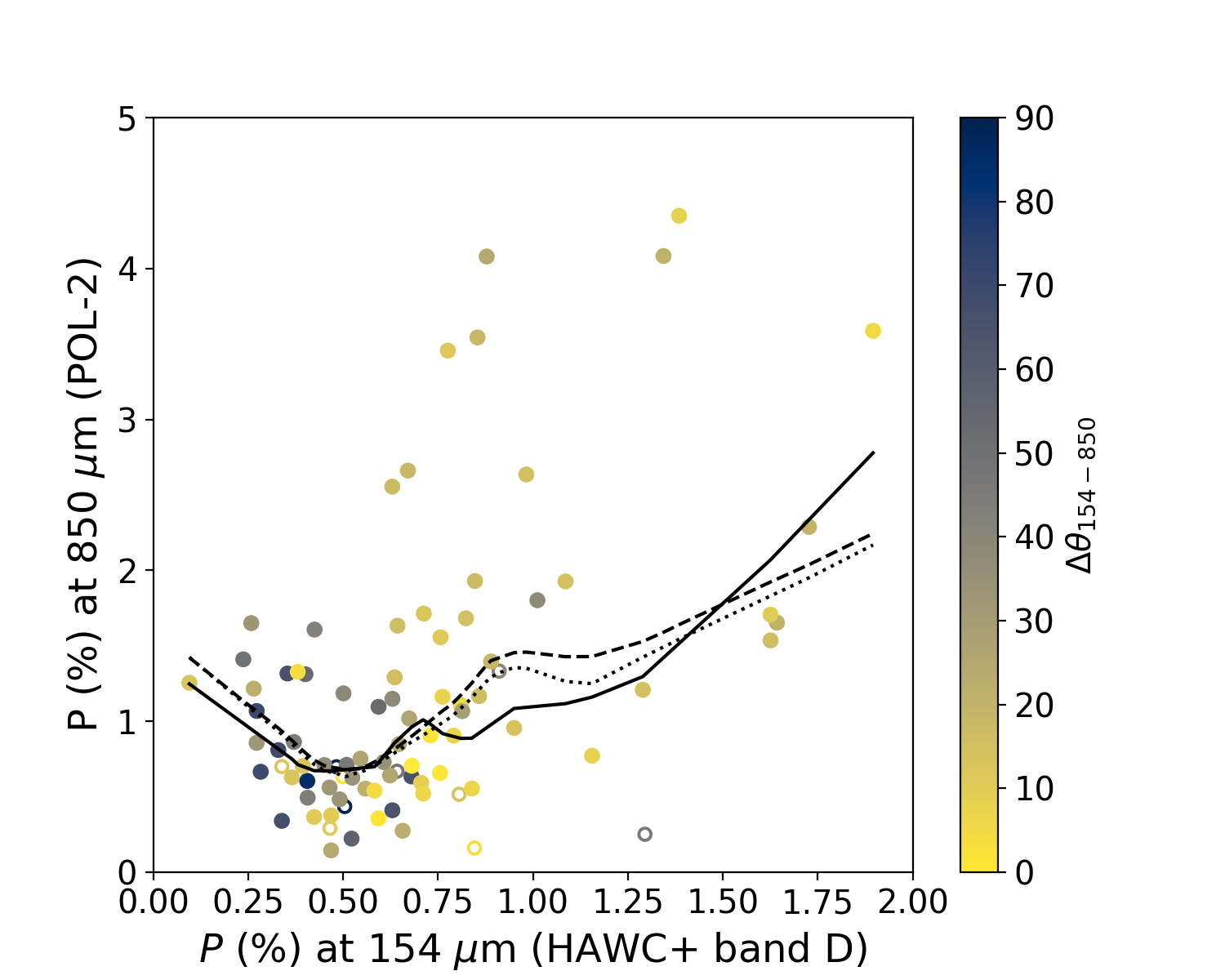}
\includegraphics[width=\columnwidth]{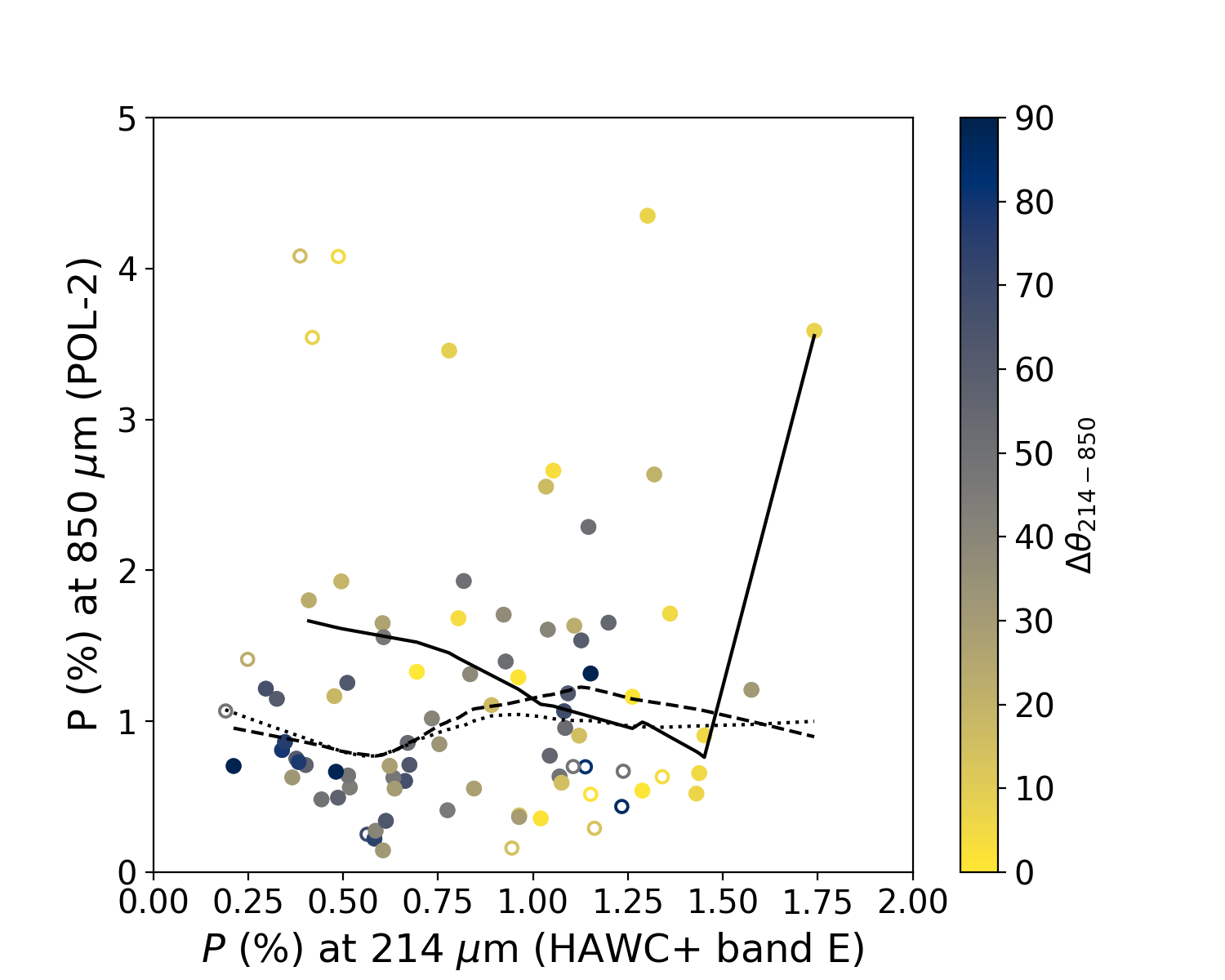}
\caption{
Comparison of polarization fraction $P$ between 154 and 214 $\mu$m (top), between 154 and 850~$\mu$m (middle) and between 214 and 850~$\mu$m (bottom), color coded by the absolute value of $\Delta \theta$ (see Section 3.2 in the main paper).
Lines show the overall trends as traced by LOWESS-smoothed data. Enlargements of the bottom-left corners are shown in an inset for the 154 and 214 $\mu$m case. Compare Figure 6 in the main paper.
}
\label{fig:PvsP_absDtheta}
\end{figure}

The data analysis in the main body of the paper shows angle differences ($\Delta \theta$) within the range ($-90^\circ-90^\circ$), with the sign providing information on which angle is larger. In this section, we reproduce key figures from the article using the range ($0^\circ-90^\circ$) instead, i.e. using only information on the absolute difference between angles. Figure \ref{fig:theta-abs_compare} shows the distribution of $\Delta \theta_{154-214}$, $\Delta \theta_{154-850}$ and $\Delta \theta_{214-850}$. 
Figure~\ref{fig:theta-abs_diff} shows maps of the various $\Delta \theta$ in the central 3$\arcmin$, with CO emission and a column density contour overplotted for comparison. Finally, Figure~\ref{fig:PvsP_absDtheta} shows all $P$ vs. $P$ plots.





\bsp	
\label{lastpage}